\documentclass{egpubl}
 \electronicVersion 

\ifpdf \usepackage[pdftex]{graphicx} \pdfcompresslevel=9
\else \usepackage[dvips]{graphicx} \fi

\PrintedOrElectronic

\usepackage{t1enc,dfadobe}

\usepackage{egweblnk}
\usepackage{cite}
\usepackage{listings}
\usepackage{tabularx}
\usepackage{booktabs}
\usepackage{color, colortbl}
\usepackage{multirow}
\usepackage{scalerel}
\usepackage{makecell}
\usepackage{xspace}

\usepackage[final]{review}
\definecolor{Gray}{gray}{0.95}
\definecolor{White}{rgb}{1,1,1}
\newcolumntype{g}{>{\columncolor{Gray}}c}
\newcolumntype{h}{>{\columncolor{Gray}}X}
\newcolumntype{i}{>{\columncolor{Gray}}l}

\title[Tasks, Techniques, and Tools for Genomic Data Visualization]{Tasks, Techniques, and Tools for Genomic Data Visualization \vspace{-0.8cm}}      

\author[S. Nusrat, T. Harbig, N. Gehlenborg]
{\parbox{\textwidth}{\centering Sabrina Nusrat$^{1,*}$ 
         Theresa Harbig$^{1,*}$ 
        and Nils Gehlenborg$^{1}$ 
        }
        \\
{\parbox{\textwidth}{\centering $^1$Department of Biomedical Informatics, Harvard Medical School, Boston, MA, USA; 
        $^{*}$Authors contributed equally to this work
       }
}
}

\begin{document}
\newcommand{\linear}{{\color{blue}\textbf{---}}}
\newcommand{\circular}{{\color{blue}$\bigcirc$}} 
\newcommand{\spatial}{{\color{blue}$\varphi$}}
\newcommand{\sf}{{\color{blue}$\bigsqcup$}} 
\newcommand{\para}{{\color{blue}$=\joinrel=$}} 
\newcommand{\ser}{{\color{blue}\textbf{-- --}}} 
\newcommand{\ortho}{{\color{blue}$|\_\_$}} 
\newcommand{\task}[1]{\textsc{#1}}

\newcommand{\symbl}[1]{\scalerel*{\includegraphics{Root/figures/tiles/#1.pdf}}{)}}
\newcommand{\symb}[1]{\protect\symbl{#1}}
\newcommand{\symbsm}[1]{\scalerel*{\includegraphics{Root/figures/tiles/#1.pdf}}{B}}

\newcommand{\tsummarize}{\symbsm{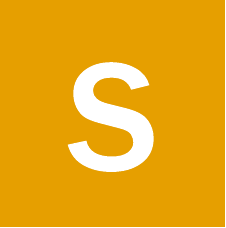}~\textsc{summarize}\xspace}
\newcommand{\tbrowse}{\symbsm{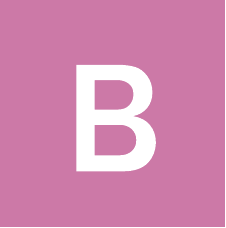}~\textsc{browse}\xspace}
\newcommand{\tlookup}{\symbsm{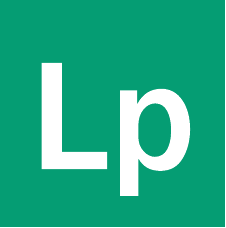}~\textsc{lookup}\xspace}
\newcommand{\tidentify}{\symbsm{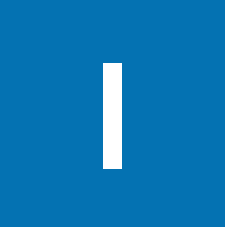}~\textsc{identify}\xspace}
\newcommand{\texplore}{\symbsm{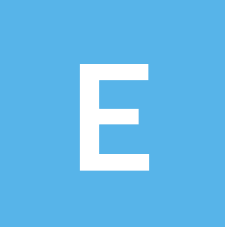}~\textsc{explore}\xspace}
\newcommand{\tlocate}{\symbsm{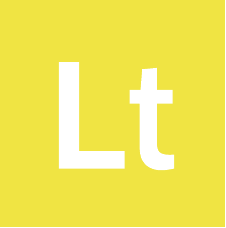}~\textsc{locate}\xspace}
\newcommand{\tcompare}{\symbsm{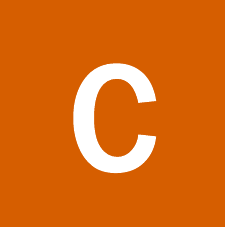}~\textsc{compare}\xspace}

 \teaser{
    \centering
    \includegraphics[width=\textwidth]{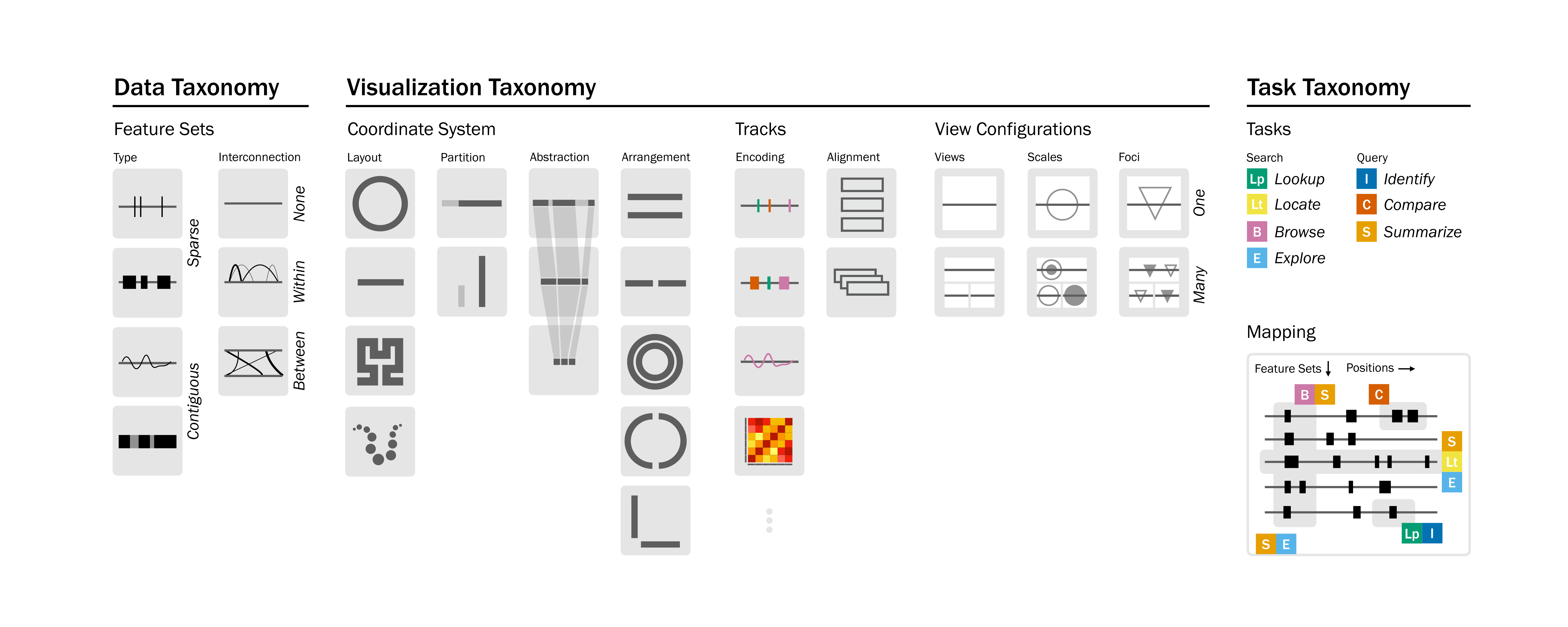}
    \caption{Data taxonomy, visualization taxonomy, and task taxonomy for genomic visualizations. The data taxonomy describes how different genomic data types can be encoded as \textbf{feature sets}. A genomic visualization contains one or multiple \textbf{coordinate systems} applying a specific \textbf{layout}, \textbf{partition}, \textbf{abstraction} and \textbf{arrangement} (in case of multiple axes) of sequence coordinates. Feature sets are encoded as \textbf{tracks} and placed on the coordinate systems. Multiple tracks can either be aligned by stacking or overlaying. A visualization can consist of one or multiple \textbf{views}, each containing a set of aligned tracks. Multiple views can show data on one or multiple \textbf{scales} and \textbf{foci}. The task taxonomy explains how different \textbf{search} and \textbf{query} tasks operate on genomic visualizations.}
    \label{fig:Overview}
}

\maketitle
\begin{abstract}
Genomic data visualization is essential for interpretation and hypothesis generation as well as a valuable aid in communicating discoveries. Visual tools bridge the gap between algorithmic approaches and the cognitive skills of investigators. Addressing this need has become crucial in genomics, as biomedical research is increasingly data-driven and many studies lack well-defined hypotheses. A key challenge in data-driven research is to discover unexpected patterns and to formulate hypotheses in an unbiased manner in vast amounts of genomic and other associated data. Over the past two decades, this has driven the development of numerous data visualization techniques and tools for visualizing genomic data. Based on a comprehensive literature survey, we propose taxonomies for data, visualization, and tasks involved in genomic data visualization. Furthermore, we provide a comprehensive review of published genomic visualization tools in the context of the proposed taxonomies.

\end{abstract}  
\section{Introduction}
A rapidly growing understanding of how the genome and epigenome of an organism control molecular function and cellular processes has revolutionized research in biology and medicine. Driven by affordable high-throughput technologies that allow scientists and clinicians to reliably obtain high quality sequence information from DNA and RNA molecules, generation and handling of genomic sequencing data are now routine aspects of many basic science and clinical research projects in biology and medicine. 

While a large amount of genomic data is produced within individual small scale projects, there are also numerous national and international efforts to generate genomic data at large scale. These kinds of projects include efforts to catalog genomic features across cell types and tissues (e.g. ENCODE Consortium~\cite{ENCODE_Project_Consortium2012-rq}), studies aimed at understanding fundamental principles of DNA architecture (e.g. 4D Nucleome Project~\cite{Dekker2017-ct}), as well as disease specific studies that aim to elucidate the molecular changes that cause diseases such as cancer (e.g. The Cancer Genome Atlas~\cite{Cancer_Genome_Atlas_Research_Network2013-mm}, International Cancer Genome Consortium~\cite{The_International_Cancer_Genome_Consortium2010-zr}). Yet other projects aim to develop new approaches for early identification of genetic risk factors (e.g. BabySeq~\cite{Holm2018-vx}). Many of these projects do not only generate genomic data for many samples, but also produce dozens of different types of genomic data. 

Visualization of genomic data is frequently employed in biomedical research to access knowledge within a genomic context, to communicate, and to explore datasets for hypothesis generation. As biomedical research is increasingly data-driven and many studies lack well-defined hypotheses~\cite{Weinberg2010-hx, Golub2010-xn}, it is a key challenge to discover unexpected patterns and to formulate questions in an unbiased manner in vast amounts of genomic and other associated data.

Over the last two decades, hundreds of visualization tools for genomic data have been published. The large number of tools are an indicator for the broad application of genomic data and a sign that visualization of genomic data is a complex problem and active research domain. 
 
Several challenges in visualizing genomic data are directly connected to how genomes are organized. Genomes are collections of one or more chromosomes, which are individual molecules that encode information as a sequence of nucleotides. Although genomic information is stored in the form of a sequence, the function of the genome is influenced by and requires various types of long- and short-range interactions between non-adjacent regions of the sequence. This includes interactions within and between chromosomes. Patterns in genomic data can be found across many scales, ranging from the size of whole chromosomes, which can span hundreds of millions of nucleotides, down to individual nucleotides. Another important aspect of many genomes is the sparse distribution of many types of patterns along the genome sequence.

Furthermore, the questions that are addressed with genomic data are aimed at the understanding of complex biological systems where all components are highly interconnected and influence each other. For example, the regulation of gene activity is controlled by the presence or absence of particular regulatory proteins, chemical modification of parts of the DNA, and the 3D structure of chromosomes, all of which are changing depending on environmental and other factors. An abundance of proteins, chemical modifications, and 3D structures can be measured comprehensively and mapped to the genome. While this is a greatly simplified view, it shows the diversity and number of data types from multiple sources that often need to be integrated into visualization in order to interpret genomic information. 

The combination of long sequences, sparse distribution of patterns across multiple scales, interactions between distant parts of the sequence, and large numbers of diverse data types pose numerous visualization challenges. \new{These require the design and development of specialized tools.} Additionally, the number of features, the size of the datasets, and the diversity of data types all make tight integration of genomic data visualization tools with algorithmic tools a requirement for efficient analysis workflows. This further complicates the design of effective visualization tools for genomic data.

As the sequential organization is a key characteristic of genomic data, we limit the scope of this survey to visualizations that incorporate one or more genomic coordinate systems and present data in the order defined by the sequence of that coordinate system. This explicitly excludes many techniques that are based on reorderable matrices and node-link diagram approaches such as visualization of gene expression levels as matrix-based, clustered heatmaps or visualization of gene regulatory networks as node-link diagrams with expression data mapped overlaid onto the nodes. Since the presence of a genome sequence is required for inclusion in this survey, we also excluded tools for genome assembly, which is the process of defining the sequence of a reference genome for a given species.

Our survey makes two major contributions: \new{In Section \ref{sec:taxonomy}}, we propose taxonomies for data, visualization, and tasks involved in genomic data visualization. \new{In Sections \ref{sec:singleSequence} and \ref{sec:MultiSequence}}, we provide a comprehensive review of published genomic visualization tools in the context of the proposed taxonomies. In addition, we discuss current challenges and research opportunities in genomic data visualization.


\section{Biological Background}\label{sec:background}
\subsection{DNA Structure}
\textit{Deoxyribonucleic acid} (DNA) is a molecule contained in living cells carrying the genetic instructions to build every biological function and molecule of an organism. DNA is studied for numerous reasons, including analyzing cancerous DNA to find treatments, finding possible risk factors for certain diseases, and comparing DNA of different species to study them in the context of evolution.

DNA consists of two complementary strands coiled up in a double helix. Each strand is composed of smaller units called nucleotides, each consisting of a base (either \textit{Adenine (A)}, \textit{Cytosine (C)}, \textit{Guanine (G)}, or \textit{Thymine (T)}), a sugar, and a phosphate group.  Biological information is stored in the order of the different nucleotides. The two strands, called forward and reverse strands, are connected at the bases. They are called complementary because an A in the forward strand corresponds to a T in the reverse strand as well as G corresponds to C and vice versa. Therefore, often only one of the strands is considered when analyzing or visualizing genomic data.
In prokaryotes (bacteria and archaea) DNA is organized in one single circular sequence, which is called a chromosome, while eukaryotic DNA is usually organized in multiple chromosomes (multiple sequences).  

\begin{figure}
    \centering
    \includegraphics[width=\linewidth]{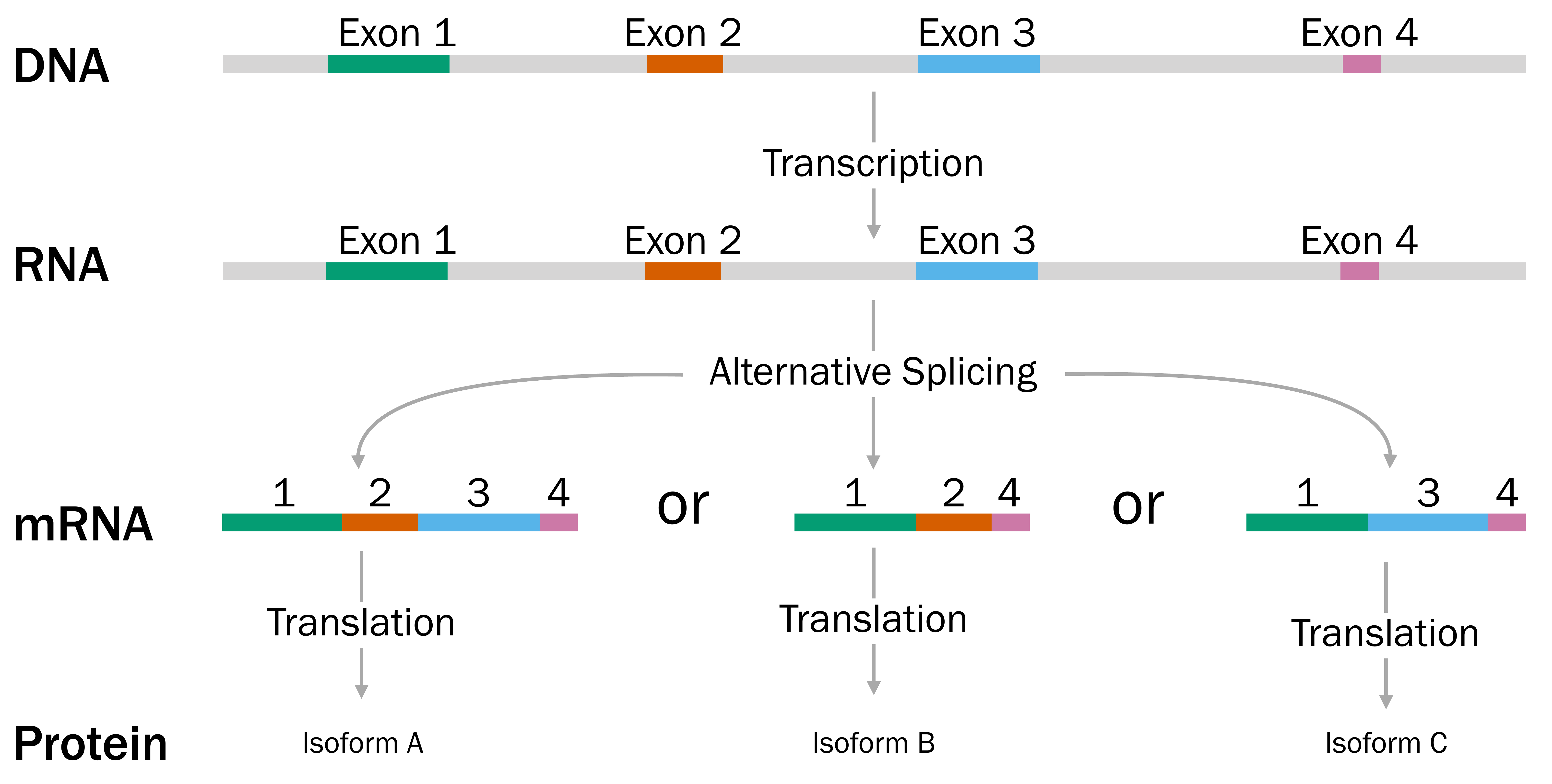}
    \caption{Genes on a DNA molecule are transcribed onto a mRNA molecule and translated into amino acids to form the final protein product. A gene consists of coding parts (exons) and non-coding parts (introns). In a process called alternative splicing exons can be combined in various ways to form different protein products.}
    \label{fig:alternativeSplicing}
\end{figure}

A gene is a sequence of nucleotides encoding for a protein, a molecule which has a biological function in the organism such as catalyzing reactions (as an enzyme) or being a building block of a tissue. In order to build a protein using the information of a gene, the DNA has to be transcribed to mRNA and translated into a protein (see Figure \ref{fig:alternativeSplicing}). The process of transcribing genes into mRNA is called gene expression. The rate at which a gene is expressed (and translated into a protein) is not the same at all times but depends on many different regulatory factors, such as molecules called \textit{transcription factors}. Certain sequences of the RNA molecule can initiate the transcription of a gene, called promotors, which are located upstream of the gene sequence. Transcription factors can bind to sequences in the promotor region to regulate gene expression (see Figure \ref{fig:epigenetics}).

In eukaryotic organisms, not the entire DNA sequence encodes for genes. Instead, the sequence consists of protein-coding parts called exons and non-protein-coding parts called introns (see Figure \ref{fig:alternativeSplicing}, top). During transcription, introns are cut out and neighboring exons are combined to form genes. In a process called alternative splicing, one gene can encode for multiple different protein products, called protein isoforms by combining different exons (see Figure \ref{fig:alternativeSplicing}, middle). The exons not needed for the formation of the protein product are spliced out. Knowledge about the abundance of isoforms is important for biologists to understand both normal processes and diseases in order to eventually improve treatment through targeted therapies.

\new{Every triplet of nucleotides of the mRNA molecule encodes for one amino acid, the basic building block of a protein. Each triplet is called a \textit{codon}.  There are 64 different codons encoding for 20 amino acids and three codons signaling a stop of translation.} 

\subsection{Mutations}
When a cell divides, each daughter cell receives a copy of the cell's DNA. This process requires copying the DNA, which can lead to errors. Moreover, DNA can be damaged by external factors such as radiation or carcinogens. Errors can be divided into small-scale and large-scale mutations. Small scale mutations include the insertion or deletion of one or multiple nucleotides and substitutions of  single nucleotides. Substitution mutations can alter an amino acid in the resulting protein or a premature termination of transcription. Insertions and deletions can change all the triplets succeeding the mutation, which often leads to a completely altered protein product.

Large-scale mutations include amplifications and deletions of entire regions on a chromosome, which can lead to an increased dosage of genes in these regions or the loss of genes \new{called \textit{copy number variation}}. Furthermore, parts of separate genes can be fused together to form a fusion gene. Another type of large-scale mutations is chromosomal rearrangement. For example, parts of DNA can be exchanged between non-homologous chromosomes or the orientation of chromosomal segments can be inverted. 
\subsection{Sequencing}
Sequencing is the process of determining the sequence of nucleotides of a DNA or RNA molecule. With most of the current techniques it is not possible to sequence an entire genome at once, but the sequence has to be broken down into little pieces which are sequenced separately \new{called sequencing reads}. To reconstruct the entire genome, DNA sequencing reads have to be puzzled together by using overlaps at the end of the reads and often an already sequenced genome (called a reference genome). The process of reconstructing the sequence from sequencing reads is called \textit{assembly}. 

DNA sequencing data is then further analyzed to find mutations or structural rearrangements or for the comparisons to other species or individuals. RNA sequencing data is usually not assembled but used to determine expression levels or patterns of alternative splicing by mapping the RNA sequencing reads to the DNA sequence.
\subsection{Alignment}\label{sec:alignement}
Sequence alignment is one of the most important operations performed in the analysis of genetic information. It is often used to find functional or evolutionary relationships between sequences stemming from different individuals/species.

\begin{figure}[t]
    \centering
    \includegraphics[width=\linewidth]{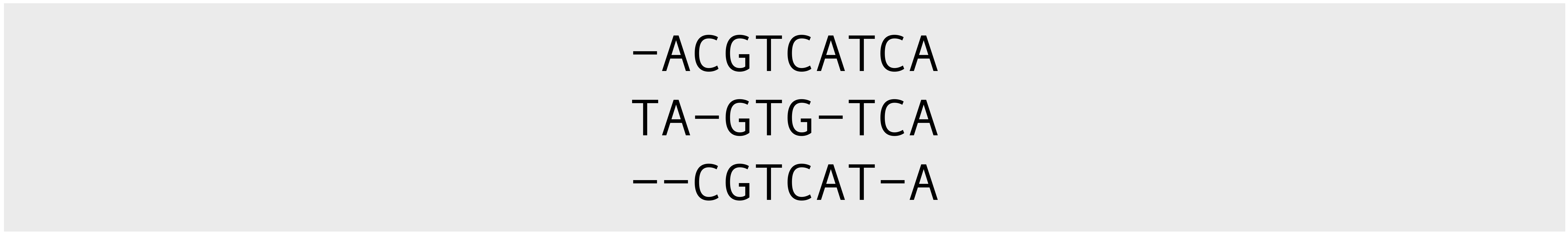}
    \caption{A multiple sequence alignment (MSA) of the sequences `ACGTCATCA', `TAGTGTCA' and `CGTCATA'.}
    \label{fig:msa-example}
\end{figure}

In order to align multiple sequences, they have to be arranged in a way that makes it possible to identify similar regions. A good alignment maximizes the number of shared symbols in one column while minimizing gaps and non-matching symbols (mismatches) and retaining the sequence order. Figure~\ref{fig:msa-example} shows a simple example of a \textit{multiple sequence alignment (MSA)}.

\subsection{Epigenetics}
Monozygotic twins are genetically identical. However, especially older twins often show significant differences in their appearance and they sometimes even have acquired individual diseases. While not all factors for this phenomenon are understood, epigenetic differences have been identified as correlating with different phenotypes in twins \cite{Haque2009-se,Fraga2005-fk}. Through epigenetic processes, genes can be turned on and off without altering the genetic sequence, often influenced by environmental factors or stochastic processes.

\begin{figure}
    \centering
    \includegraphics[width=0.9\linewidth]{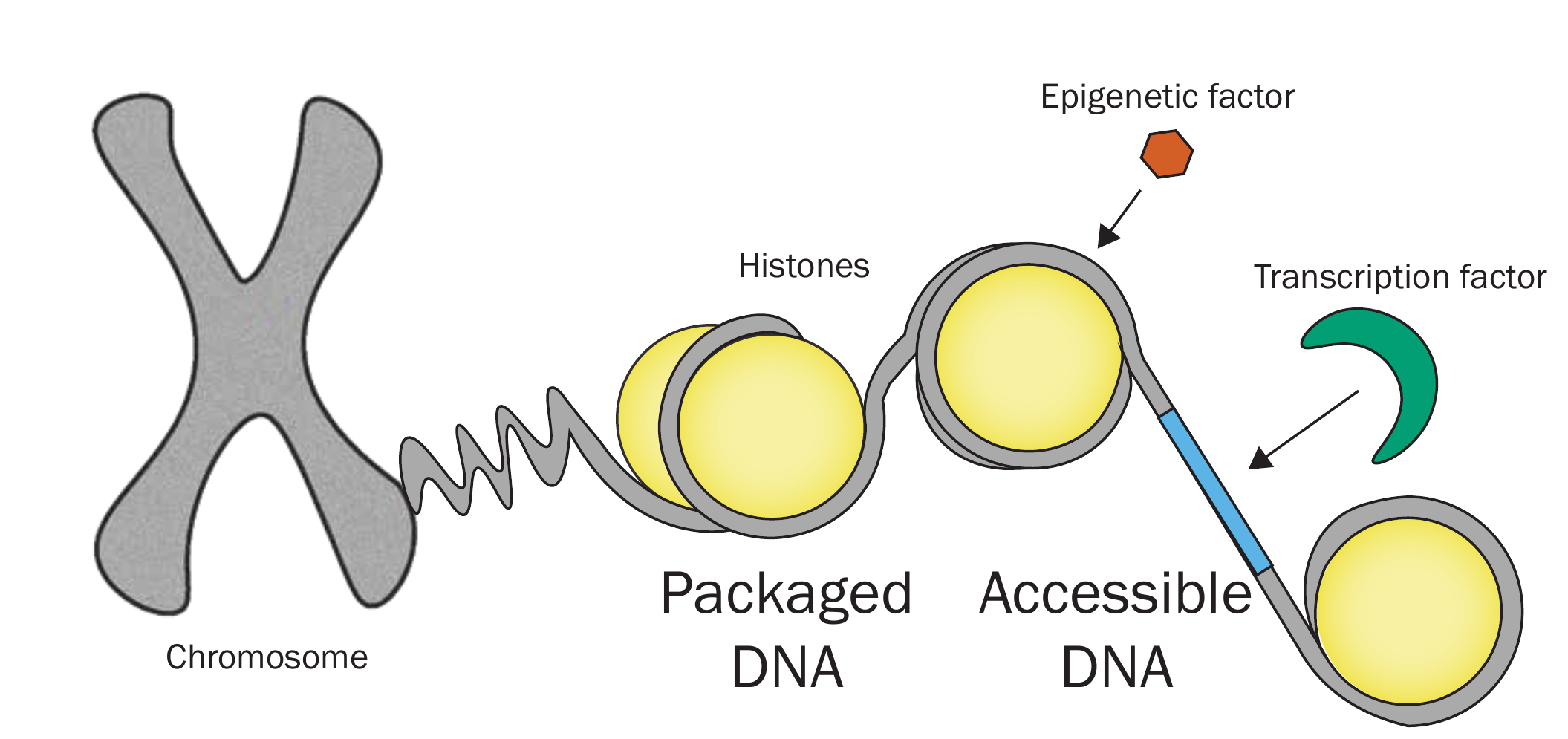}
    \caption{DNA is organized in histones. When the DNA is wound up around the histones transcription cannot occur. When it is unwound transcription factors can bind and initiate gene expression. Epigenetic factors control to which extent the DNA is wound up.}
    \label{fig:epigenetics}
\end{figure}

In order to compact and organize chromosomes, eukaryotic DNA is wound up around proteins called histones. Compacted DNA cannot be transcribed since it is not accessible for transcription factors (See Figure \ref{fig:epigenetics}). Epigenetic processes such as DNA methylation or histone modification can control the extent to which DNA is wound up and unwind parts of it to make it accessible to transcription factors. Therefore, expression and protein synthesis can be controlled without changing the underlying DNA sequence. These changes can be inherited by daughter cells, as well as by the descendants of the organism. Since the described changes do not affect the DNA sequence, they are called epigenetic modifications (from Greek epi meaning ``over, outside of, around'').
\subsection{Chromosome Conformation Capture}
Another aspect to consider when studying different phenotypes is the 3D structure of DNA. Nucleotides separated by many positions in the sequence can be in very close proximity in 3D space, as indicated by Figure \ref{fig:epigenetics}. Nucleotides can be close in 3D space when they are wound up around the same histone, or when they are part of a loop that controls transcription.

With chromosome confirmation capture techniques, the interactions of genomic loci in 3D space can be quantified. One way of doing this is measuring the interactions between fragments of the genome. With the Hi-C technique the interactions between all possible non-overlapping fragments of a genome can be determined resulting in a matrix of interaction frequencies (also called contact frequencies). The higher the interaction frequency of two fragments, the smaller the distance between them in 3D space.
\subsection{Genome Evolution}
Environmental pressures during the evolution of species lead to changes in sequence and composition of the species' genomes. Not every region of the genome changes at the same rate. Sequences that have a function in fundamental processes (genes, exons) are more similar or more \textit{conserved} across different species. Moreover, the order, orientation and location of subsequences can change over time. A gene shared between two species can be at different chromosomes in a different genomic neighborhood. 
Genomic \textit{synteny} refers to the order of conserved blocks within two sets of chromosomes that are being compared. Two sequences are called syntenic if they contain similar blocks of genes in the same relative positions in the compared genomes.

\new{\subsection{Previous Literature Surveys}
The following literature surveys focusing on genomic data visualization are all aimed at audiences in the bioinformatics and biology research communities. Therefore, their review of visualization tools is often more focused on features available to the users than on formalizing the description of common tasks and techniques. }

\new{Nielsen et al.~\cite{Nielsen2010-by} reviews the techniques and challenges in visualizing genomes with a focus on three core user tasks: (i) analyzing sequence data, both in the context of de novo assembly and of re-sequencing experiments, (ii) browsing annotations and experimental data mapped to a reference genome and, finally, (iii) comparing sequences from different organisms or individuals. They review several stand-alone and web-based tools and compare their cost, operating systems, and compatibility. Despite technical advancement, several challenges for analysis and visualization of genomics data remain due to the volume and heterogeneity of these data. The authors also recommend ways of improving the design of these visualization tools. For example, a high-level overview of data, or providing recommendations for where to look at, can improve user efficiency. In addition, Nielsen et al. suggest that new genome browsers should build on the successes of earlier tools, allowing easy cross-platform access, customization of data and display, and the ability to perform on-the-fly computation within the visualization. The authors point out that although several successful visualization tools are used for specialized analysis demands of the users, there is a great need to improve the integration among tools and ease the transition from one analysis to another. }

\new{Schroeder et al.~\cite{Schroeder2013-rg} review common visualization techniques for exploring oncogenomics data and compare several existing tools. They describe genomic coordinates that help researchers find answers to questions about alterations tied to genomic loci, or to inspect some particular genomic locus. Heatmaps are frequently used to describe transcriptomics and genomics data stored in the form of matrices. Node-link diagrams are used to visualize functional relationships between different entities, such as genes. 
Qu et al.~\cite{Qu2019-dz} also review visualization methods for oncogenomic data, such as scatterplots, networks, heatmaps, clusters and the combination of machine learning and visualization. Moreover, they discuss future trends in this field.
Pavlopoulos et al.~\cite{Pavlopoulos2015-vf} conducted a comprehensive review of general genome visualization tools and summarize them into four categories: genome alignment visualization tools, genome assembly visualization tools, genome browsers, and tools to directly compare different genomes with each other for efficient detection of genomic variants.}

\new{Yardimci et al.~\cite{Yardimci2017-ln} reviewed five visualization tools for genomic interaction data generated using chromosome conformation capture approaches. They characterized the visualization functionality of those tools based on available visualization types and also discussed integration of supplementary views and data handling capabilities. They categorized visualization techniques based on whether they are more suitable for short-range interactions or long-range interactions. Goodstadt et al.~\cite{Goodstadt2017-zc} reviewed visualization challenges for 3D genome architecture and provide a taxonomy of tasks outlining essential features
of 3D genome visualization. These tasks and challenges including data representation, data refining, and data interaction.}

\new{O'Donoghue et al.~\cite{ODonoghue2018-pg} surveyed how visualization
is being used in a broad range of data-driven biomedical research areas. The authors reported on current visualization techniques and challenges in genomics and epigenetics, RNA biology, protein structures, systems biology, cellular and tissue imaging, and populations and ecosystems. They also pointed out the limitations in popular tools such as the widespread use of rainbow color maps and recommended that tailored visualization methods and tools are necessary for advancements in biomedical insights.}

\new{Unlike the surveys mentioned in the previous section, this survey is aimed at the visualization community and bioinformatics researchers who develop visualization tools and focuses on visualization tasks, techniques, and tools. A secondary goal of this survey is to take a step toward bridging the gap between research in the visualization and bioinformatics communities and to highlight the promising research areas in this emerging cross-disciplinary field. }
\section{Process}

We searched both the PubMed database (\url{https://pubmed.gov}) and Google Scholar (\url{https://scholar.google.com}) for tools related to genomic visualization using the following keywords: ``visualization of genomic data'', ``genome data visualization'', ``genomic sequence visualization'', and ``transcriptome data visualization''. These searches resulted in a seed collection that we considered relevant for this survey based on their titles and abstracts. \new{To this seed collection, we also added paper describing tools and methods that we were familiar with but had not been returned by the search.}

Using this seed collection, we identified several tool categories such as genome browsers, multiple sequence alignment tools, and others, that we used for more focused searches to expand our collection. In the next step, we removed manuscripts that did not fit our scope of sequence-based visualization, resulting in a total of 111 papers. We further removed papers that only marginally mentioned visualization or did not present a tool or technique. We also removed papers that described re-implementations of a particular technique, such as Circos or multiple sequence alignments, if there was no novel aspect to the visualization. In such cases, we focused on those papers with a higher citation count. 

Ultimately, the collection surveyed for this review contains 83 papers describing genomic visualization techniques or tools. Of the 83 surveyed papers, 7 papers (8\%) were published in visualization venues (all IEEE TVCG) and 76  manuscripts (92\%) were published in bioinformatics and biology venues. 

\section{Taxonomy}\label{sec:taxonomy}
In this section, we will provide an abstract understanding of genomic data and the basic parameters and techniques for its visualization. We introduce three taxonomies: A data taxonomy characterizing genomic features, a task taxonomy categorizing the most important tasks for genomic visualizations and a visualization taxonomy which we use to categorize tools.
\subsection{Genomic Features}
\subsubsection{Types of Features}

A genomic feature is a data point of \textit{measured} or \textit{knowledge-based} data that can be mapped to genomic coordinates and has an extent of one or more nucleotides. \new{Knowledge-based data represents the knowledge we have about a genome without conducting new experiments, which includes a reference genome with a known sequence and annotations. For example, gene annotations and functional annotations. The reference genome represents the known sequence of a genome of a species. It is not the genomic sequence of one individual, but it is derived from a group of individuals. An analogy for knowledge-based data can be found in the visualization of geological maps. Usually, geological structures such as mountains and rivers are named or elevations are indicated by numbers, which can be understood similarly to the annotation of a reference genome. In the context of maps, measured data can be for example, traffic data, the size of cities, population data or election results. For genomic data, measured data is anything that can be measured about a genomic sequence or is derived from that measurement. Examples of data that can be measured include the sequence of DNA in a sample, epigenetic signals, and contact frequencies. Derived data is often created by setting measured data in context with the knowledge based data. For example, by comparing sequencing data of a cancerous sample to a reference genome, mutations and genomic rearrangements can be deducted.}

\new{Depending on properties of the underlying data, features can be of different types.} Features that only cover one nucleotide are defined as \textit{point features}, while features covering more than one nucleotide are \textit{segment features}. Features can be associated with zero, one or multiple attributes which can be quantitative, ordinal or categorical. A feature with zero attributes only shows position and extent, such as the position and extent of a gene. If the gene is associated with other data, such as names, functions and expression levels, the feature is associated with multiple attributes.

\subsubsection{Feature Sets}

A set of features belonging to the same biological entity, such as the set of all genes, the set of all expression levels belonging to the same sample or the set of all mutations are called a \textit{feature set}. 
Inspired by a publication by Gundersen et al.~\cite{Gundersen2011-fq}, we discriminate two types of genomic features sets: sparse feature sets (\symb{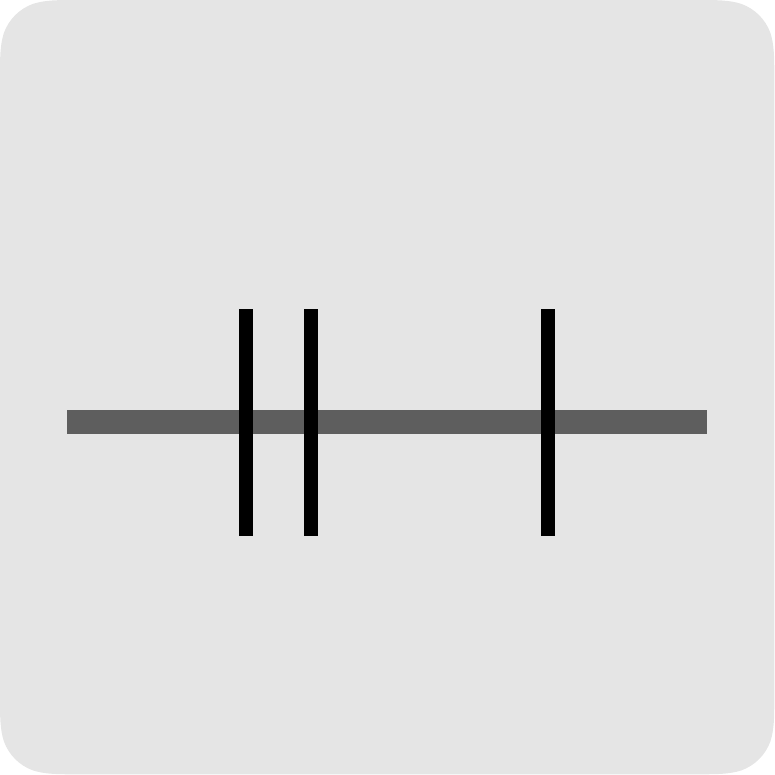},\symb{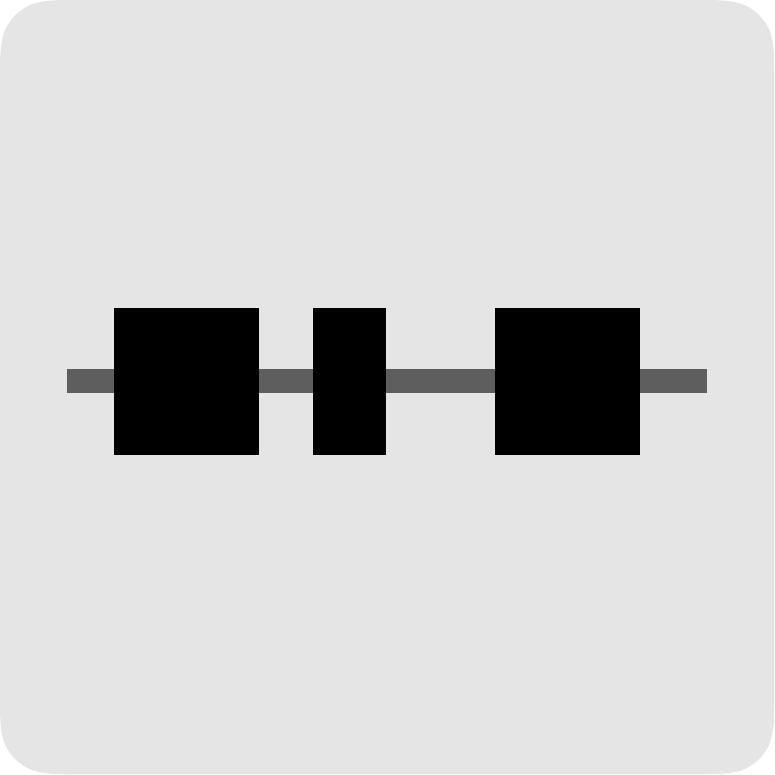}) and contiguous features sets (\symb{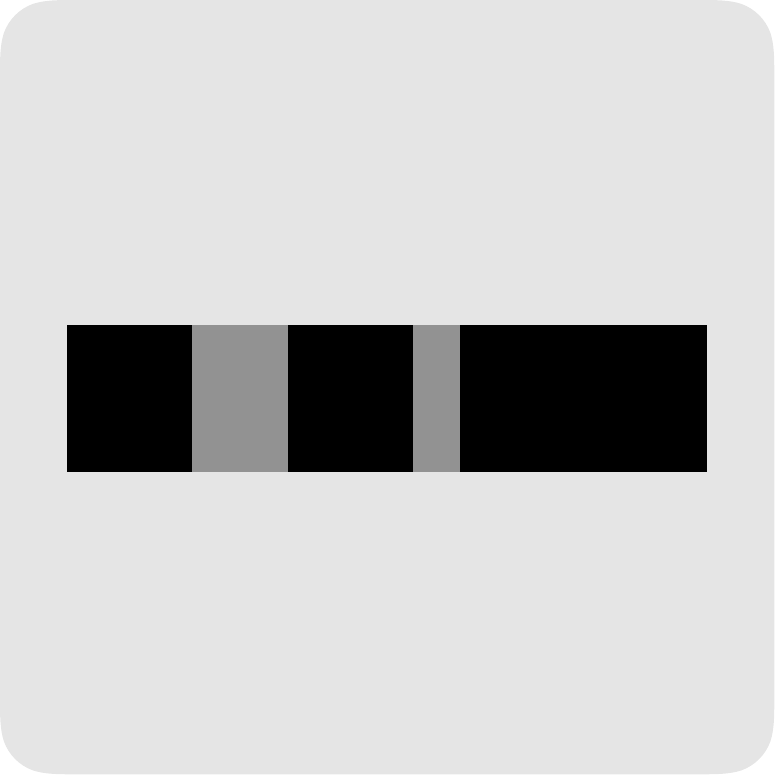},\symb{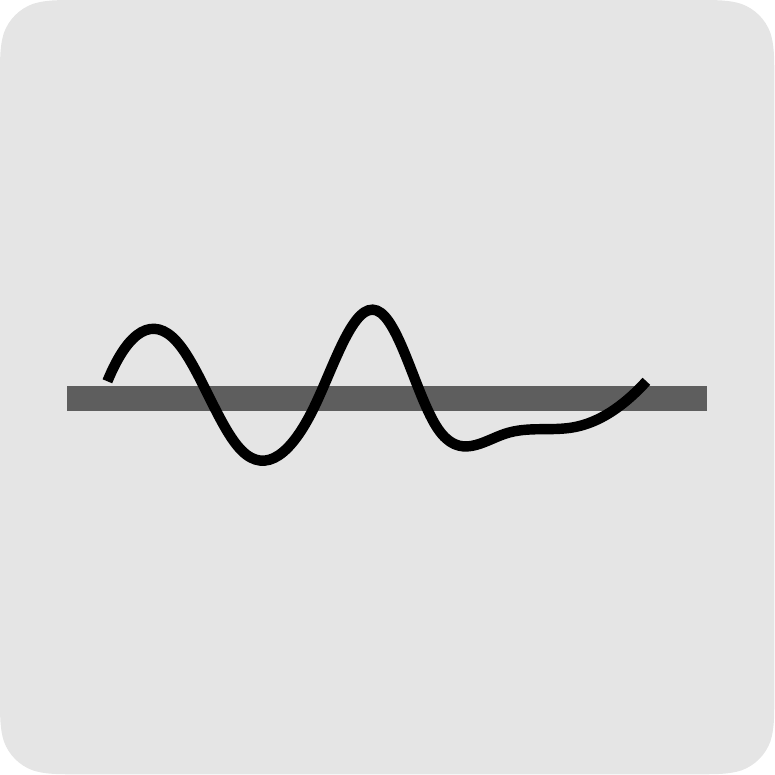}, see Figure~\ref{fig:Overview}, feature sets). While there can be gaps between features in sparse sets, contiguous features sets cover the entire genome. Features in feature sets usually are of the same type, which means they are either point or segment features and are associated with the same attributes. Contiguous feature sets can, for example, encode for the partitioning of a sequence into ``coding'' and ``non-coding'' regions. Also features with ordinal or quantitative attributes can be encoded, such as copy number levels by contiguous sets of features of different extents associated with an ordinal value, or epigenomic data by contiguous sets of non-overlapping features of equal size. 
The reference sequence itself corresponds to a contiguous set of valued point features since each point is associated with a nucleotide.

Feature sets can be combined by intersecting or uniting them. For example, consider the combination of a set of sparse features, such as coding region, with a set of contiguous features encoding for an epigenetic signal across the sequence. Features in the resulting set are associated with two attributes: the epigenetic signal and a Boolean attribute that encodes whether they are in a coding region.

Pairs of features can be connected via \textit{interconnection features}. In a map, an interconnection feature can be understood as a link between regions. For example, a link showing the number of daily transatlantic flights between Europe and the US. In a genomic sequence context, interconnection features can for example encode for chromosome rearrangements: If two segments of two chromosomes are swapped (translocation) it can be encoded with an interconnection feature which links the two subsequences. Like the other features, an interconnection feature can also be associated with attributes, such as the type of the interconnection (``translocation'') or a quantitative value. Interconnection features can connect features within a sequence \symb{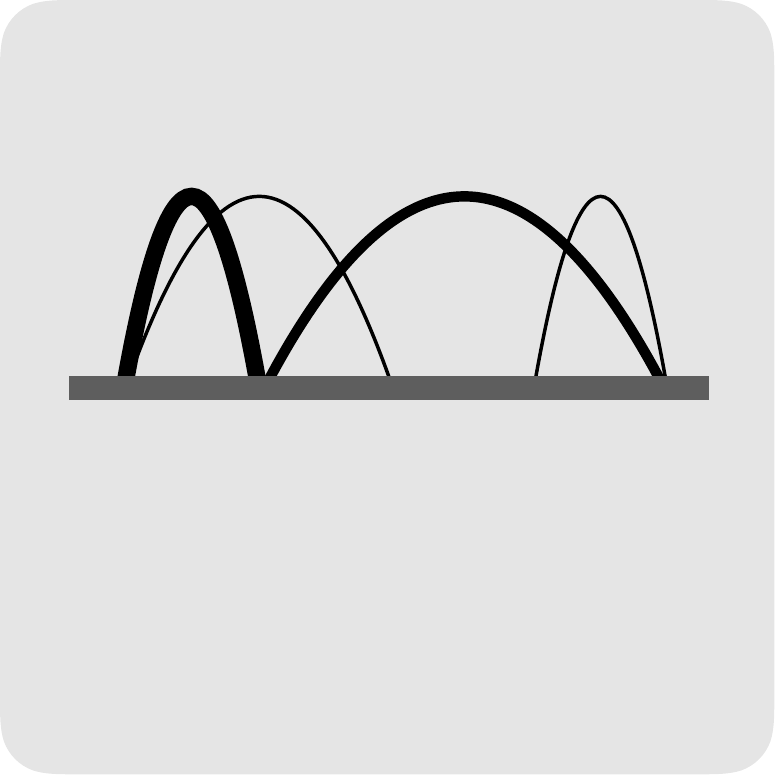} or between sequences \symb{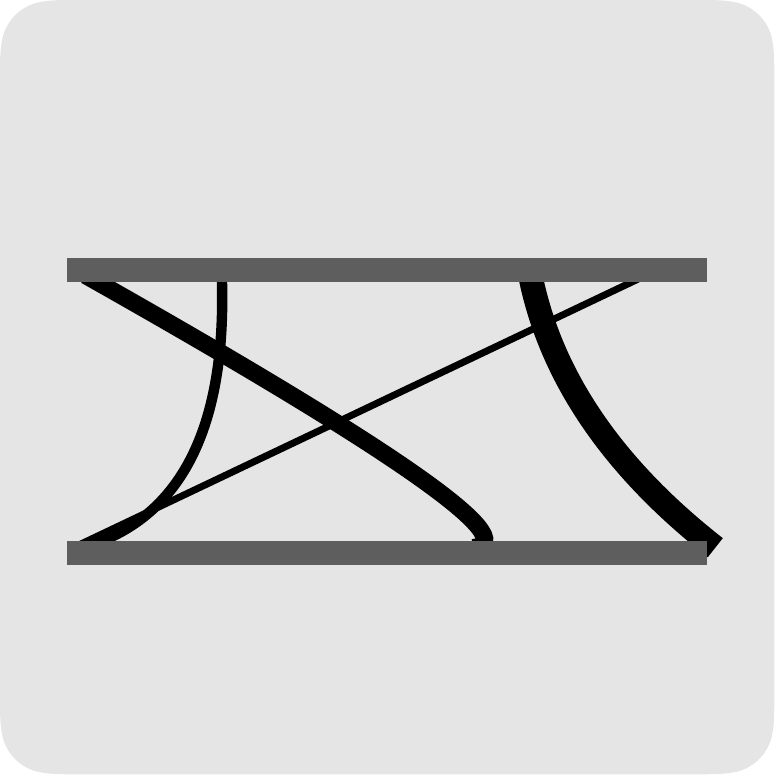} (\symb{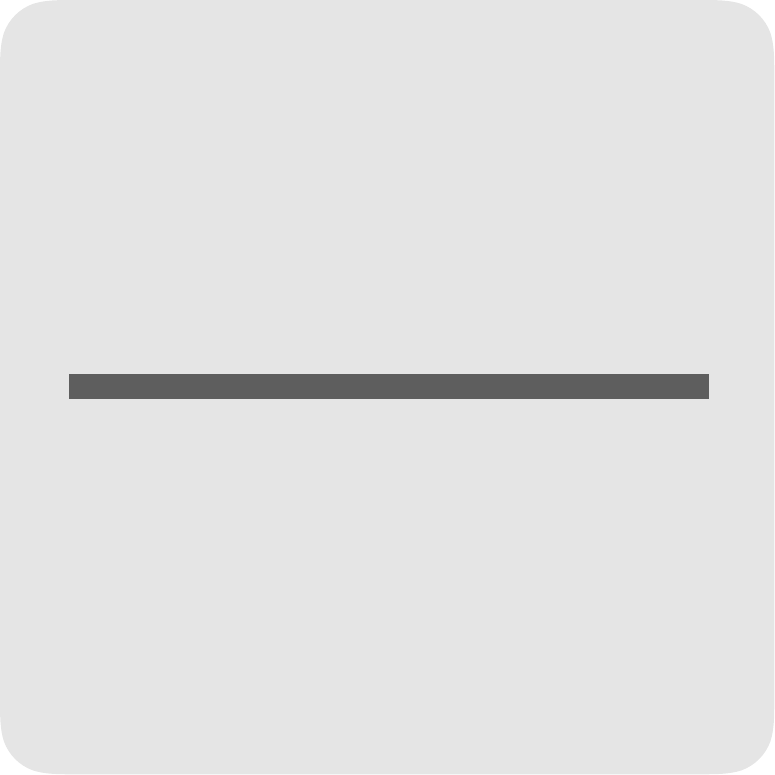} refers to no interconnections).

\subsubsection{Meta data}

\new{Feature sets are usually associated with meta data. For example, a feature set containing expression data of a patient sample can be associated with data about the sample itself and the sample donor, such as the date when the sample was taken, the type of tissue, if the donor has cancer or if he is a smoker. Similarly, in our map example, it could be the date when the map was created or the type of the map. Meta data helps put the data into context. For example, it can help identify possible correlations between different phenotypes and measured data.}

\subsection{Visualization}\label{sec:LAC}

\subsubsection{Sequence Coordinate Systems}
\begin{figure}
    \centering
    \includegraphics[width=\linewidth]{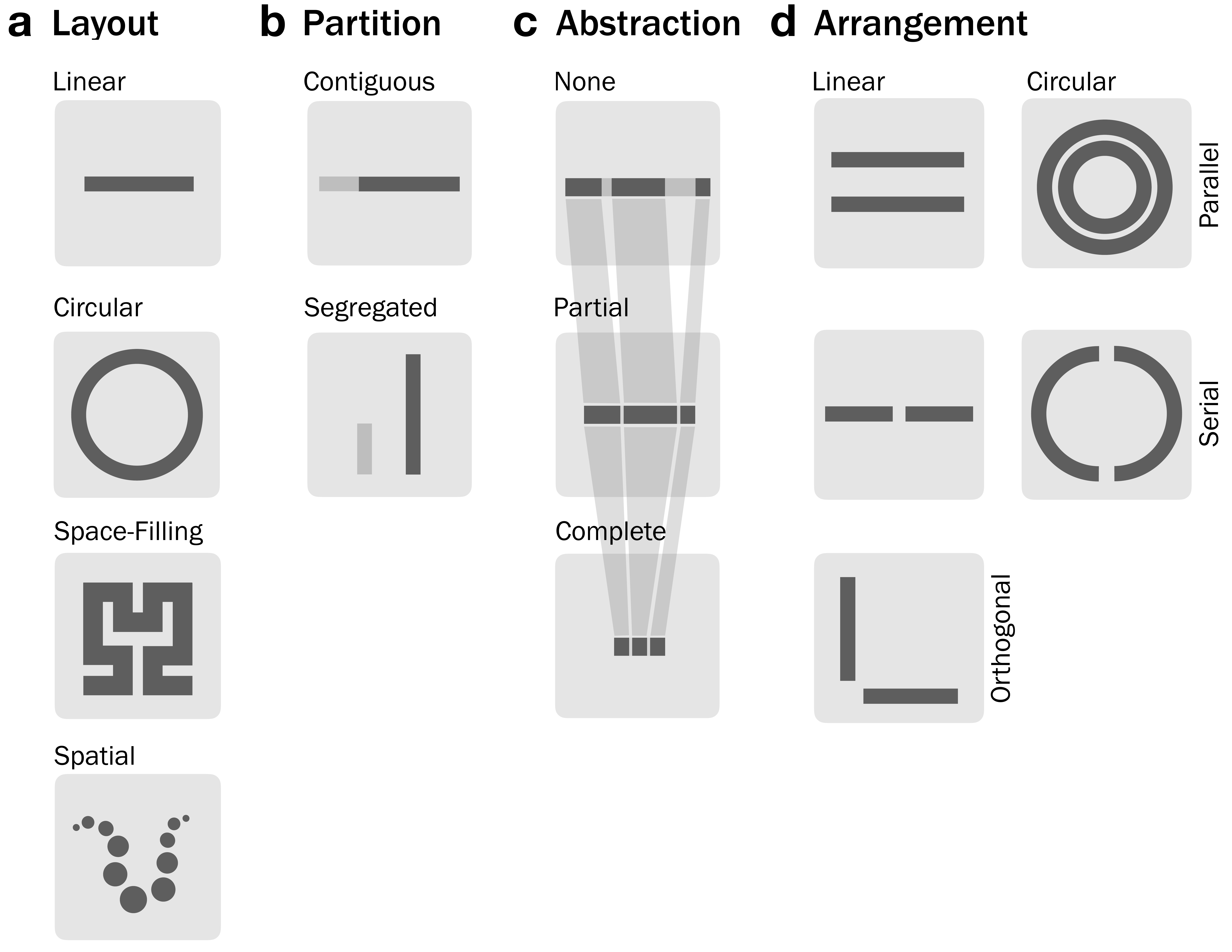}
    \caption{Layout, abstraction, partition and arrangement of sequence axes. (a) A sequence axis can be displayed in a linear, circular, space-filling or spatial layout. (b) A sequence axis can be visualized completely or some parts or the entire sequence can be abstracted. (c) Distinct sequence parts can be visualized in as a whole (contiguous) or  segregated. (d) Two sequence axes can be arranged in different ways.}
    \label{fig:basicConcepts}
\end{figure}
Theoretically, a sequence can be visualized in any layout preserving the sequential nature. In practice, most genomic visualizations display sequential data either in a linear or a circular fashion (see Figure~\ref{fig:basicConcepts}a). The genomic coordinates of a sequence correspond to a sequence axis, which is visualized in a layout. A sequence axis is a coordinate system for genomic features. (see Figure~\ref{fig:basicConcepts}d).

\paragraph*{Layout} Linear layouts (\symb{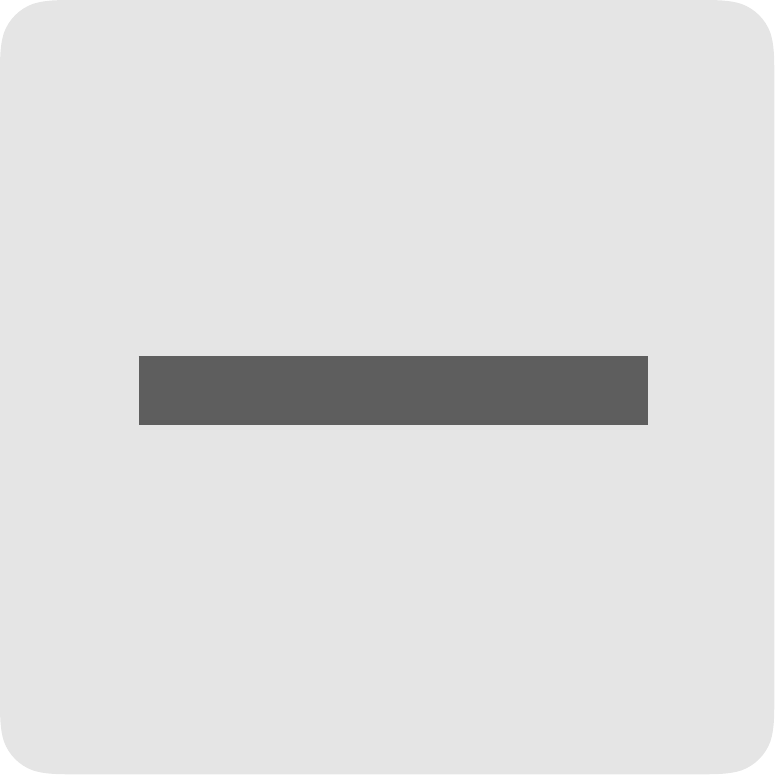}) are intuitive since they are easy to read (usually from left to right). However, since genomic sequences can be extremely long, zooming and panning is often required. Circular layouts (\symb{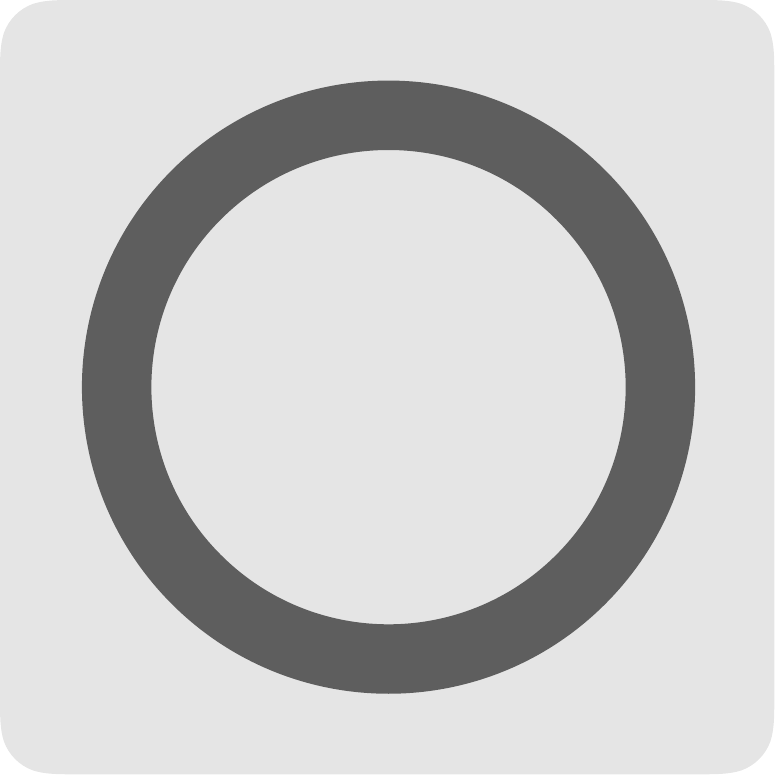}) are mainly used for three reasons: (i) the displayed sequence itself is circular, (ii) a non-circular sequence is displayed in a space-saving way, or (iii) interactions between different parts of the sequence(s) are shown using a chord diagram. A type of layout even more space efficient than a circular layout is a space-filling curve (\symb{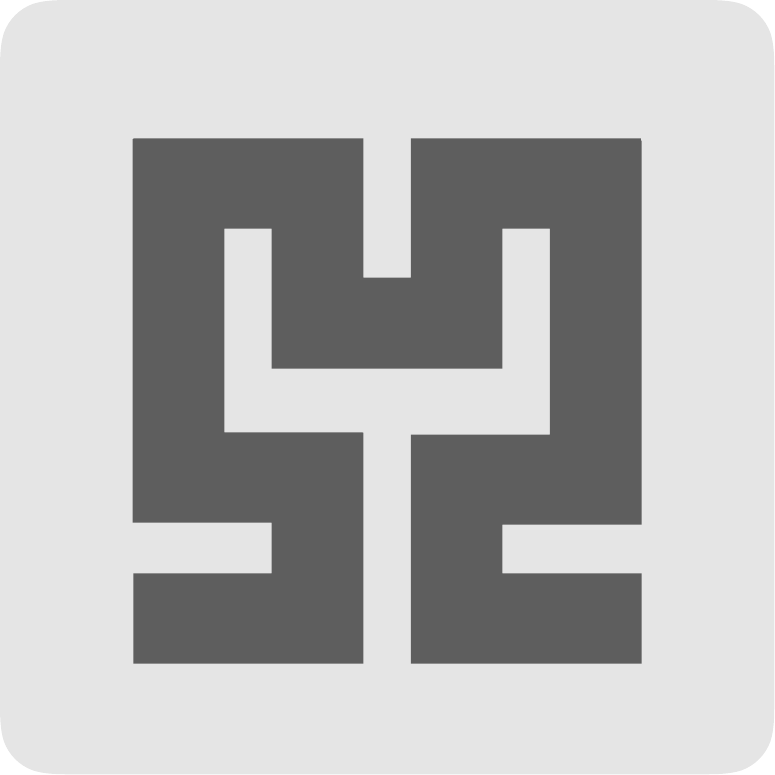}), such as Hilbert curves, which are often used to display a global overview of the genome while maintaining the spatial distribution of features. However, space-filling curves can only show one feature set and it is hard to visually estimate the distances between two positions of the sequence. A sequence can also be displayed in a spatial layout (\symb{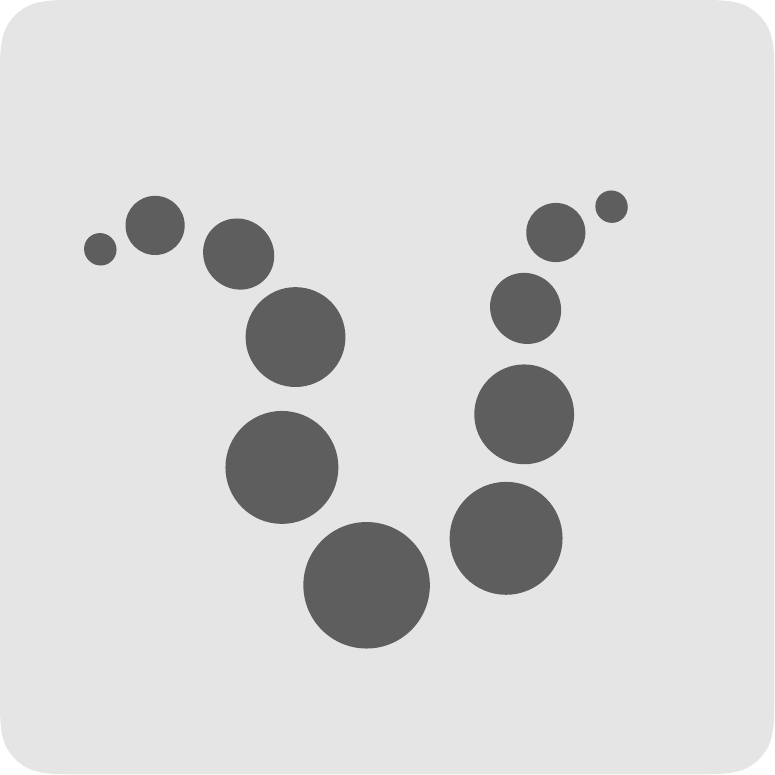}), where the 3-dimensional structure can be shown. \new{A spatial visualization requires three spatial axes}, yet there is only one sequence axis.

\paragraph*{Abstraction} A way of reducing the space of a sequence in order to concentrate on specific regions is abstraction (see Figure~\ref{fig:basicConcepts}b), which means replacing parts of the sequence (\symb{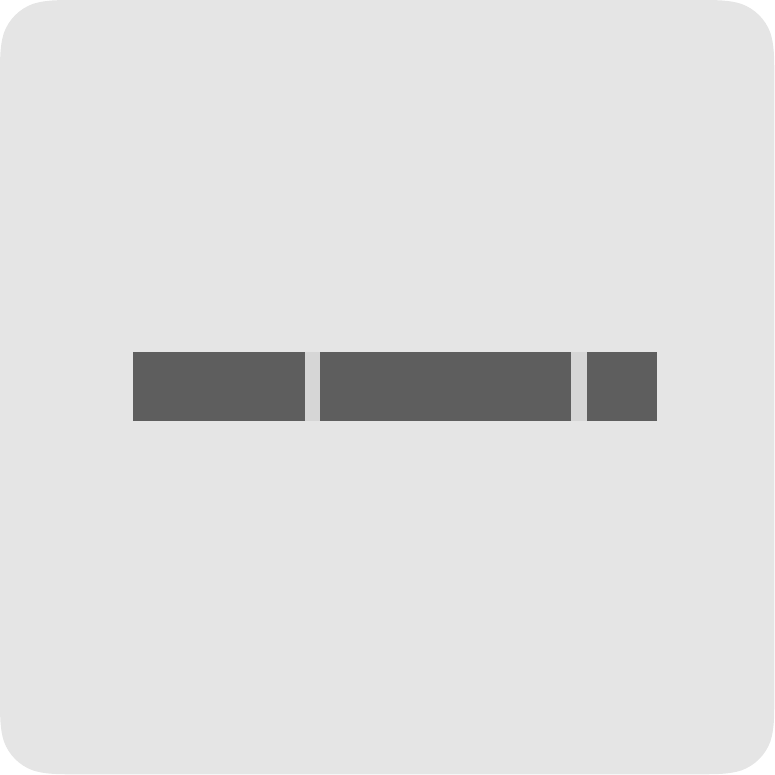}) or the entire sequence by abstraction elements (\symb{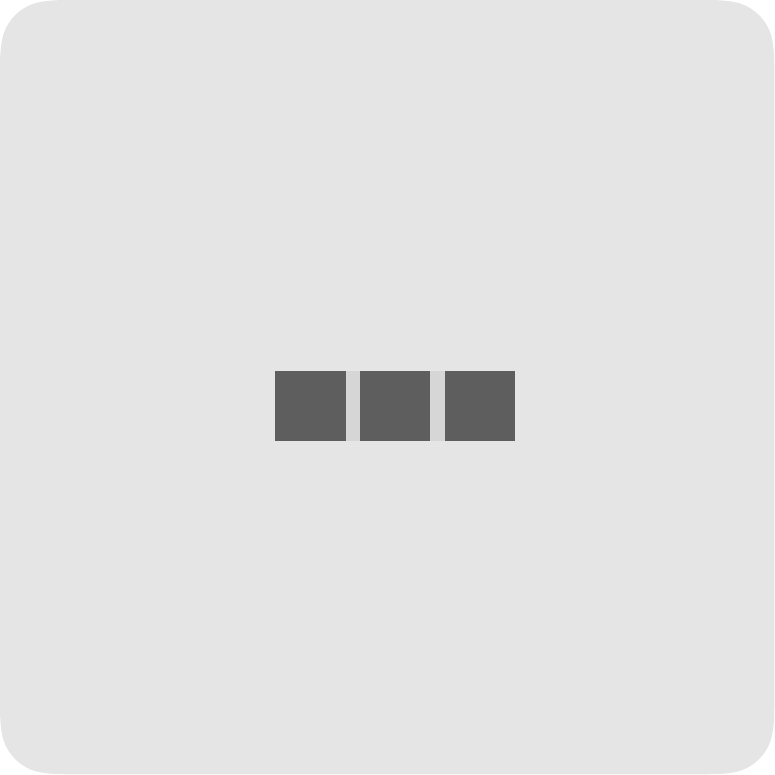}), such as symbols, while maintaining the order of the elements. \new{For example, when we are only interested in the exons, not the entire genome,} the introns can be abstracted to a gap or a symbol or completely filtered out. Therefore, non-adjacent parts of the sequence are next to each other in the visualization. We could not identify a tool applying a complete abstraction in our literature research, which could correspond to replacing both exons and introns with symbols. 

\paragraph*{Partition} Eukaryotic genomes often consist of multiple chromosomes, which are distinct sequences. However, they are often visualized as one contiguous sequence by placing them end-to-end of each other (see Figure~\ref{fig:basicConcepts}c, top \symb{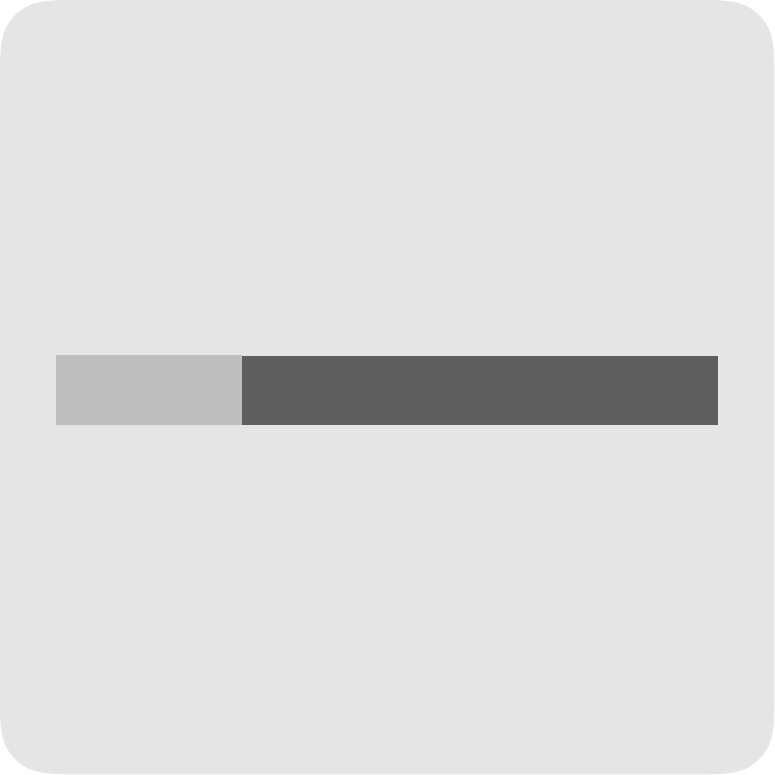}). Some visualizations, especially when comparing genomes, treat chromosomes as separate elements (see Figure~\ref{fig:basicConcepts}c, bottom ,\symb{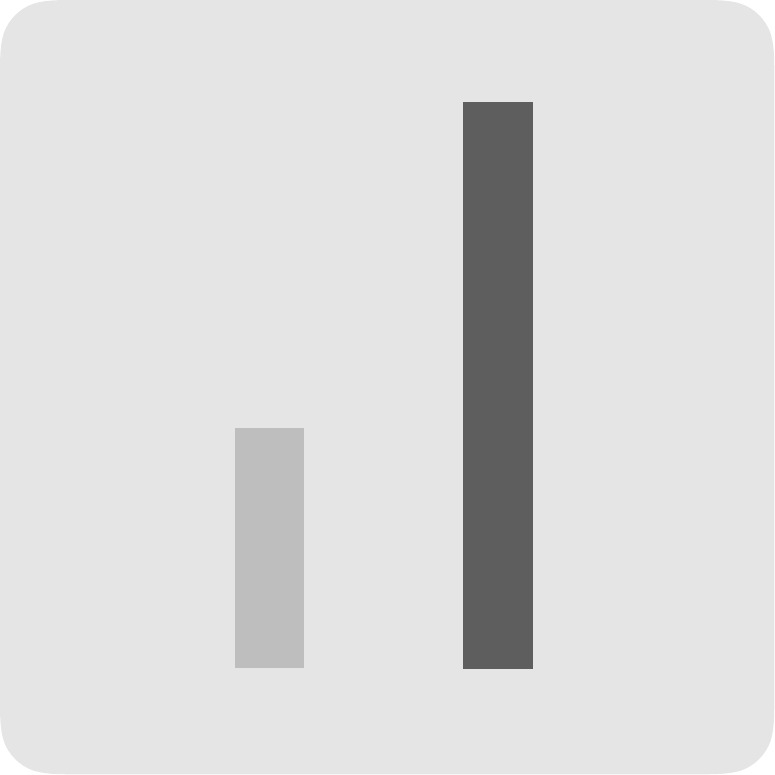}). While different chromosomes are the most common reason to display a sequence in separate parts, one could imagine partitioning a sequence based on other factors too, such as partitioning it in equally sized subsequences to show the entire sequence in multiple rows (similar to space-filling layouts).

\paragraph*{Arrangement} Axes of the same layout can be arranged in different ways, as shown in Figure~\ref{fig:basicConcepts}d, which is derived from Meyer et al.~\cite{Meyer2009-bi}. Axes in a linear and circular layout can be displayed in a parallel (\symb{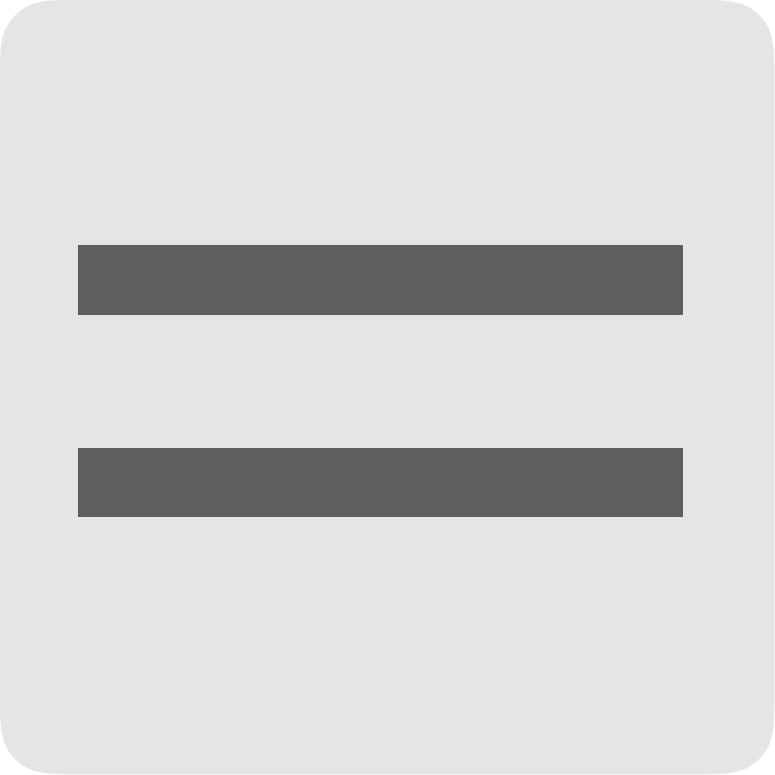} \symb{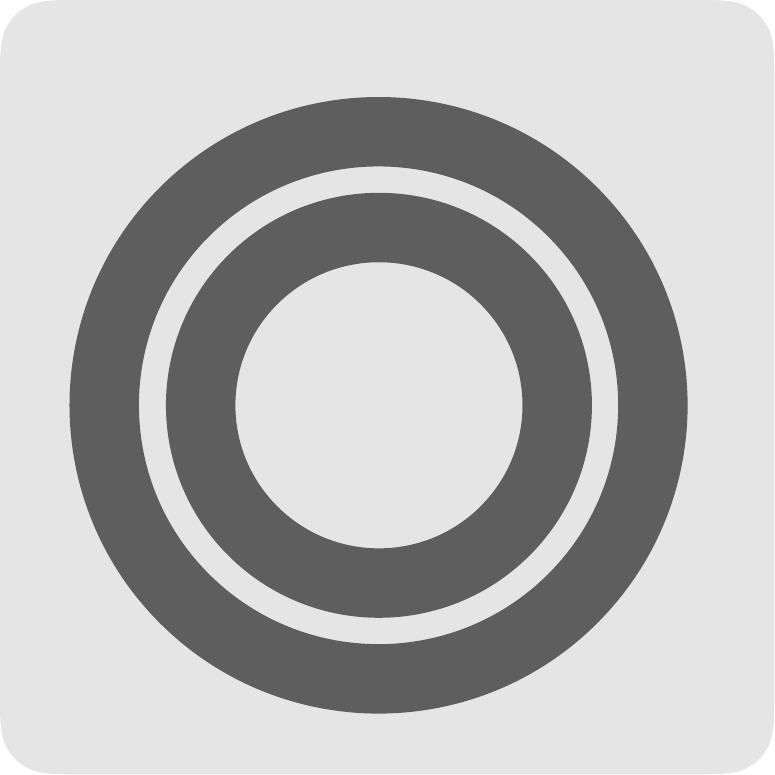}) or serial (\symb{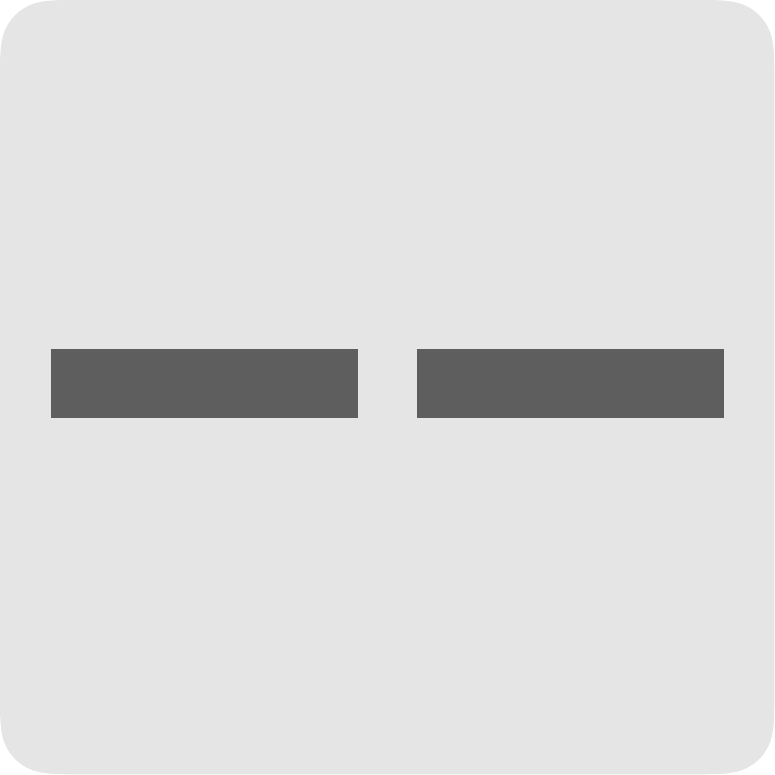} \symb{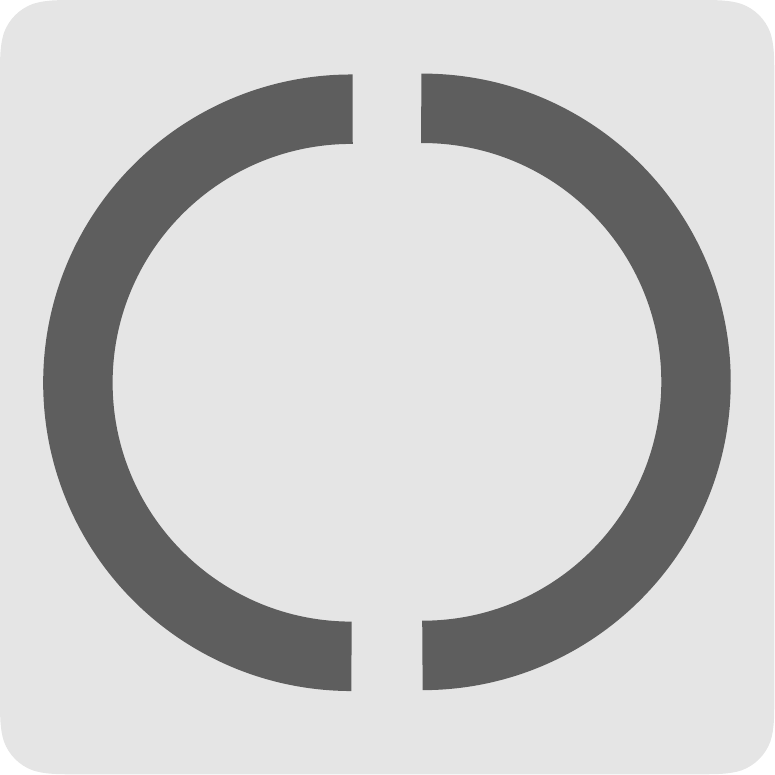}) arrangement. A serial circular arrangement corresponds to two sequence axes in a ``half-circle'' layout. Additionally, axes in linear layout can be visualized in a orthogonal arrangement (\symb{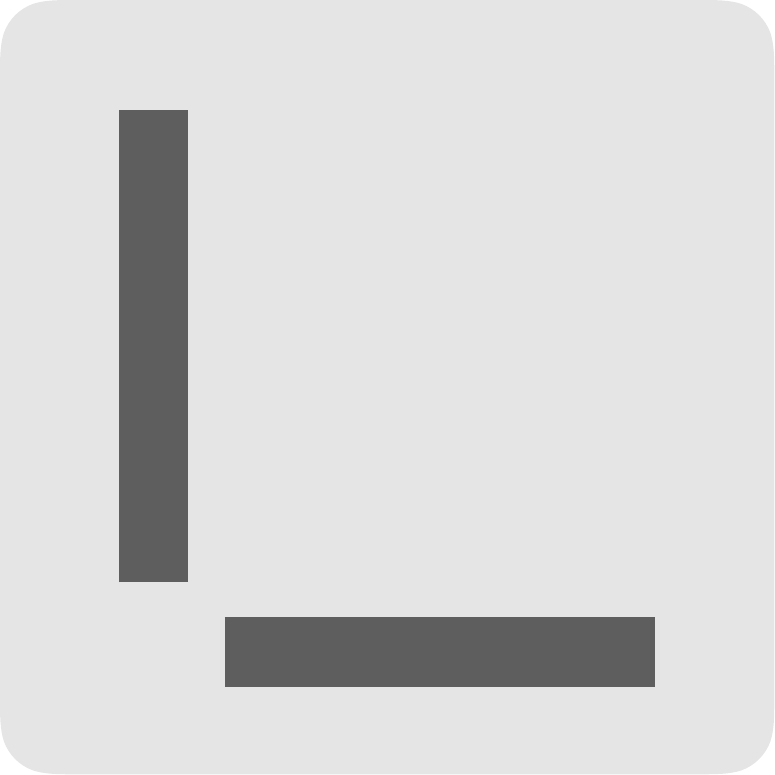}). The different arrangements can be used to (i) visualize interconnection features or (ii) compare two different sequence in the context of comparative genomics. 

\subsubsection{Genomic Tracks and Matrices}\label{sec:TrackMatrix}

For genomic data it is important to put multiple different data types into context in order to draw conclusions. For example, when analyzing mutations in a genome it is useful to visualize them together with gene annotations to estimate their functional impact. In order to analyze multiple data types at the same time, a genomic visualization often contains multiple \textit{tracks}. A track is a visual representation of genomic data with one or multiple parallel sequence axes showing one feature set. Typically, tracks are oriented horizontally, but in some tools they can also be oriented vertically. The annotations and mutations in the stated example are two features sets that are represented by two separate tracks. 

\paragraph*{Track Types} For each type of feature sets a separate track type can be defined as proposed by Gundersen et al.~\cite{Gundersen2011-fq}. According to the authors seven different basic track types can be defined (excluding interconnections). For sparse feature sets they developed four track types. A track for a sparse feature set with only features of length 1 with zero attributes is called a \textit{point track}. For example, the positions of all substitution mutations could be displayed with this track type. When the features in a point track are associated with an attribute, such as the substituted nucleotide, the track corresponds to a \textit{valued point track}. Respectively, \textit{segment tracks} can encode for the position and extent of genes, while \textit{valued segment tracks} can additionally show an attribute, such as the gene expression or gene name. For contiguous feature sets the track types correspond to ungapped versions of the previously described types. Contiguous feature sets containing features with a length produce a \textit{genome partition track}. If the features are associated with an attribute, the track corresponds to a \textit{step function}. A valued point track without gaps corresponds to a \textit{function}.

Gundersen et al.~\cite{Gundersen2011-fq} also propose eight extended track types which can additionally encode for interconnection features. Seven correspond to the previously described types, but pairs of features are associated with interconnections which can be directed and/or have a weight. In scope of this review paper we allow the association of multiple attributes or complex attributes to track elements and interconnections. For example, a track should be able to encode for the distribution of different nucleotides at point mutations, which corresponds to a valued point with a complex attribute.

Theoretically, all of the described feature set types can be encoded with these track types. However, especially Hi-C data is usually visualized in arrangements using more than one sequence axis. In Hi-C data all pairs of contiguous segments of a specific size in the genome (bins) are associated with an interaction frequency, which represents an undirected weighted interconnection. Since this kind of data is usually easier to display using two axes we distinguish tracks showing features on a single axis (one-axis tracks) and tracks showing data on two axes (two-axes tracks). A matrix is a special form of a two-axis track showing data on two sequence axes that are arranged orthogonally.

\begin{figure}
    \centering
    \includegraphics[width=0.9\linewidth]{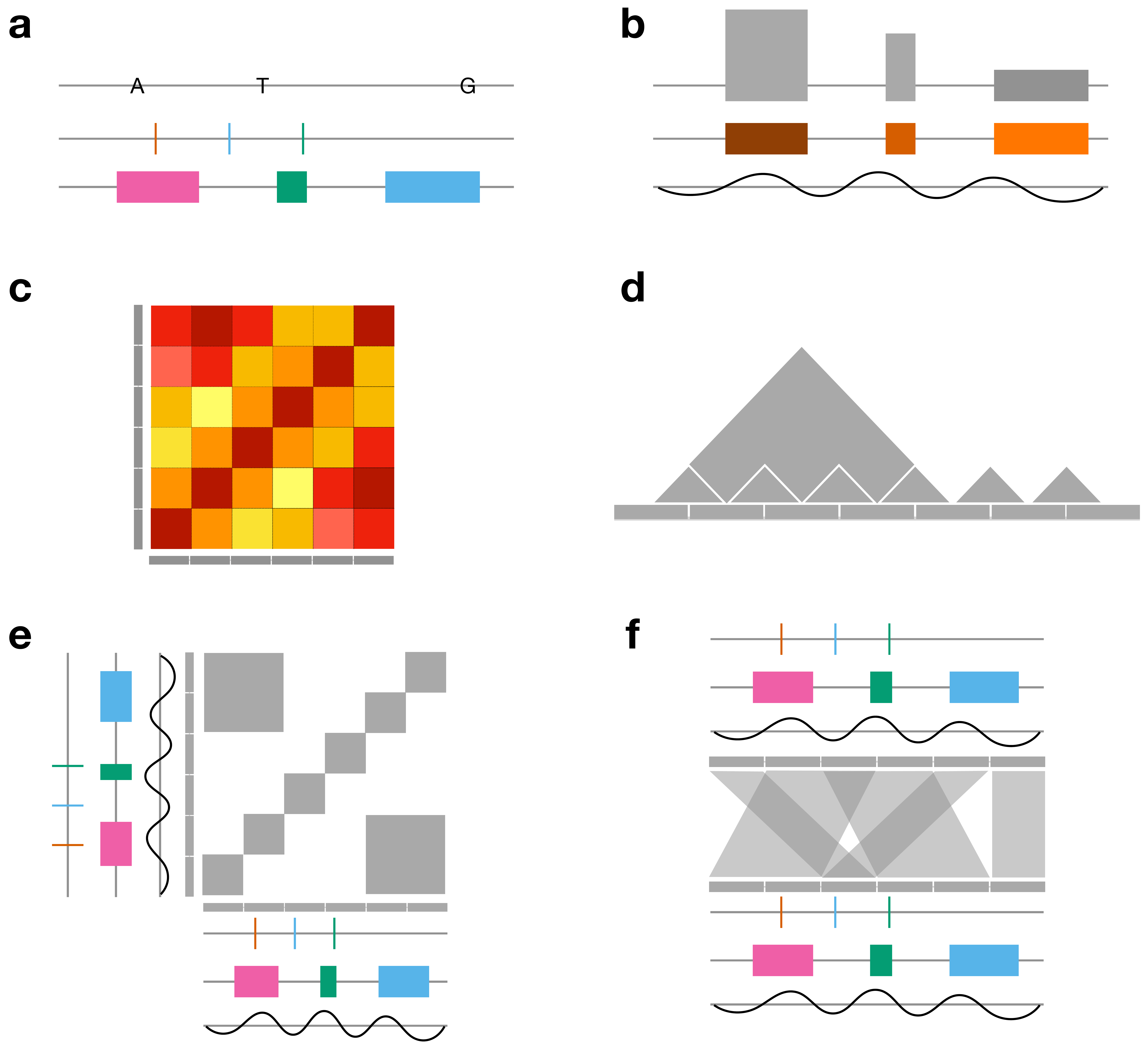}
    \caption{Examples of visual encodings of feature sets and arrangement of tracks. Features can be encoded through color, height of blocks and positions (a,b,c,d). \new{A two-axes track can be arranged with multiple one-axes tracks (e,f)}.}
    \label{fig:featureEncoding}
\end{figure}
\paragraph*{Visual Encoding and Track Alignment} Some common visual encodings of tracks and matrices can be seen in Figure~\ref{fig:featureEncoding}. One of the most commonly used encodings is color. For example, valued point tracks showing variants can be encoded by coloring corresponding to the variant type, valued segments showing genes can be colored by functional category or gene expression (continuous color scale).

A sparse set of segment features can contain segments that can overlap. Read data is an example of this type of feature. Sequencing reads can be mapped to long sequences and usually overlap, yet they do not always cover the entire genome. Overlapping segment features can be stacked in a way that avoids visual overlaps without introducing unnecessary white space.

Categorical attributes are often also represented with symbols, such as the encoding of substitution mutations with the letter of the altered nucleotide or symbols for deletions and insertions (see Figure~\ref{fig:featureEncoding}a, top). Often, nucleotides are encoded using both color and symbol. Features with continuous attributes can also be encoded using heights. Segments can be displayed as blocks with varying heights depending on the attribute value (see Figure~\ref{fig:featureEncoding}b). The function track type is usually encoded by using a line chart which spans the entire sequence. 

As previously described, matrices can be used to display interconnection features. Figure~\ref{fig:featureEncoding}c shows a continuous attribute of an interconnection feature encoded using a heatmap. The colors encode the attribute value, i.e. the weight of the interconnection. In contrast, the matrix in Figure~\ref{fig:featureEncoding}e only shows which features are interconnected. 

Undirected interconnection features (such as Hi-C data) can also be encoded on a one-axis track. For sets of undirected interconnection features, a matrix representation is symmetrical. Therefore, the matrix can be cut in half along the diagonal, rotated, and mapped to sequence coordinates forming a track. The value of an interconnection can be retrieved by following imaginary lines originating at two features in 45-degree angles. Like matrices, this type of track encoding can show interconnection without a continuous attribute as seen in Figure~\ref{fig:featureEncoding}d.

Often, multiple tracks are displayed in one visualization in order to correlate multiple feature sets. Tracks using the same sequence axis can be stacked (\symb{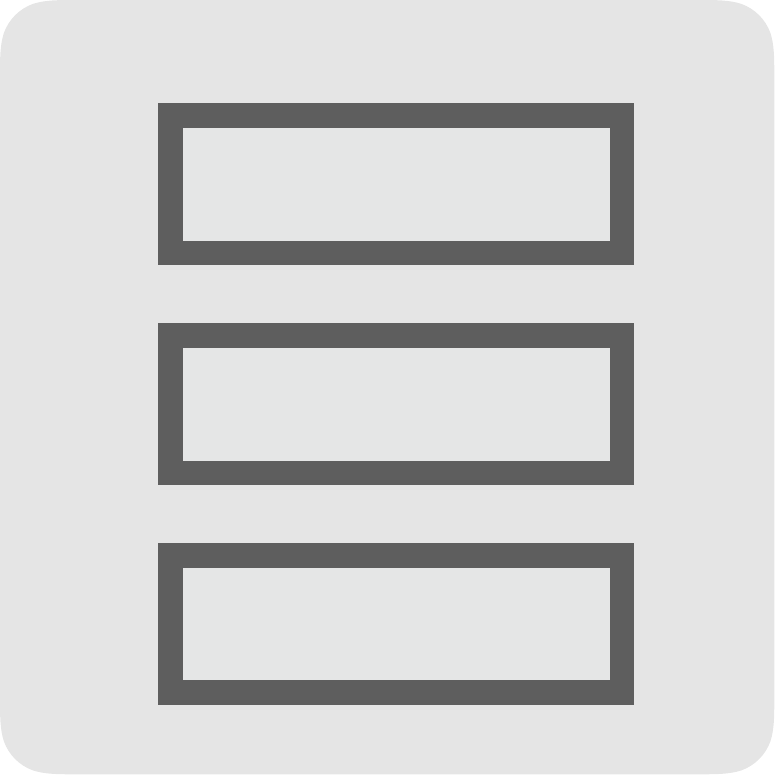}) or overlaid (\symb{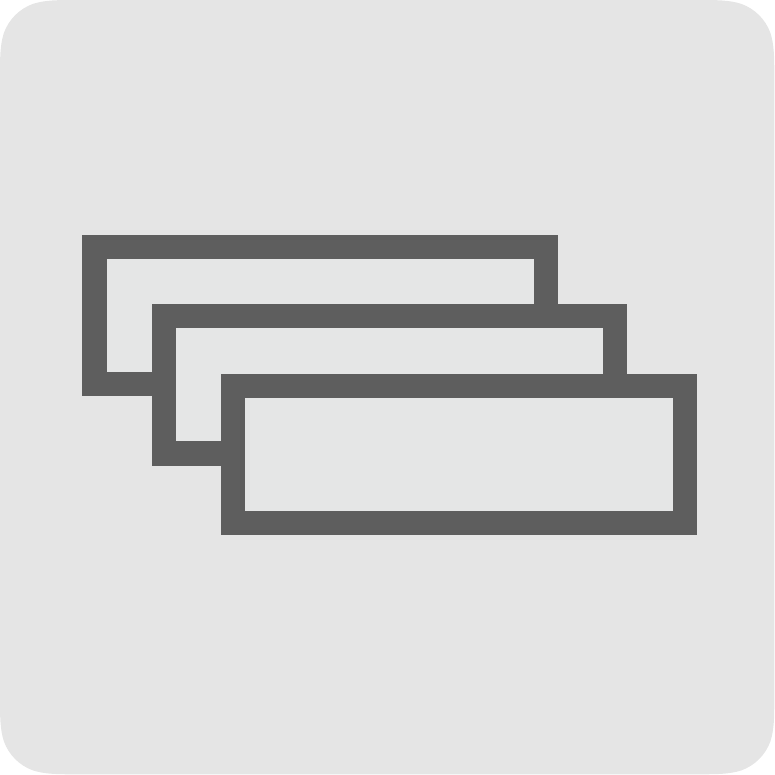}) and aligned by sequence coordinates. Figure~\ref{fig:featureEncoding}e shows an arrangement of an orthogonal two-axes track aligned with multiple tracks in a linear layout. Since orthogonal arrangements contain two axes, they can be arranged with multiple other tracks and matrices vertically and horizontally. Figure~\ref{fig:featureEncoding}f shows a parallel two-axes track, where each axis is aligned with multiple one-axis tracks. Note that the two coordinate systems of the axes in the two-axes track do not have to be aligned, but can show different regions of the sequence to show interconnection features.

Aligning multiple space-filling tracks is rather uncommon and limited. Since space-filling layouts use space most effectively, it is hard to arrange tracks in parallel. Moreover, a parallel arrangement complicates the identification of the same coordinate across tracks. For this reason overlaid track alignments are more common. In order to avoid hiding features \new{transparent} colors can be used. Another possibility is combining feature sets instead of overlaying two tracks. 

\subsubsection{\new{Multiple Sequences}}
\new{In the previous sections, we described how features on a single sequence can be visualized using one or multiple sequence axes. Yet in the field of \textit{comparative genomics}, multiple sequences or reference genomes are analyzed to study genome evolution. Two or more sequences are compared} to find blocks of high similarity on the sequence level (conservation) and to analyze if the location, order, proximity, and orientation of these blocks is similar in the compared genomes (synteny). The goal of the visual encoding of \new{sequence} comparisons is connecting sequence coordinate systems to show regions of high similarity of the sequences. Similar regions can be visualized using different techniques and visual encodings. The size of sequences to be compared can vary greatly from small regions, such as genes to entire chromosomes or entire genomes.

\new{There are three basic techniques for visualizing sequence comparisons usually applied for genomic visualization: (i) comparison by alignment, (ii) comparison by connecting conserved blocks, and (iii) comparison by using dot plots. Especially for the visualization of many small sequences, alignment-based techniques are used, with which shared nucleotides are algorithmically aligned as described in Section~\ref{sec:alignement}. This represents a construction of a shared coordinate system and sequence axis. Each sequence corresponds to a valued segment or point track that is aligned to the coordinate system. 

Another way of visualizing comparisons is to keep the different coordinates for each sequence separate and visualize the comparison in one or multiple two-axes tracks. Interconnection features between segments in two sequences can encode for the positions and extents of syntenic regions, which can be encoded by connecting them with lines or bands or by applying the same color. Meyer et al.~\cite{Meyer2009-bi} describe which combination of encoding, layout and arrangement is most effective depending on the length of the displayed sequence. They recommend using bands and parallel linear or serial circular arrangements for shorter sequences (up to chromosomes) to avoid too many crossing lines and color encoding for whole genome comparisons in parallel arrangements. 

Orthogonal arrangements of sequence axes can show the similarity between every position or bin of positions of one sequence to every position or bin of positions in the other sequence. The two sequences are arranged in a 90 degree angle spanning a comparison matrix. If nucleotides or bins match between positions of the two genomes, a dot is drawn in the corresponding cell. Similar regions form diagonal lines of dots. With this technique insertions, deletions and inversions can be identified.}

\subsubsection{View Configurations for Genomic Visualizations}
\begin{figure}
    \centering
    \includegraphics[width=0.9\linewidth]{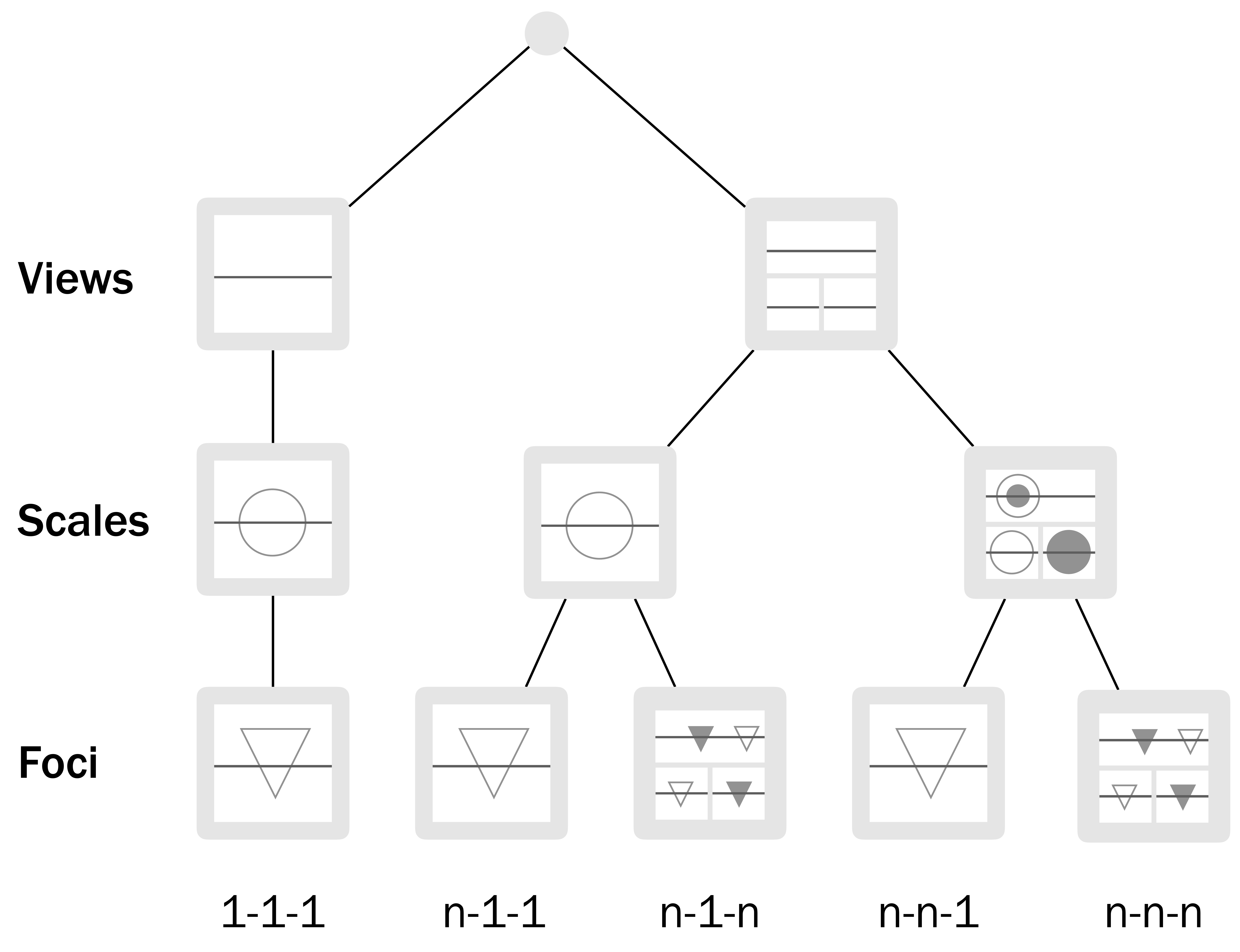}
    \caption{The number of views, scales and foci are important parameters of a genomic visualization. The combination of these parameters results in five different view configurations.}
    \label{fig:ParameterTree}
\end{figure}

In order to categorize tools and techniques, we define three parameters of a genomic visualization (see Figure~\ref{fig:ParameterTree}): (i) the number of views that show data mapped to genomic coordinates, (ii) the number of scales used to analyze the data at the same time and (iii) the number of foci, i.e. non-adjacent genomic segments that can be viewed independently.

We restrict a view to a set of one or multiple aligned tracks, which contains at most one two-axis-track. A visualization can consist of one or multiple views displaying features mapped to genomic coordinates, which can be linked or independent. 

A genome can be very large and analyzing it on different scales, such as the whole genome or single genes, can be of great value. Similar to a map: A view of the entire world provides a useful overview, but we cannot analyze the street structure of New York. While some genome visualizations only allow visualizing sequences on one scale at a time, others provide multiple views to visualize the data on different scales. In our taxonomy, a multi-scale visualization visualizes the data on multiple scales at the same time in multiple different views. However, also single-scale visualizations can provide a zooming interaction.

A focus can be understood as a sliding window across the genome. Only the region in this window can be analyzed. Multiple foci enable users to look at distinct segments of a genome in parallel and compare features which are dispersed across the genome. In the map analogy this would be, for example, comparing the street structure of New York to that of London with Google Maps. If we zoom out of the map, we can only see the names of the two cities; if we zoom in, we can only see one city at a time. We need to open a second browser window to view them in detail at the same time.
Note that foci can be defined in a flexible way and are visualized in separate views, while sequence abstraction is usually done in a previous step, where in the case of filtering non-adjacent sequences are ``glued'' together.

\begin{figure}[t]
    \centering
    \includegraphics[width=0.6\linewidth]{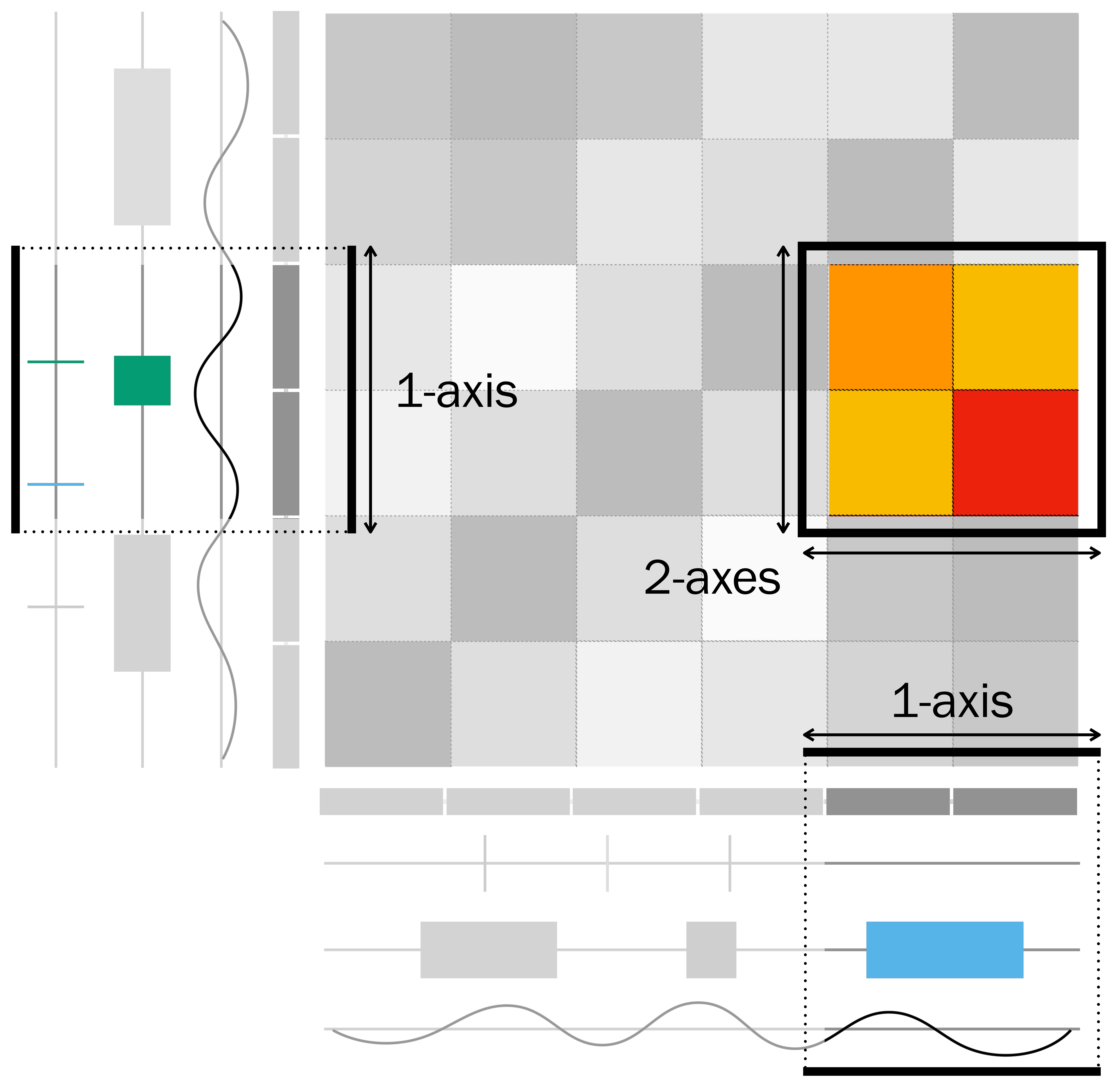}
    \caption{A visualization with multiple axes can have foci in multiple dimensions. This orthogonal arrangement of axes contains two one-axis foci and one two-axes focus.}
    \label{fig:2AxesFocus}
\end{figure}
Sometimes, arrangements of two sequence axes corresponding to the same sequence are used to facilitate the visualization of interconnection features. For arrangements of multiple axes, we have to add another dimension for scale and focus. Figure~\ref{fig:2AxesFocus} shows an example of foci in an arrangement with two axes. Two different foci are shown in the tracks aligned to the linear axes. However, the matrix spanned by the orthogonal arrangement can only visualize one focus of two-dimensional data. Similarly, two orthogonal axes can be on different scales, leading to two one-axis scales and one two-axes scale.

Combining these parameters in all possible ways results in five basic view configurations of genomic visualizations as seen in Figure~\ref{fig:ParameterTree}. Note that if a visualization contains two axes, foci and scale refer to two-axes foci and two-axes scales. 

\new{As we restrict a view to  a  set  of  one  or  multiple  aligned  tracks, with  at  most  one  two-axis track,} it is not possible to visualize multiple scales and foci (of the highest dimension) in one view with our taxonomy. \new{Therefore,} the left branch only consists of one path. The other view configurations are combinations of multiple views with one or more scales and foci. 
We categorize our tools in Sections \ref{sec:singleSequence} and \ref{sec:MultiSequence} using these five basic view configurations. For visualizations with multiple axes we additionally show the number of one-axis foci and one-axes scales.

\subsubsection{Linking Views}

Genomic visualizations often incorporate multiple views that can be (i) \textit{independent}, (ii) \textit{weakly linked}, (iii) \textit{medium linked} or (iv) \textit{strongly linked}. While independent views are not connected in any way, weakly linked views are linked by brushing and linking. Medium linked views share navigation, for example two views are at different scales, but zooming always affects both. Strongly linked views share genomic coordinates at one axis and can also share tracks that are aligned to the axis.

Utility views provide information about (i) tracks, (ii) features, or (iii) genomic coordinates. The property that distinguishes these views from e.g. detail views is that they never show genomic coordinates directly but only meta data or derived data. These types of visualizations can be either aligned with the genomic visualization, or \new{weakly linked}. As an aligned visualization, consider a view showing meta data about tracks which is situated on the left side of each track. \new{In case of sequencing data or expression data of individuals, this could be, for example, phenotypic information about the sample donor.} Since this kind of visualization is in direct association with the genomic data we call it a \textit{strongly connected utility visualization}. On the other hand, consider a visualization that is connected with a table through brushing and linking. Only by clicking on an element in the visualization is the corresponding element in the table highlighted and vice versa. Since the views are not aligned we call this type of visualization a \textit{weakly connected utility visualization}.
\subsection{Tasks}
Visualization tasks represent actions that users may perform on their data~\cite{Brehmer2013-tm}. These can be both low-level operations or high-level user intents while interacting with a system. Visualization tasks have been defined and classified, often depending on the context and scope of the tasks. A common feature of most genomic coordinate visualizations is that they visualize one or multiple types of features at their corresponding positions, therefore the tasks that different tools and techniques help to solve are often similar. In this section we describe the most common tasks performed using genomic visualizations.

A typology of abstract visualization tasks proposed by Brehmer and Munzner~\cite{Brehmer2013-tm} focuses on three questions: {\em why} is a task performed, {\em what} are the inputs and outputs, and {\em how} is the task performed. What is particularly useful in this typology is that it distinguishes between \new{user intents} (that answer why) and \new{interactions} (that answer how) and provides a link between the two questions. In this section, we summarize and categorize the ``Why'' task for genomic visualizations. Moreover, we give an overview of \new{common interactions (``How'') and inputs and outputs (``What'').} 

\subsubsection{Why?}
We adapt the \new{task topology} described by Brehmer et al.~\cite{Brehmer2013-tm} to genomic visualizations, starting with slight adjustments to the described search tasks \new{\tlookup, \tlocate, \tbrowse and \texplore}. In our task taxonomy \tlookup refers to viewing features at one position, for example by navigating to a known gene and a specific feature set, such as gene expression. \new{\tlocate} refers to finding one or multiple features with desired properties, such as locating peaks of an epigenetic signal or highly expressed genes in a single feature set. \tbrowse is similar to \tlookup, yet while the position is known, the feature set is unknown. For example, different feature sets, such as expression and mutations can be browsed at the position of a known gene. \texplore refers to a very broad task. Neither the position nor the feature set are known, \new{therefore} \new{multiple} feature sets at \new{different} position in multiple loci are explored. Exploring corresponds to a combination of other tasks. In order to explore, we repeatedly browse features at positions or locate features in a feature set. However, the characteristics that we look for or the positions that are browsed are not predefined and can change during the process of exploration.

\begin{table}[t]
\centering
\caption{Categorization of ``Why'' Tasks}\label{tab:tasks}
\begin{tabularx}{\linewidth}{llhh}
\toprule
\rowcolor{White}
 & Task type & Single Feature Set & Multi Feature Sets  \\
\midrule
\rowcolor{White}
\makecell[tl]{Single \\ Locus} & Search & \tlookup & \tbrowse \\
\cmidrule(r){3-4}
\rowcolor{White}
 & Query & \tidentify & \makecell[tl]{ \tcompare*, \\ \tsummarize } \\
\cmidrule(r){2-4}
\rowcolor{White}
\makecell[tl]{Multi \\ Locus} & Search & \makecell[tl]{\tlocate, \\ \texplore } & \texplore \\
\cmidrule(r){3-4}
\rowcolor{White}
& Query & \makecell[tl]{ \tcompare, \\ \tsummarize } & \tsummarize \\
\bottomrule
\end{tabularx}
\end{table}

We categorize the described tasks plus the query tasks \tidentify, \tcompare and \tsummarize proposed by Brehmer et al.~\cite{Brehmer2013-tm} in single feature set, multi feature set, single locus and multi locus tasks (see Table~\ref{tab:tasks}). The task taxonomy illustration in Figure~\ref{fig:Overview} shows how these tasks operate on genomic visualizations. \tlookup and \tidentify correspond to single feature set, single locus tasks. While \tlookup aims to find the desired feature set at a locus, \tidentify characterizes the feature attributes. These tasks are often paired, for example after looking up an epigenetic signal at the position of a gene, we can identify the actual value of the signal.

\tlocate and \tcompare are single feature set, multi locus task. \tlocate refers to finding positions, while \tcompare finds a relationship between \new{features at the located positions}. For example, the expression levels of two genes can be compared. This involves two of the previously described tasks: The expression feature set of each gene has to be looked up and the expression value has to be identified before it can be compared. If feature sets are of the same type, for example expression data for two different samples, \tcompare can also be applied across feature sets at the same locus.

Exploring and summarizing both can refer to multiple loci and multiple feature sets. While \texplore \new{is a search task}, the goal of \tsummarize is \new{to find data patterns and trends. This task provides an overview or a ``big picture'' of the data}, such as ``expression levels of \new{genes in this pathway} are high''. Therefore, summarizing can also be done for a single feature set or a single locus. After exploring the data or after browsing a specific locus and identifying the feature attributes, the patterns in a single locus can be summarized, for example summarizing that a highly expressed gene has mutations in its promoter region. As an example for summarizing a single feature set, consider locating features with interesting attributes in a feature set while exploring the data. By summarizing attributes of features in one set, the distribution of attributes can be characterized.

\subsubsection{How?}
``How'' refers to the methods with which the ``why'' tasks can be solved using interaction. Independently of the described view configurations, tools for the visualization of genomic data often differ in their level of interactivity. Many tools can only plot data as a static image. Different datasets, different visual representations, regions and zoom levels can often be chosen as parameters for the plot, yet there is no interaction with the visualization itself.

\new{Interactive tools often offer navigation interactions to \task{navigate} along the sequence axis via zooming panning or jumping to regions. Navigation is essential for most genomic tools due to the immense size of genomes, especially for the search tasks \tlookup and \tlocate. Moreover \task{selection} interactions are often implemented for highlighting features or selecting them in order to \task{derive} a new visualization or feature. Some tools allow \task{rearranging} views and tracks and \task{changing the visual encodings} of features. In general, flexible interactions enable a more in-depth exploration of the data, and provide the users with details on demand~\cite{Shneiderman2003-ar}.}

As described in section~\ref{sec:LAC} a sequence can be \task{abstracted}. Many tools offer sequence abstraction as an interaction, most commonly by filtering introns or abstracting them using gaps. Additionally, it is often possible to filter out other regions that are not of interest. \new{Abstraction, especially by filtering introns helps users to \texplore the parts of the sequence that are the most informative for their problems.}

\task{Aggregation} is often implemented together with zooming. Features are encoded differently depending on the amount of space that is available. For example a multiple sequence alignment can be displayed by showing every nucleotide individually when zoomed in and as blocks showing conserved parts when zoomed out. \new{By applying aggregation while zooming, the visualization remains informative on different scales and features can be \texplore-d and \tsummarize-d on multiple levels.}

\subsubsection{What?}
The question ``What'' refers to the input and output of a task. Naturally, the input and output depend heavily on the tool itself, yet a few general statements can be made. Depending on the type of the task, the input can consist of one or multiple feature sets. While the tasks \tlookup, \tidentify, \tlocate and \tcompare have a single feature set as input, \tbrowse works with multiple feature sets. \tsummarize \new{and \texplore} can have one or multiple sets as input. 

Similarly, the outputs can be \new{described}. For single feature set tasks the outputs are a combination of features positions or feature attributes. The output of the search tasks \tlookup and \tbrowse is a feature at a position, while the output of \tlocate is one or multiple positions of features of interest. \tidentify returns the attributes of a feature, \tcompare a relationship between two (or more) attributes. 

The outputs of \texplore and \tsummarize are not as easily defined. Exploring can return everything starting from one feature at a position to multiple positions, patterns of different types of features or correlations. \tsummarize returns a statement about the data, such as the distribution of the attributes in a set of features, or the relation between multiple sets of features.

\subsubsection{High-level vs. Low-level Tasks}
\new{For characterizing genomic visualization tools it is important to differentiate between low-level and high-level tasks. Low-level tasks help us model how a tool works, while high-level tasks correspond more to biological questions. The tasks described are low-level tasks and most of the tools in Sections~\ref{sec:singleSequence} and~\ref{sec:MultiSequence} support these tasks for exploring the data. Yet, the tools differ in the biological questions that users want to solve with a tool. Questions can range from a very broad question, for example, ``How does sequencing data from a cancer sample compare to the reference genome?'' to a very specific question, such as ``Is TP53 mutated in this sample?'' The biological question determines the low-level tasks. In order to answer the first question, the users have to \texplore the data by \tbrowse-ing positions of interest and \tlocate-ing peaks in tracks. While exploring, users \tsummarize their insights. This is done by \task{navigating} along coordinates, \task{arranging} tracks, \task{changing encodings}, \task{filtering} introns and other user interactions depending on the tool. The biological questions depend heavily on the input data as well as the user intent.}

\section{Single Genomic Coordinate System}\label{sec:singleSequence}
\subsection{Genome-Scale Visualizations}
Genome-scale visualizations display one or multiple regions of a genome on absolute coordinates. We further categorize genome-scale visualizations based on the type of the features that they are specialized on. They can be focused on displaying (i)  \new{non-interconnected feature sets, (ii) sparsely interconnected feature sets and (iii) densely interconnected feature sets}.
\subsubsection{Non-Interconnected Feature Sets}

\begin{table*}[t]
\centering
\caption{Layouts (linear \protect\symb{layout-linear}, circular \protect\symb{layout-circular}, space-filling \protect\symb{layout-spacefilling}) and view configurations of tools for the visualization of non-interconnected features. Tools marked with * can apply a form of abstraction.}\label{tab:FeatureViewer}
\begin{tabularx}{\textwidth}{ccccX}
\toprule
Layout & Views & Scales & Foci &  \\
\midrule
\symb{layout-linear} & 1 & 1 &  1 &  UCSC~\cite{Kent2002-fm}*, NCBI Genome Data Viewer~\cite{NCBI_Resource_Coordinators2017-sl}, GenomeView~\cite{Abeel2012-rp}, JBrowse~\cite{Buels2016-qp}, EpiViz~\cite{Chelaru2014-sn}, 3d Genome Browser \cite{Wang2017-cl} , Dalliance~\cite{Down2011-fw}, EaSeq~\cite{Lerdrup2016-cs}, Wash U Epigenome Browser~\cite{Zhou2011-px}*, GView~\cite{Petkau2010-ya}, MGV~\cite{Kerkhoven2004-kt}, MGcV~\cite{Overmars2013-zz}, DNAPlotter~\cite{Carver2009-va}, ggBio~\cite{Yin2012-eh}, ReadXplorer~\cite{Hilker2014-fv}, GenPlay~\cite{Lajugie2011-ef}\\
 \cmidrule(r){2-5}
 & n & 1 & 1 & Savant Genome Browser 2~\cite{Fiume2012-hb} \\
  \cmidrule(r){2-5}
 &  n & n &  1 &  GBrowse~\cite{Donlin2009-pk}, MochiView~\cite{Homann2010-kt}, Ensembl~\cite{Hubbard2002-ar}, IGB~\cite{Freese2016-vl}(s) \\
 \cmidrule(r){2-5}
 & n & n & n &  NCBI Sequence Viewer~\cite{NCBI_Resource_Coordinators2017-sl}, ABrowse\cite{Kong2012-wp}, IGV~\cite{Thorvaldsdottir2013-ym}, CEpBrowser~\cite{Cao2013-wt}, \new{Xena~\cite{Goldman2018-iq}},
 \new{Island Viewer 3 \cite{Dhillon2015-ij}}\\
 \midrule
\symb{layout-circular} &  1 &  1 &  1 &  CGView~\cite{Stothard2005-bw}, GView\cite{Petkau2010-ya}, MGV~\cite{Kerkhoven2004-kt}, CiVi~\cite{Overmars2015-ks}, DNAPlotter~\cite{Carver2009-va}, Edgar Circular Plot~\cite{Blom2016-gk}, CGView server~\cite{Grant2008-tk} \\
 \cmidrule(r){2-5}
\new{& n & n & n & Island Viewer 3}~\cite{Dhillon2015-ij} \\
\midrule
\symb{layout-spacefilling} & 1 & 1 & 1 &  HilbertVis~\cite{Albers2011-hg}, HilbertCurve~\cite{Gu2016-pd}\\
\bottomrule
\end{tabularx}
\end{table*}
Often non-interconnected feature sets are visualized using tools that consist of one or multiple parallel tracks and visualize many kinds of different features. They display features using linear, circular and space-filling layouts (see Table~\ref{tab:FeatureViewer}). While certain data types are very common, some tools are more specialized on the visualization of a specific type of genomic data, such as the Savant Genome Browser 2~\cite{Fiume2012-hb}, which is specialized on showing structural variation or HilbertVis~\cite{Anders2009-sr} and HilbertCurve~\cite{Gu2016-pd}, which are especially useful to display epigenetic data. Tools in this section are specialized on visualizing many sets of non-interconnected features, yet they can sometimes visualize local interconnections.

\begin{figure}
    \centering
    \includegraphics[width=0.9\linewidth]{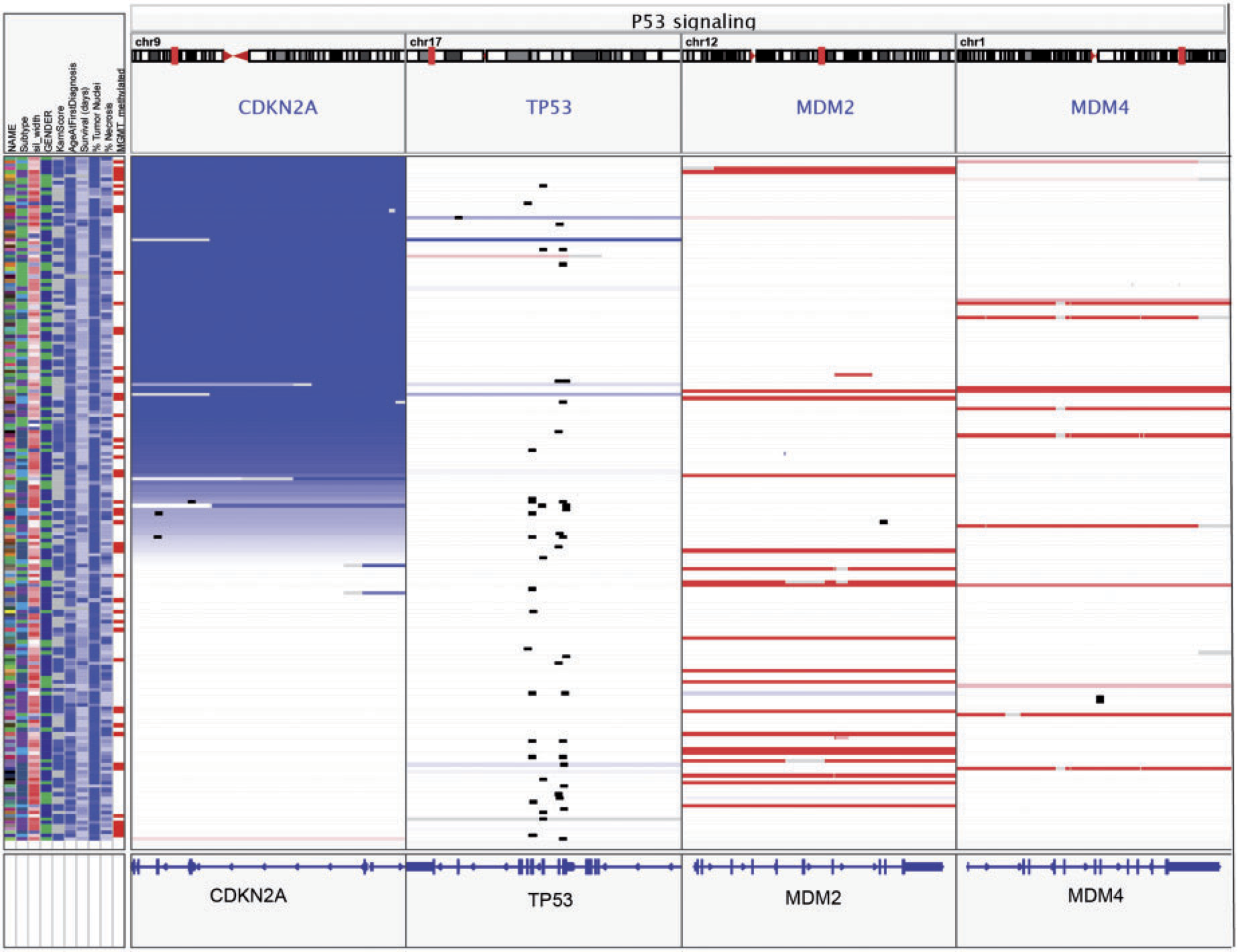}
    \caption{The Integrative Genomics Viewer (IGV) is an example of a genome browser that can show multiple scales and foci in separate views.  Moreover, the visualization includes a strongly linked utility view in form of a column next to the tracks showing meta data. Figure from~\protect\cite{Thorvaldsdottir2013-ym}.
    \protect\\Data: $D=$ \symb{type-contiguous-point}, \symb{type-contiguous-segment}+\symb{interconnection-none}; 
    Coordinate System: $C=$ \symb{layout-linear}+\symb{partition-contiguous}+\symb{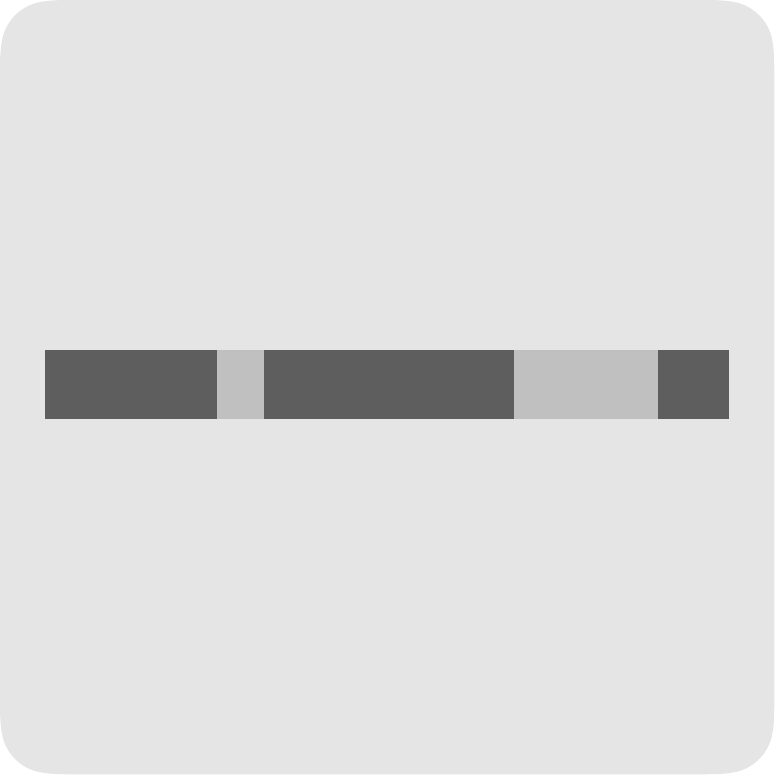};
    Tracks: $T=$ \symb{alignment-parallel},\symb{alignment-overlaid};
    Views: $V=$ \symb{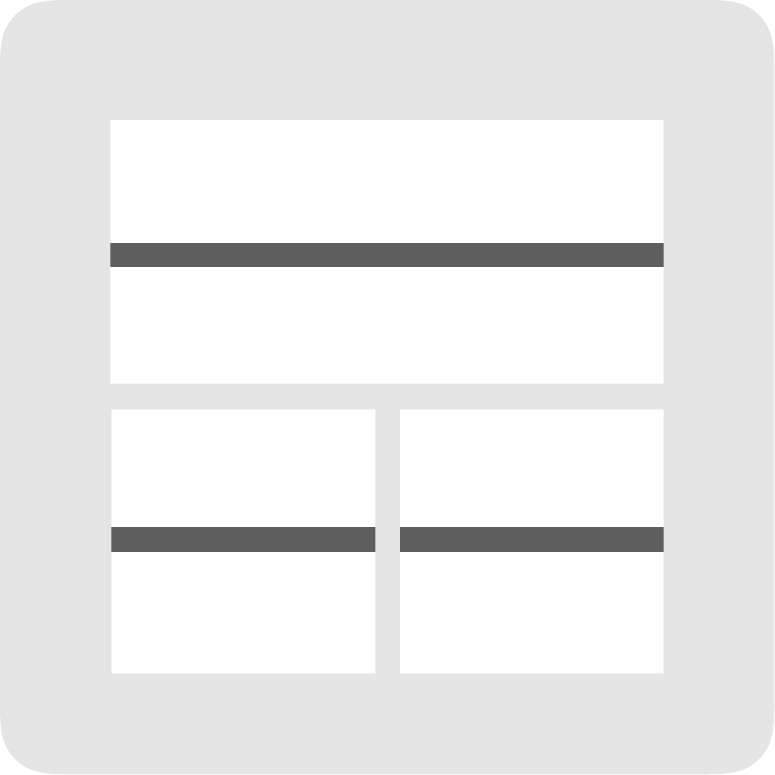}+\symb{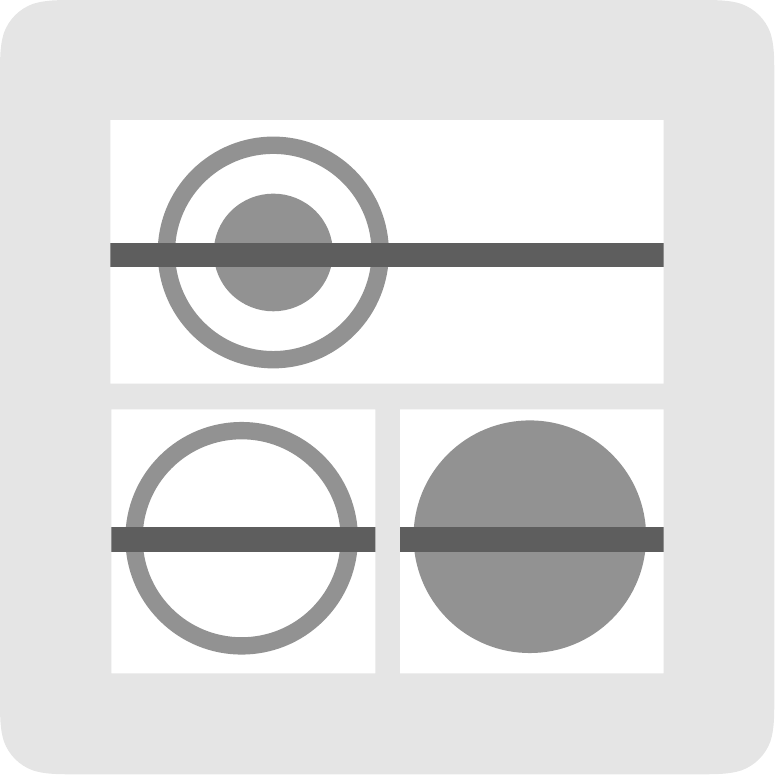}+\symb{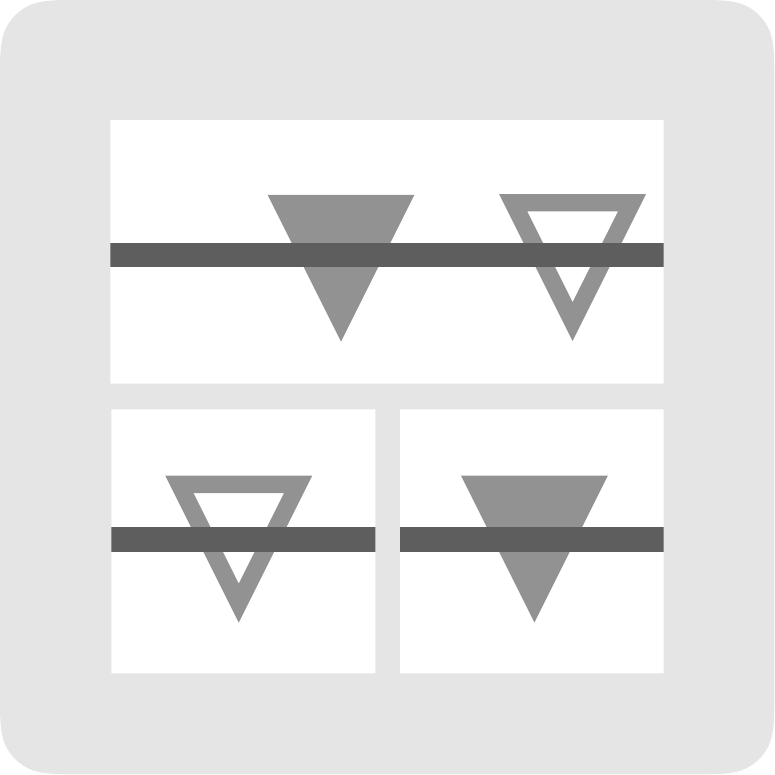}}
    \label{fig:IGV}
\end{figure}
\paragraph*{Linear Layout}
A tool group known as ``genome browsers'' usually displays multiple parallel tracks in a linear layout. A genome browser commonly consists of three components: A reference genome, annotations, and one or multiple tracks; see Figure~\ref{fig:IGV} from the Integrative Genomics Viewer (IGV)~\cite{Thorvaldsdottir2013-ym}. \new{Genome browsers are used for exploring a reference genome together with other knowledge-based data or for comparing experimental results, such as sequencing data, to the reference genome.}

Genome browsers usually enable the visualization of a small window of the genome and allow navigation such as zooming and panning. Most of them are not suited for \new{meaningful} overview visualizations of whole genomes, since the data must be extremely aggregated to fit on the screen. Most genome browsers are limited to the visualization of single chromosomes~\cite{Kent2002-fm, NCBI_Resource_Coordinators2017-sl}. Some browsers apply a predefined minimum zoom level\new{~\cite{Down2011-fw}}, others show empty tracks \new{for certain feature sets} if the visualization is zoomed out too much~\cite{Thorvaldsdottir2013-ym}. In terms of the defined visualization parameters, genome browsers often consist of a single view and can therefore only visualize the data at one scale and focus at a time (1-1-1 view configuration, see Table~\ref{tab:FeatureViewer}). Consequently, features can only be \new{explored}, compared, and summarized in a relatively small window \new{which complicates, for example, analyzing gene co-expression in non-adjacent regions}. To counter this, some tools provide methods of data abstraction, with which introns can be filtered out or specific regions of interest can be placed next to each other~\cite{Kent2002-fm, Zhou2011-px}. However, the regions are still visualized in the same view and to change the borders of the regions they have to be redefined. 

The Savant Genome Browser~\cite{Fiume2012-hb} is the only feature viewer that was grouped into the multiple views, single scale, and single focus (n-1-1 view configuration, see Table~\ref{tab:FeatureViewer}). It consists of two visualizations both showing features on genomic coordinates: a classic genome browser view and an additional view which can display population data with different visualizations. For example, it can display a heatmap which shows the patterns of alterations of single bases in multiple samples. Columns correspond to the positions of the alterations and rows correspond to samples and cells show the type of  alteration. Although the heatmap shows the data on a different scale, we associated it with a single scale configuration, since it does not display the same data (the same tracks) on a different scale but constitutes an entirely different visualization.

Another category of genome browsers are ``overview-detail'' browsers~\cite{Donlin2009-pk, Hubbard2002-ar, Freese2016-vl} (n-n-1 view configuration, see Table~\ref{tab:FeatureViewer}). Like single view browsers they show one region of the genome (which means that they have one focus) but also have an additional detailed view for a part of this region. Navigation is usually linked in these browsers to ensure that the detailed view is always part of the chosen region. \new{An advantage of this genome browser layout is that features can be analyzed in a small window as well as in their global neighborhood simultaneously}.

Few genome browsers provide a very high level of flexibility by enabling the visualization of multiple focus regions on different scales~\cite{NCBI_Resource_Coordinators2017-sl, Kong2012-wp,Thorvaldsdottir2013-ym, Cao2013-wt} (n-n-n view configuration, see Table~\ref{tab:FeatureViewer}). Initially, they visualize single regions, but multiple regions can be selected or added through different mechanisms. IGV allows selecting a region with a user dialog~\cite{Thorvaldsdottir2013-ym}, while the NCBI Sequence Viewer enables the creation of new views by highlighting a region and selecting ``create new panel'' in a context menu~\cite{NCBI_Resource_Coordinators2017-sl}. Regions can be selected with different extents, yet they receive the same amount of screen space. Therefore, the two region views can show features on different scales. Region views can be either arranged horizontally (IGV)~\cite{Thorvaldsdottir2013-ym} or vertically (NCBI Sequence Viewer)~\cite{NCBI_Resource_Coordinators2017-sl}. While a horizontal arrangement facilitates comparing tracks, the available horizontal space for each region is smaller. \new{The increased flexibility of genome browsers in an n-n-n view configuration often goes along with a more complicated user interface and more possible interactions. For problems that do not require viewing different scales and regions at the same time, a genome browser in a less flexible configuration can facilitate the exploration process.}

Often genome browsers contain utility views that are strongly connected to the tracks and show different kinds of meta data. When the different tracks correspond to data obtained by analyzing different samples, for example with IGV~\cite{Thorvaldsdottir2013-ym}, they can display data about the sample donor, such as gender, race, ethnicity and many more.

Xena~\cite{Goldman2018-iq} is another type of visualization for non-interconnected feature sets, which can be seen as a ``population browser.'' Unlike a genome browser, it enables the exploration of entire patient cohorts. Xena displays population data in a column-based layout, where each column corresponds to a view showing a different type of feature. Sample data is displayed as parallel tracks in each column. Tracks are sorted by columns, while the order is preserved across columns. Columns can either show meta data about the samples or genomic data, such as gene expression, copy number alterations, or somatic mutations for small regions such as genes. Columns can be added, sorted, and rearranged interactively. 
In contrast to genome browsers, the focus of Xena is not on visualizing data on genomic coordinates, but on stratifying and characterizing populations. Therefore, what is considered meta data and utility views in \new{our taxonomy}, \new{can be seen as} main data types and main views in Xena.

\paragraph*{Circular Layout}
\begin{figure}
    \centering
    \includegraphics[width=0.6\linewidth]{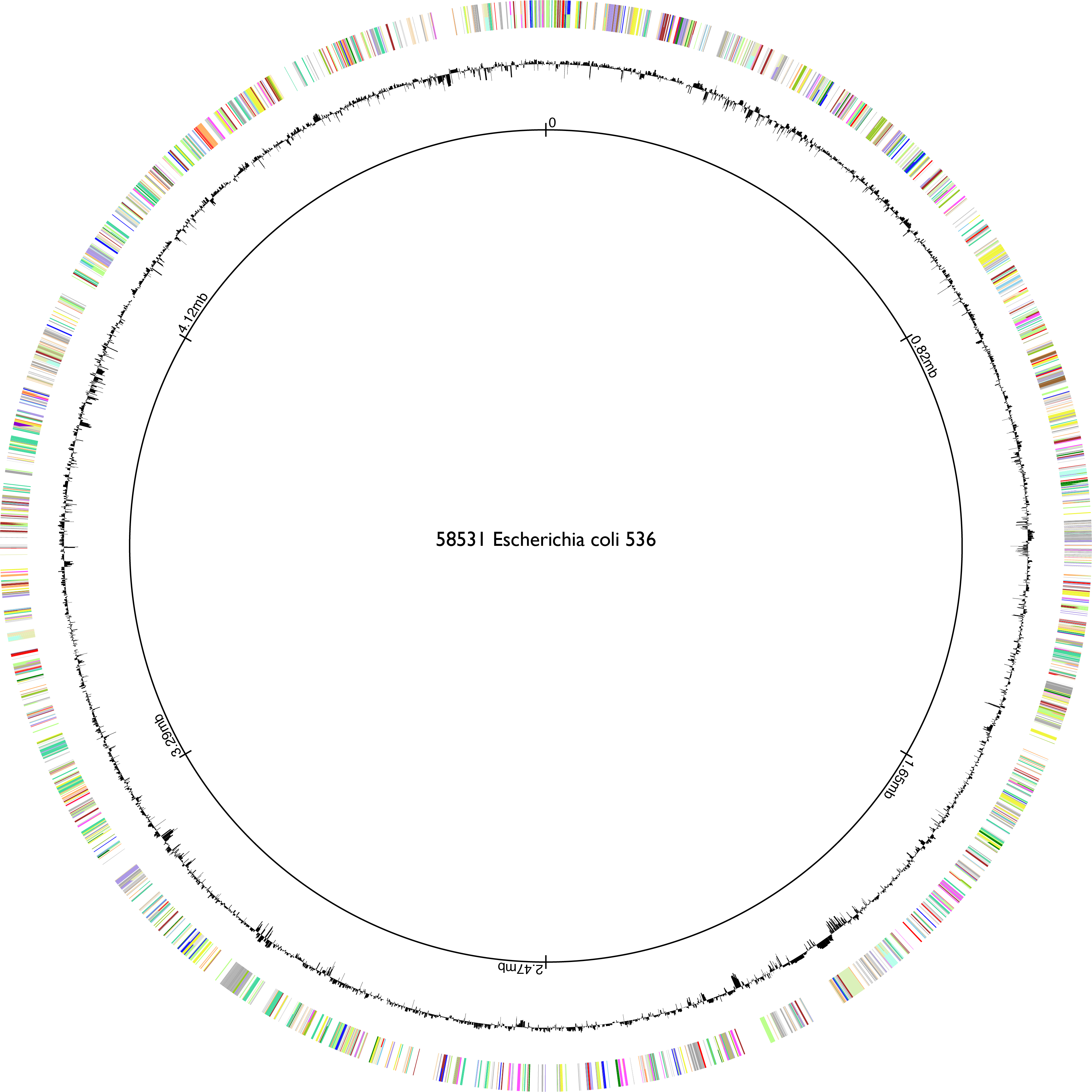}
    \caption{Circular visualization of Escherichia coli with CiVi~\protect\cite{Overmars2015-ks}. The tracks show genes and GC content. \protect\\$D=$ 
    \symb{type-sparse-segment},\symb{type-contiguous-point}+\symb{interconnection-none};
    $T=$
    \symb{alignment-parallel};
    $C=$ \symb{layout-circular}+\symb{partition-contiguous}+\symb{abstraction-none};
    $V=$ \symb{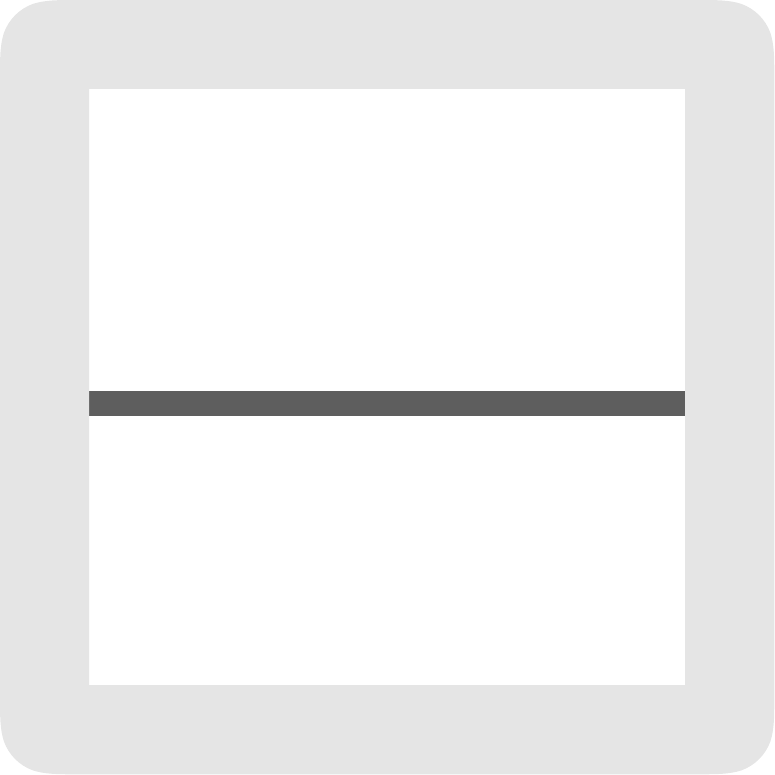}+\symb{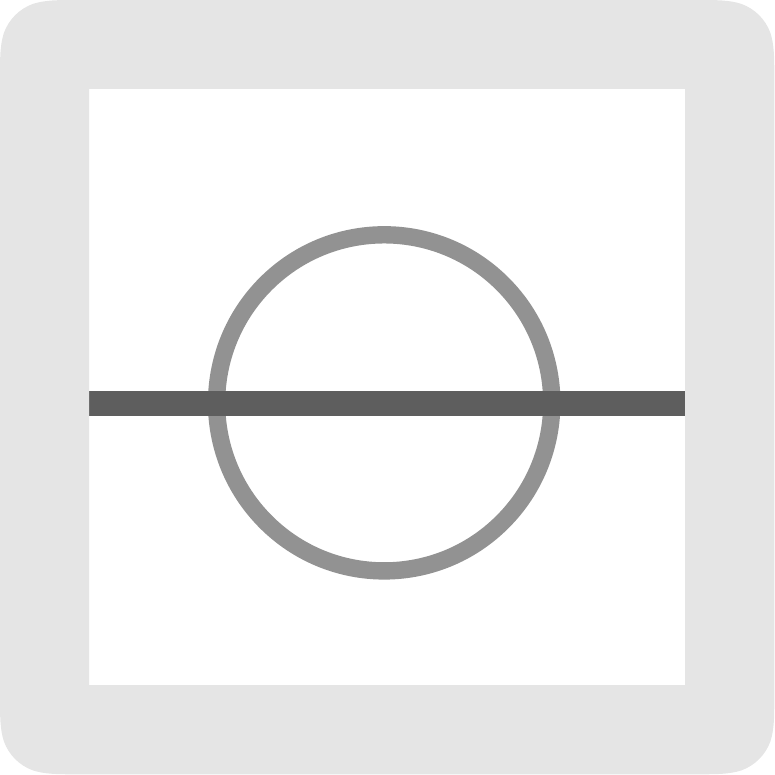}+\symb{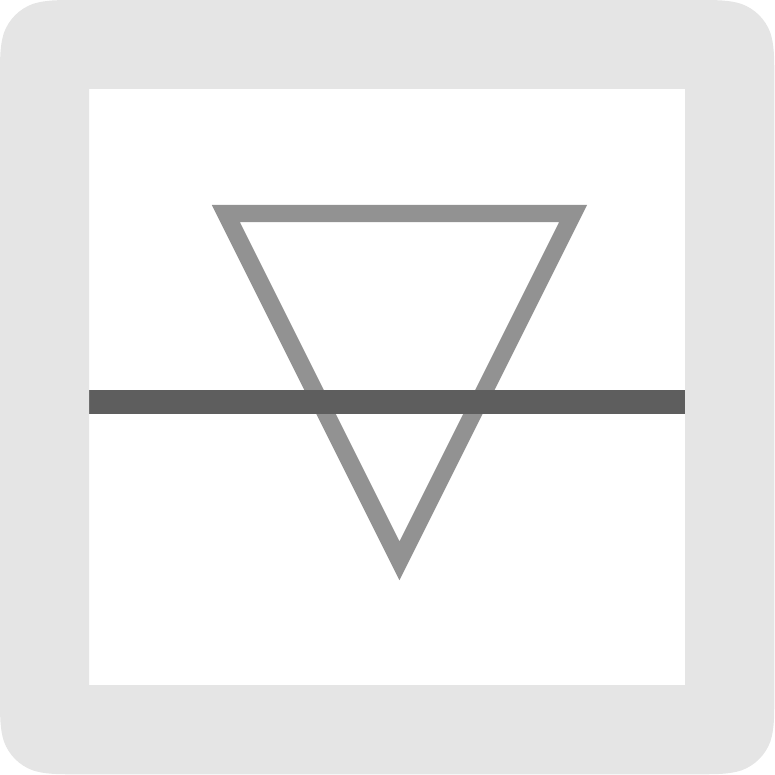}}
    \label{fig:civi}
\end{figure}
Circular layouts are often used for the visualization of non-interconnected feature sets. These visualizations are commonly known as ``genome maps''~\cite{Petkau2010-ya, Kerkhoven2004-kt, Carver2009-va, Stothard2005-bw, Overmars2015-ks, Blom2016-gk, Grant2008-tk}. In contrast to linear visualizations, circular genome maps provide an overview of the entire genome \new{and therefore allow summarizing genomic features on a global scale}. Usually they visualize prokaryotic genomes, which are circular and much smaller than eukaryotic genomes. Figure~\ref{fig:civi} shows an example of a genome map created with CiVi~\cite{Overmars2015-ks}. Despite  smaller genomes, data still has to be aggregated to be visualized. Even the smallest known prokaryotic genome is larger than 500 kilo bases~\cite{Fraser1995-gy}. Contiguous feature sets with quantitative attributes are averaged for windows of equal size, and only segments bigger than a minimum size are displayed. For a more detailed visualization, many tools provide the possibility to visualize the entire genome~\cite{Carver2009-va} or a small region~\cite{Kerkhoven2004-kt, Petkau2010-ya} in a linear layout. Most Genome Maps are static plotting tools and do not provide direct interactivity~\cite{Carver2009-va}. Others allow zooming and navigating the genome map~\cite{Stothard2005-bw, Petkau2010-ya}. 

\new{An example of a more interactive prokaryotic genome visualization is Island Viewer 3~\cite{Dhillon2015-ij}. It can visualize one or two prokaryotic genomes in parallel in an overview-detail configuration (n-n-1). It uses a circular layout for the genome overview and a linear layout for the details. In contrast to the other described circular visualizations, Island Viewer is specialized. It focuses on a specific data type (called genomic islands) and compares different prediction methods for this data.}

\begin{figure}
    \centering
    \includegraphics[width=0.9\linewidth]{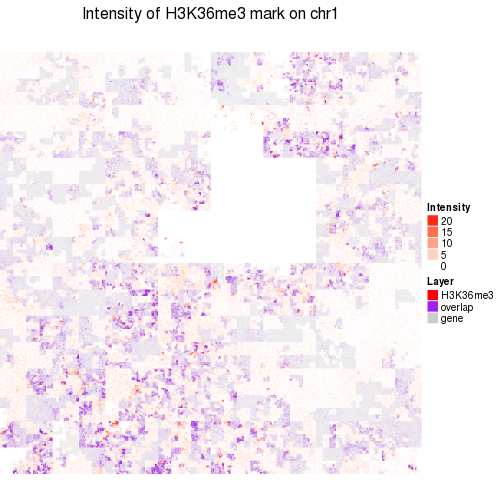}
    \caption{Hilbert curve visualization for an epigenetic marker created with the R package HilbertCurve described in~\protect\cite{Gu2016-pd}. \protect\\$D=$ \symb{type-sparse-segment},\symb{type-contiguous-point}+\symb{interconnection-none};
    $C=$ \symb{layout-spacefilling}+\symb{partition-contiguous}+\symb{abstraction-none};
    $V=$ \symb{view-one}+\symb{scale-one}+\symb{focus-one}}
    \label{fig:hilbert}
\end{figure}

\paragraph*{Space-Filling Layout} 
A visualization technique for chromosome- or genome-wide overviews of features, especially epigenomic marks~\cite{Kharchenko2011-qr} is based on a genome layout that uses a space-filling curve rather than a linear or circular genome layout. A desirable property of space-filling curves, such as the Hilbert curve~\cite{Hilbert1935-wg}, is that they arrange 1D sequence information on a 2D grid so that features close to each other in 1D, are close to each other in 2D. Such plots can be created with HilbertVis~\cite{Anders2009-sr} and the HilbertCurve~\cite{Gu2016-pd} R package (see Figure~\ref{fig:hilbert}). Unlike linear and circular genome layouts, parallel arrangements of tracks are not feasible for space-filling curve layouts, limiting the approach to single or a limited number of overlaid tracks. 

\subsubsection{Sparsely Interconnected Feature Sets}

\begin{table*}[t]
\centering
\caption{Layouts (linear \symb{layout-linear}, circular 
\symb{circular-layout} and view configurations of tools for the visualization of sparsely distributed interconnected features. Tools marked with * apply a form of abstraction.}\label{tab:SegmentInterconnectivity}

\begin{tabularx}{\textwidth}{lcccX}
\toprule
Layout & Views & Scales & Foci &  \\
\midrule
\symb{layout-linear} &  1 &  1 &  1 &  SpliceSeq~\cite{Ryan2012-xr}, SpliceGrapher~\cite{Rogers2012-pv}, SashimiPlots (IGV)~\cite{Katz2015-zo}, Vials~\cite{Strobelt2016-bz}*, ggBio~\cite{Yin2012-eh}, SplicePlot~\cite{Wu2014-la} \\
\cmidrule(r){2-5}
 &  n &  1 &  1 &  Savant Genome Browser 2~\cite{Fiume2012-hb} \\
\cmidrule(r){2-5}
 & n & n & 1 & Gremlin~\cite{OBrien2010-zc} \\
 \midrule
\scalerel*{\includegraphics{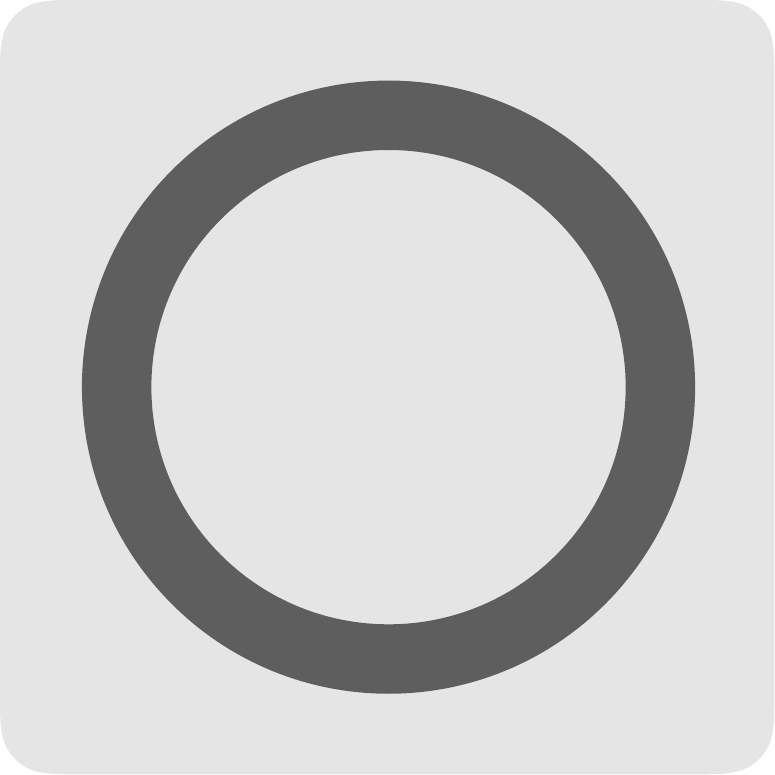}}{)} &  1 &  1 &  1 & Circos~\cite{Krzywinski2009-yv}, ClicO Free Service~\cite{Cheong2015-lz}, J-Circos\cite{An2015-iv}\\
\bottomrule
\end{tabularx}
\end{table*}

\new{Sparsely interconnected feature sets correspond to a set of segment features with interconnections that are sparsely distributed across the genome.} Researchers often deal with this kind of data when studying structural variation or alternative splicing.
Both fields of research are based on mapping reads to the genome and deducting patterns from this mapping. Patterns of alternative splicing can be found by mapping reads obtained with RNA sequencing to a DNA sequence. Based on the number of reads per exon and reads that contain sequences of multiple exons (junction reads), splicing patterns can be deducted. Genomic rearrangements are obtained by mapping DNA sequencing reads to the genome. Rearrangements can be found by analyzing read depth, paired-end reads, and reads containing sequences of distant regions (split reads). For a more exhaustive description of the biological background see section \ref{sec:background}. Sparsely interconnected feature \new{sets} are commonly visualized in a linear or circular layout (see Table \ref{tab:SegmentInterconnectivity}). \new{The main goals for visualizing these data types are exploring the data as well as comparing patterns across different samples and conditions. Especially for genomic rearrangement, both the global distribution of the arrangements, as well as patterns on the sequence level are of interest for the exploration.}

\paragraph*{Linear Layout}
Alternative splicing is usually visualized in a linear layout and is sometimes displayed as a track in a genome browser \cite{Thorvaldsdottir2013-ym}. In principle, alternative splicing data corresponds to a \new{set of segment features}, where each segment encodes for an exon. Exons are connected with interconnection features, which show which exons are contained in the final protein product. More specifically, an interconnection shows that there exists at least one read that contains parts of both exons. Often, exon segments are associated with the number of reads mapped to the exon as well as interconnection features are associated with a quantitative attribute that represents the number junction reads. 

This type of feature is commonly visualized in a ``splice graph'', where exon segments are displayed as nodes and interconnection features as segments, as first proposed in 2004 \cite{Xing2004-cd}.  This technique is applied by many tools including SpliceSeq \cite{Ryan2012-xr} and the R package SpliceGrapher \cite{Rogers2012-pv}. 
A disadvantage of splice graphs is that they do not visually encode for the number of reads mapped to the exons and splice junctions but only use labels. Therefore, SpliceGrapher provides additional tracks for supported junctions and a view visualizing read coverage with an area chart.

\begin{figure}
\centering
\includegraphics[width=0.9\linewidth]{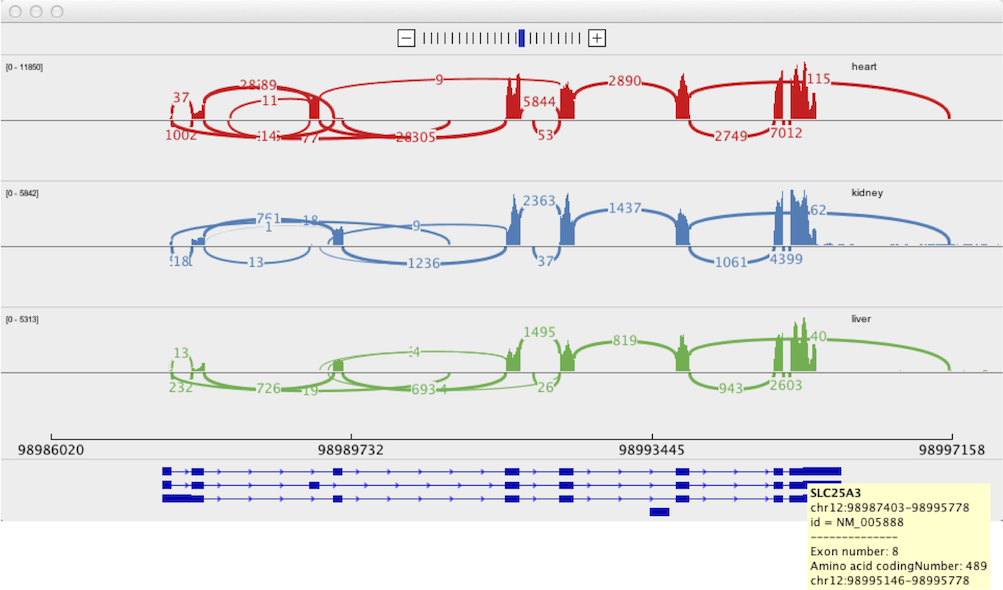}

\caption{
Sashimi plots from IGV~\protect\cite{Broad_Institute_and_the_Regents_of_the_University_of_California_undated-mf} enable users to analyze splicing patterns of different samples. The bottom view shows known isoforms, the top view shows splicing patterns of three samples. At exons read density is displayed with area charts, junction reads are displayed as arcs labeled with the number of reads. \protect\\$D=$ \symb{type-sparse-point}+\symb{interconnection-within};
    $C=$ \symb{layout-linear}+\symb{partition-contiguous}+\symb{abstraction-none};
    $T=$ \symb{alignment-parallel};
    $V=$ \symb{view-one}+\symb{scale-one}+\symb{focus-one}
}
\label{fig:sashimi}
\end{figure}
A commonly used technique uniting splice graphs and quantitative data in one visualization are ``Sashimi plots''~\cite{Katz2015-zo}. Figure \ref{fig:sashimi} shows the splicing patterns of three samples with Sashimi plots created with IGV \new{\cite{Broad_Institute_and_the_Regents_of_the_University_of_California_undated-mf}}. A Sashimi plot is displayed using an interconnected valued point track. Each nucleotide in an exon is associated with a quantitative attribute, the read depth, which is encoded using an area chart. Nucleotides that form junction sites are connected by interconnection features with quantitative attributes, encoded as weighted arcs. The view on the bottom of Figure \ref{fig:sashimi} shows known isoforms using a segment track by encoding sequences contained in the isoform as blocks and spliced-out sequences as lines. \new{In IGV, multiple Sashimi plots can be visualized in separate tracks, which allows the comparison of splicing patterns across samples or conditions.} In order to summarize splicing patterns at a population level, the command line utility SplicePlot \cite{Wu2014-la} produces averaged Sashimi plots to find splicing patterns that manifest as phenotypic differences between population groups. 

Sashimi plots often show numerous interconnections, which complicates the identification of isoforms and comparisons of multiple plots. Vials~\cite{Strobelt2016-bz} was designed to address the shortcomings of Sashimi plots and to visualize and compare data from multiple samples at once, by distributing the data across multiple tracks showing junctions, abundances, and expression levels. Elements in the tracks corresponding to the same population are linked by brushing and linking. As Vials allows visualizing one gene at a time, it is a one view, one scale, one focus visualization. Yet, it applies a method of abstraction. Introns can be displayed in their full length or abstracted to equally sized gaps. In comparison to Sashimi plots, Vials provides a more abstract form of alternative splicing visualization which can make it harder \new{for novices} to interpret.

\begin{figure}
\centering
\includegraphics[width=0.9\linewidth]{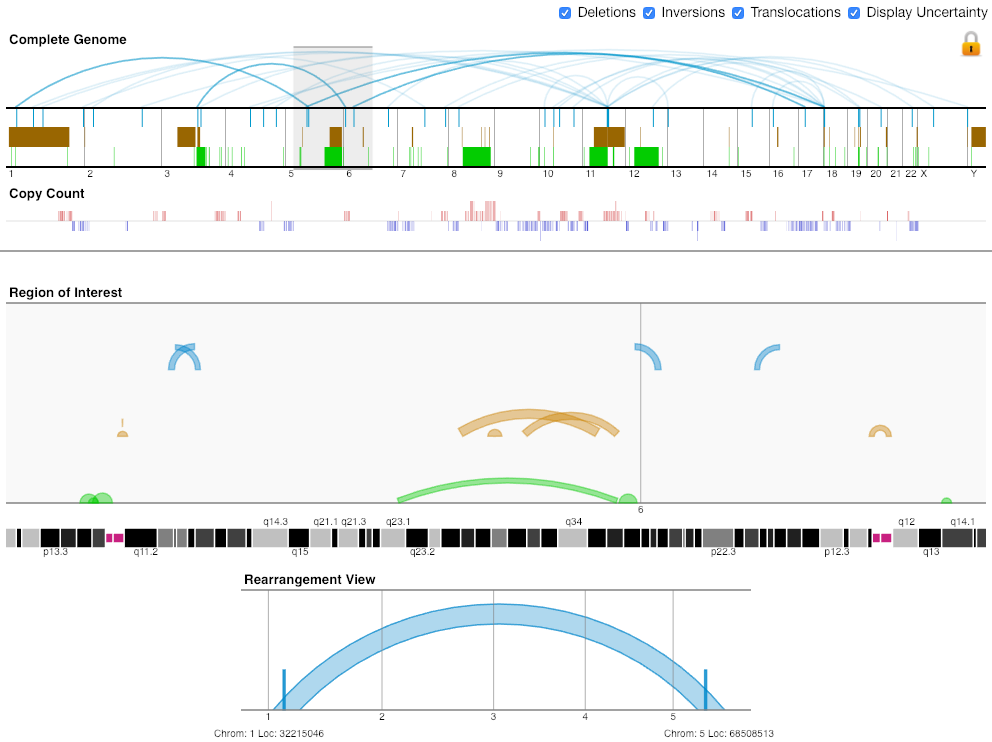}
\caption{
Gremlin~\protect\cite{OBrien2010-zc} enables users explore the high-level, complete genome perspective as well as low-level, structural rearrangement view. \protect\\$D=$ \symb{type-sparse-segment}+\symb{interconnection-within};
    $C=$ \symb{layout-linear}+\symb{partition-contiguous}+\symb{abstraction-none};
    $T=$ \symb{alignment-parallel};
    $V=$ \symb{view-many}+\symb{scale-many}+\symb{focus-one}
}
\label{fig:gremlin}
\end{figure}
Similar to alternative splicing patterns, genomic rearrangements can be displayed as arcs \new{on a sequence axis in a linear layout}. 
Gremlin ~\cite{OBrien2010-zc} is an overview-detail visualization that can show rearrangement events at three separate scales, informing users about the context of where the breakpoints of a rearrangement occur (see Figure~\ref{fig:gremlin}).  It allows navigation across the genome by selecting regions of interest with a sliding window. Within these regions, single arcs corresponding to structural rearrangements can be selected and viewed in even more detail. In the complete genome view, as well as in the region of interest, tracks display the location and extent of deletions, inversions and translocations and copy number data. With multiple linked views, it supports both global trend analysis and local feature detection. As a result, this visualization enables users explore the high-level, complete genome perspective as well as low-level, structural rearrangement view.

\paragraph*{Circular Layout}

One of the most common ways to represent \new{structural rearrangements} spanning large genomic distances is by displaying the chromosomes in a circular layout and drawing arcs between connected regions with tools like Circos~\cite{Krzywinski2009-yv}.
\new{Like Gremlin, Circos can encode translocations, inversions, and deletions using a combination of interconnections and segment tracks. Tracks showing other genomic features can be stacked and can apply visual encodings such as scatterplots and heatmaps.} Depending on the size of the depicted region, Circos-style plots can show smaller or larger relationships between distant genomic regions.  Though widely used, such depictions are limited in their ability to show data at varying scales. Zooming on a circular genomic representation is unintuitive and seldom implemented. ClicO Free Service~\cite{Cheong2015-lz} is an online web-service which provides a user-friendly, interactive, web-based interface with configurable features to generate Circos circular plots.

\subsubsection{Densely Interconnected Feature Sets}
\begin{table*}[t]
\centering
\caption{Layouts (linear~\symb{layout-linear}, circular~\symb{layout-circular}, spatial~\symb{layout-spatial}), Arrangements (A) (parallel~\symb{arrangement-parallel-linear}, orthogonal~\symb{arrangement-orthogonal}) and view configurations of tools for the visualization of densely interconnected feature \new{sets}. Tools marked with * can apply a form of abstraction.}\label{tab:SequencInt}

\begin{tabularx}{\textwidth}{lccccX}
\toprule
L & A & Views & Scales (1 Axis/2 Axes) & Foci (1 Axis/2 Axes) & \\
\midrule
\symb{layout-linear} &  - & 1 & 1/- & 1/- & HUGIn~\cite{Martin2017-zb}, 3d Genome Browser~\cite{Wang2018-vi} \\
\cmidrule(r){2-6}
& \symb{arrangement-orthogonal} & 1 & 1/1 & n/1 & Juicebox~\cite{Durand2016-cp}, HiCExplorer~\cite{Wolff2018-ni} \\
& &  n &  1/1 & n/1 &  Juicebox.js \cite{Robinson2018-ls}\\
& &  n &  1/1 & n/1 &    my5c~\cite{Lajoie2009-fc} \\
& &  n &  n/n & n/n & HiGlass~\cite{Kerpedjiev2018-pk},  HiPiler~\cite{Lekschas2018-yp} \\
\cmidrule(r){2-6}
& \symb{arrangement-parallel-linear} & 1 & n/1 & n/1 & GIVE~\cite{Cao2018-sx} \\
\midrule
\symb{layout-circular} & - & 1 & 1/- & 1/-  & Rondo~\cite{Taberlay2016-jn} \\
\midrule
\symb{layout-spatial} & - & 1 & 1/- & 1/- & 3DGB~\cite{Butyaev2015-vm}, \new{Hi-C3d} Viewer~\cite{Djekidel2017-gg}\\
\bottomrule
\end{tabularx}
\end{table*}

Densely interconnected feature \new{sets} correspond to \new{a set of segment features} with interconnections that are densely distributed \new{across the sequence}. This type of feature set can encode interaction frequencies, which are a measure for the spatial distance between two segments. It is possible to measure the interaction frequencies of all possible segment pairs of a genome (Hi-C) or only interaction frequencies between one segment and all the other segments (4C).  In terms of the previously defined feature types, it corresponds to a contiguous set of segment features of equal size that are interconnected with undirected interconnection features with a quantitative attribute. \new{Exploring this kind of data can lead to valuable insights concerning the 3D structure of genomes under different conditions. Patterns of interest include compartmentalization for identifying active/inactive regions of the genome and topologically associating domains (TADs)}. \new{We found visualizations of this type of feature in all kinds of layouts and parallel and orthogonal arrangements (see Table \ref{tab:SequencInt}).}

\paragraph*{Linear Layout, Orthogonal Arrangement}

From the beginning, orthogonal arrangements have been the standard in displaying genome-wide interaction frequencies. Each axis represents the same sequence and the spanned heatmap matrix encodes the interaction frequencies with color. By varying the bin size, heatmaps can represent the data and relevant features at different scales and different resolutions.

\begin{figure}
\centering
\includegraphics[width=0.9\linewidth]{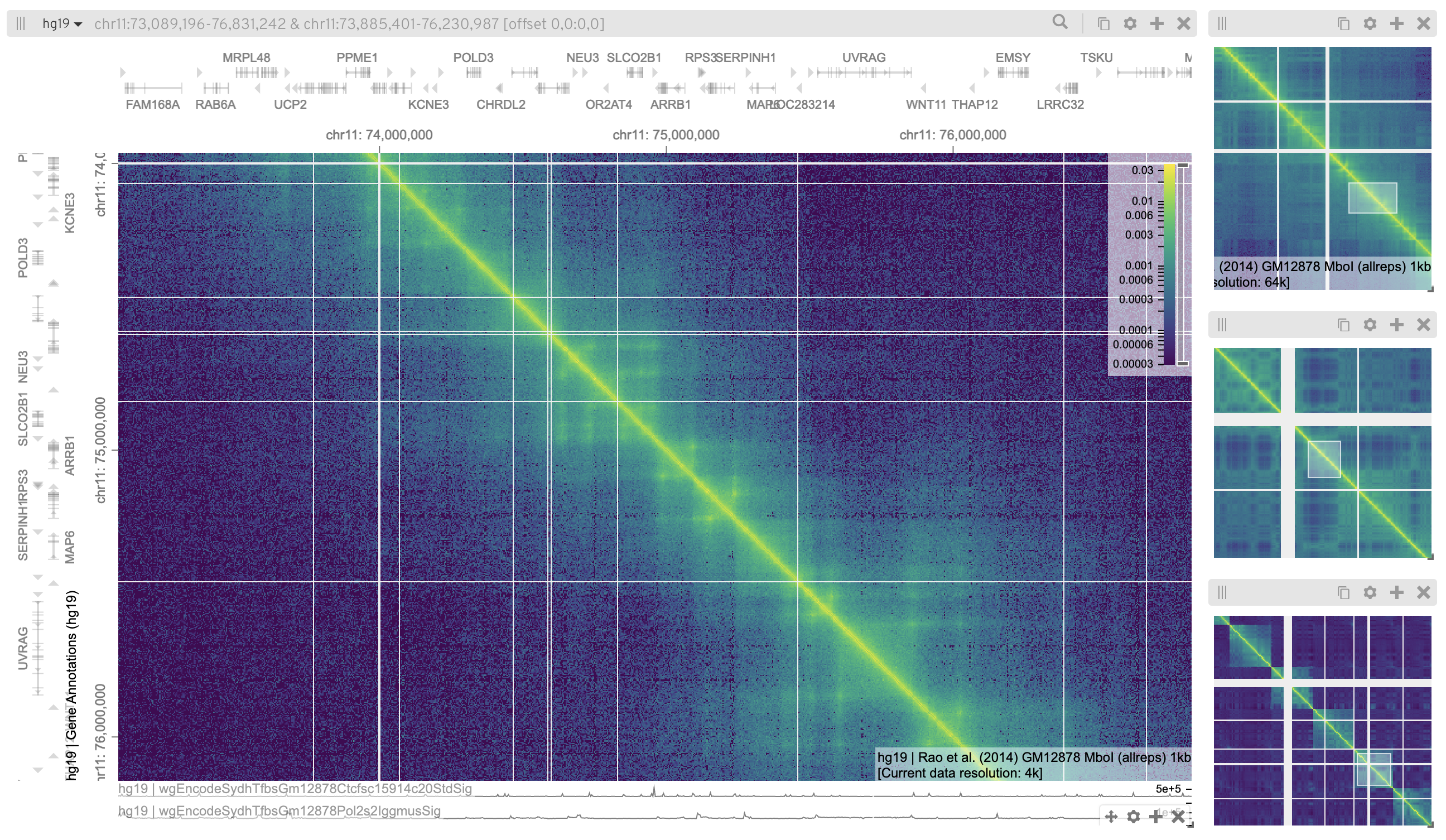}\hspace{.9cm}
\caption{
Orthogonal arrangement in HiGlass~\protect\cite{Kerpedjiev2018-pk}. Interactions are encoded using a heatmap encoding. Additional tracks are aligned on both sequence axes.\protect\\$D=$ \symb{type-sparse-segment},\symb{type-contiguous-segment}+\symb{interconnection-within};
    $C=$ \symb{layout-linear}+\symb{partition-contiguous}+\symb{abstraction-none};
    $T=$ \symb{alignment-parallel};
    $V=$ \symb{view-many}+\symb{scale-many}+\symb{focus-many}
}
\label{fig:heatmap-arc}
\end{figure}

An interconnectivity visualization in an orthogonal arrangement can be understood as a genome browser-like tool with two axes instead of one to display interconnection features more effectively (see Figure \ref{fig:heatmap-arc}). Similar to genome browsers, these tools can arrange multiple tracks in parallel, such as tracks showing gene annotations, epigenetic signals, or gene expression. As orthogonal layouts contain two axes, there can be one-axis foci, which correspond to the focus on each axis, and two-axis foci, which correspond to the foci in the matrix spanned by the axes. Table \ref{tab:SequencInt} shows both kinds of foci and scales, yet the main view configuration corresponds to the two-axes scales and two-axes foci.

Tools such as Juicebox and HiCExplorer are single view, single focus, single scale visualizations \cite{Durand2016-cp, Wolff2018-ni} (see Table \ref{tab:SequencInt}). While HiCExplorer is a plotting tool which offers multiple different visualizations, Juicebox is more interactive and allows unlimited zooming. The web application Juicebox.js \cite{Robinson2018-ls} allows the visualization of multiple matrices using linked panning and zoom (n-1-1 view configuration).

\begin{figure}
\centering
\includegraphics[width=0.9\linewidth]{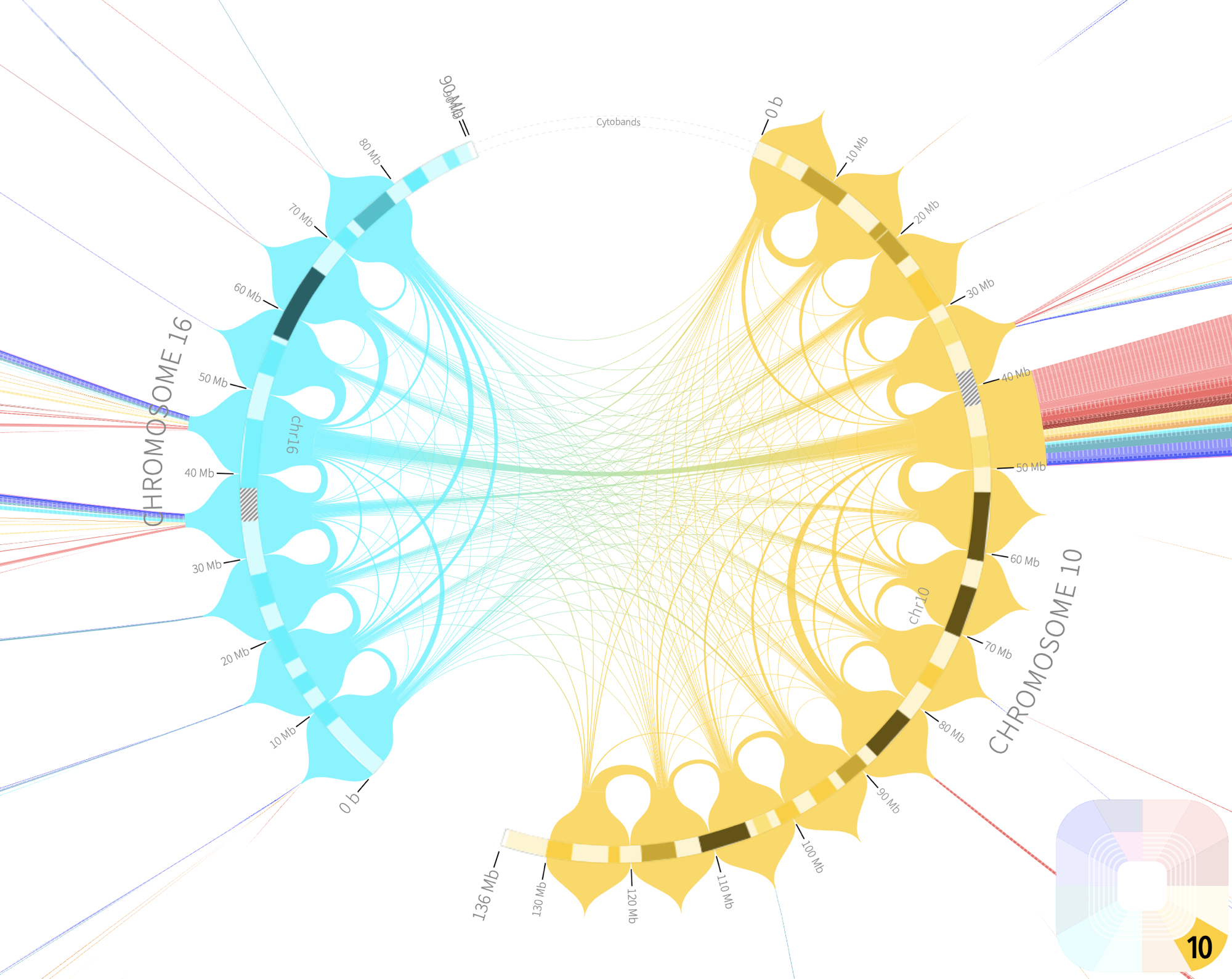}
\caption{
Circular layout in Rondo~\protect\cite{Taberlay2016-jn}. Chromosomes are encoded using color, bands represent interactions. \protect\\$D=$ \symb{type-sparse-point}, \symb{type-contiguous-segment}+\symb{interconnection-within};
    $C=$ \symb{layout-circular}+\symb{partition-contiguous}+\symb{abstraction-none};
    $T=$ \symb{alignment-parallel};
    $V=$ \symb{view-one}+\symb{scale-one}+\symb{focus-one}}
\label{fig:rondo}
\end{figure}

The tool HiGlass is very flexible in terms of creating different views and offers unlimited zooming~\cite{Kerpedjiev2018-pk} (n-n-n view configuration). Multiple matrices can be placed next to each other and navigation can be linked to compare features at different conditions. Moreover, it is possible to create overview-detail visualizations, where users can move and change the size of a sliding window placed in the main matrix to view the contents of the window in a detailed view. Multiple sliding windows can be created to visualize multiple foci. 

HiPiler is a tool building on HiGlass for the analysis of regions of interest (ROIs) in genome interaction matrices~\cite{Lekschas2018-yp}. It displays ROIs as \textit{snippets}, which are small, detailed regions of the matrix.  Similar snippets are overlaid with a visualization of the average snippet on top of the pile. The arrangement of views and tracks of HiPiler is unique in our visualization taxonomy. Every pile of snippets represents a separate view, whereas each pile contains overlaid tracks, with a special track showing the aggregation on top. The snippet views are arranged by a clustering algorithm or a scatterplot. Snippets and the main matrix visualization are connected via brushing and linking. \new{In contrast to the other presented matrix visualization, HiPiler combines pattern extraction with visual analysis in order to assess the quality of the extracted patterns.}

\paragraph*{Linear Layout, Parallel Arrangement}
In our taxonomy the tool GIVE~\cite{Cao2018-sx} represents a unique arrangement of axes. Two axes are displayed in parallel and the same set of tracks is stacked on each axis. Interconnections are displayed as bands between the two axes in the two-axes track. Each axis can be navigated independently, therefore the visualization can display a different focus and scale on the single axes (see Table \ref{tab:SequencInt}). Since interconnections can only be visualized between the two one-axis foci, there is only one focus and scale in the two-axes track. \new{A possible reason why this arrangement of axes is rarely implemented for the analysis of Hi-C data is that when viewing the arrangements on a global scale, many bands overlap and it is hard to identify regions of interest to investigate further. }

\paragraph*{Linear Layout, Single Axis}

Interconnectivity can be visualized in a linear layout with a single axis and is implemented in some genome browsers such as the WashU Epigenome Browser~\cite{Zhou2011-px} \new{and the 3d Genome Browser~\cite{Wang2018-vi}}. As described in section~\ref{sec:TrackMatrix}, symmetrical matrices can be cut in half and rotated in order to display them as a track. Usually not the entire matrix is shown, but only interactions close to the diagonal to reduce the size of the track. Additional to its matrix visualization HiGlass also provides this kind of track~\cite{Kerpedjiev2018-pk}. \new{While this is useful to analyze short-range interactions in a genome browser like setting, usually long-range interactions cannot be analyzed with this type of visualization.}

\new{The 3d Genome Browser additionally implements} another way of displaying Hi-C data in a linear layout~\cite{Wang2018-vi}, \new{which is similarly implemented by HUGIn~\cite{Martin2017-zb}}. Instead of showing interaction frequencies between all segments, they only show interaction between one segment and all other segments which corresponds to a virtual 4C plot. This is especially useful to visualize interactions between a biological relevant region, such as a gene, and all the other segments. \new{However, selecting the biological relevant region requires previous knowledge of the users and only a part of the feature set can be explored at a time}. \new{Both tools} encode Hi-C data as a line chart, where the y-axis corresponds to the interaction frequency. \new{Moreover, other tracks showing additional data can be added and arranged in parallel.}


\paragraph*{Circular Layout} 

A circular layout for the visualization of Hi-C data is implemented by Rondo~\cite{Taberlay2016-jn}, which uses a chord diagram to display interactions. As there are interactions between all segments, Rondo clusters individual interactions into larger groups. Every chromosome is encoded with a different color (see Figure~\ref{fig:rondo}). Rondo corresponds to a one view, one scale, one focus configuration and does therefore not allow viewing multiple circular plots of different datasets next to each other. However, it can encode for the comparison of two datasets that use the same coordinate system (both of the same species) by encoding the arcs corresponding to the different datasets with color and introducing a color for arcs in which the datasets overlap. \new{In contrast to visualizations with an orthogonal axis arrangement it is not possible to compare more than two conditions.} Rondo offers navigation and selection interactions. When zooming in, interaction clusters are progressively separated to show the interactions in more detail. 

\paragraph*{Spatial Layout}

A 3D representation of genome structure displays reconstructions of the 2D profiles obtained using interaction-based methods such as Hi-C. They are deceptively reminiscent of protein structure visualization, depicting a single conformation of nuclear DNA. Tools such as HiC-3DViewer and 3DGB let users pan and zoom around a reconstruction and observe the proximity of likely neighbors~\cite{Djekidel2017-gg, Butyaev2015-vm}. They allow for the overlay of a gene annotation track and the extraction of data associated with genomic locations. Despite their capacity for displaying seemingly faithful representations of DNA structure in the nucleus, 3D visualization tools are severely limited by their reliance on algorithmic reconstructions of that structure. This hides information about heterogeneity and ambiguity in the underlying data and presents one of potentially many solutions to the constraints provided by contact mapping experiments.

\subsection{Feature-Scale Visualizations}
\begin{table*}[t]
\centering
\caption{Layout (linear \symb{layout-linear}) and view configurations for feature-scale visualizations}\label{tab:aggregation}
\begin{tabularx}{\textwidth}{lcccX}
\toprule
Layout & Views & Scales & Foci &  \\
\midrule
\symb{layout-linear} &  1 &  1 &  1 &  WebLogo~\cite{Crooks2004-js}, pLogo~\cite{OShea2013-xh}, Two Sample Logo~\cite{Vacic2006-hy}, Sequence Bundles~\cite{Kultys2014-xf}, Alvis~\cite{Schwarz2016-vu}, Deep Motif Dashboard~\cite{Lanchantin2017-pm}, Lollipop Plot cBio~\cite{Cerami2012-uw}, Variant View~\cite{Ferstay2013-sj}, ggBio~\cite{Yin2012-eh} \\
\cmidrule(r){2-5}
 &  n & 1 & 1 & MAGI~\cite{Leiserson2015-vz}, deepTools Heatmap\cite{Ramirez2016-xk}, EnrichedHeatmap~\cite{Gu2018-zy}\\
 \cmidrule(r){2-5}
 &   n &  n &  1 &  MochiView~\cite{Homann2010-kt}, MEME~\cite{Bailey2006-tr}\\
 \bottomrule
\end{tabularx}
\end{table*}

Features of the same type at regions within or across sequences can be summarized in order to find and visualize patterns. Three types of features are commonly summarized: (i) contiguous numerical features for the visualization of epigenetic signals at regions of interest, (ii) point features across samples for the visualization of genetic variants, and (iii) subsequences of the same length within a sample for the visualization of motifs\new{, which are short, reoccurring subsequences that have a biological significance, such as a transcription factor binding site.} 

Summarizing features can decrease the amount and length of tracks that are displayed. Instead of locating, comparing, and browsing features within or across multiple tracks, the regions of interest are found by computational methods and displayed in one visualization. All summary visualizations displayed in Table~\ref{tab:aggregation} apply a linear layout, presumably since they usually show short sequences (single genes or only few bases) and do not display interconnection features, which often require the use of circular layout or an orthogonal arrangement of axes.

\subsubsection{Non-Aggregated Feature Summaries}
A common task when analyzing regions of biological importance, such as binding sites or genes, is exploring and summarizing the epigenetic signals around all instances of a particular type of a genomic feature, such as transcription start sites, protein-binding sites, or across the coding sequence of genes. 

Feature-by-position heatmaps are a commonly used visualization technique that supports this task. They are constructed by assigning sequence windows around feature instances to tracks and arranging the tracks in parallel. The windows can be stretched to normalize across sequences of unequal length, such as coding regions for genes. Sequence positions are relative to the genomic position of each feature instance. The strength of the quantitative attribute is mapped to a sequential or diverging color map. Tracks can be ordered by mean or median signal across the columns of positions and can be clustered as well. A line chart, called a \textit{gene body plot} often summarizes the heatmap, by showing an average value for each column of the heatmap. The stretching of sequence coordinates represents an exception to the track stacking described in section \ref{sec:TrackMatrix}, since the tracks are not aligned by sequence coordinates but stretched to align starts and ends of segment features of unequal size. 

As each heatmap can only represent one epigenomic mark, multiple juxtaposed heatmaps are needed if more than one mark of interest should be visualized. Visualization tools that can generate these types of plots are deepTools~\cite{Ramirez2016-xk}, which is both available through a Galaxy-based user instance and as a stand-alone command line tool, as well as the EnrichedHeatmap~\cite{Gu2018-zy} R package (n-1-1 view configuration, see Table \ref{tab:aggregation}).

\subsubsection{Aggregated Feature Summaries}
\paragraph*{Summarizing Point Features}


\begin{figure}[t]
    \centering
    \includegraphics[width=\linewidth]{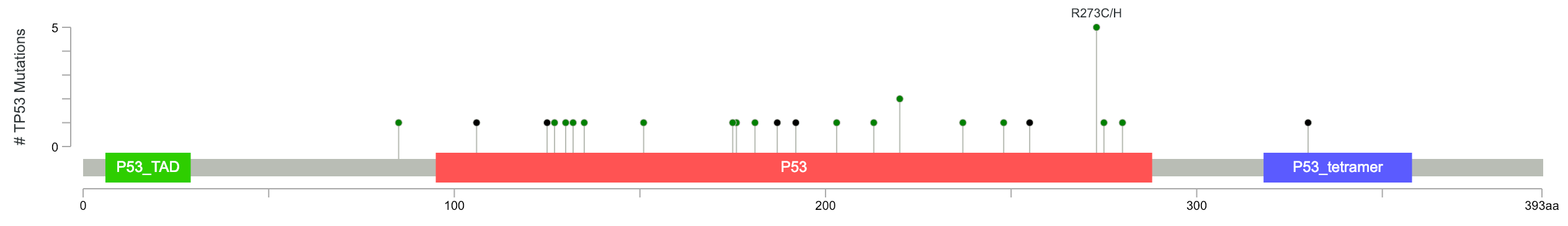}
    \caption{Lollipop plot created using the cBio Portal~\protect\cite{Cerami2012-uw}. A lollipop plot is an example for the aggregation of sets of point features. Each dot represents a mutation at a given position in the amino acid sequence of the translated gene sequence (x-Axis) and with a corresponding abundance (y-Axis). The colored boxes correspond to protein domains.\protect\\ $D=$ \symb{type-sparse-point},\symb{type-sparse-segment}+\symb{interconnection-none};
    $C=$ \symb{layout-linear}+\symb{partition-contiguous}+\symb{abstraction-none};
    $T=$ \symb{alignment-overlaid};
    $V=$ \symb{view-one}+\symb{scale-one}+\symb{focus-one}}
    \label{fig:lollipop}
\end{figure}
A genetic variant is a base alteration that can be evident for a specific phenotype of cancer. The basic goals of variant visualizations are exploring variant patterns together with protein domains, comparing the variants at different regions of the gene and correlating them with different disease phenotypes.

A commonly used visualization are lollipop plots which aggregate variants of a gene for an entire cohort (see Figure~\ref{fig:lollipop} \cite{Cerami2012-uw}). 
A lollipop plot consists of a valued point track and a valued segment track that are overlaid. The valued point track shows the number of variants (y-axis) at each position (sequence axis) for the entire cohort. For each mutation a dot is drawn at the corresponding position and abundance and the dot is connected to sequence axis with a line. Often the dots are annotated with the resulting protein change of the variant. The valued segment track shows different annotated protein domains.

A stand-alone tool for visualizing variants is Variant View~\cite{Ferstay2013-sj}. In principle, it is similar to lollipop plots, but it shows additional tracks, such as further protein annotations. Moreover, it does not show the frequency of the variants. Each variantis shown separately with a detailed annotation, such as the mutation type and the change in the amino acid. In addition to the main visualization, it contains a weakly linked utility view in form of a table that contains further information about the displayed variants.

The tool MAGI is a multi view, single scale, single focus visualization (n-1-1, see Table \ref{tab:aggregation}) \new{for population data}, which shows mutation data and copy number data on sequence coordinates, as well as numerous other views for the visualization of non-sequence related data, such as meta data like age and gender \cite{Leiserson2015-vz}. For the visualization of mutations, MAGI implements a visualization similar to a lollipop plot. \new{Mutations are encoded by symbols and stacked at the corresponding positions. For each mutation, the shape of the symbol and its color, as well as its placement above or below the axis, encodes for information about the mutation type.}

\paragraph*{Summarizing Sequences}

\begin{figure}[t]
    \centering
    \includegraphics[width=0.4\textwidth]{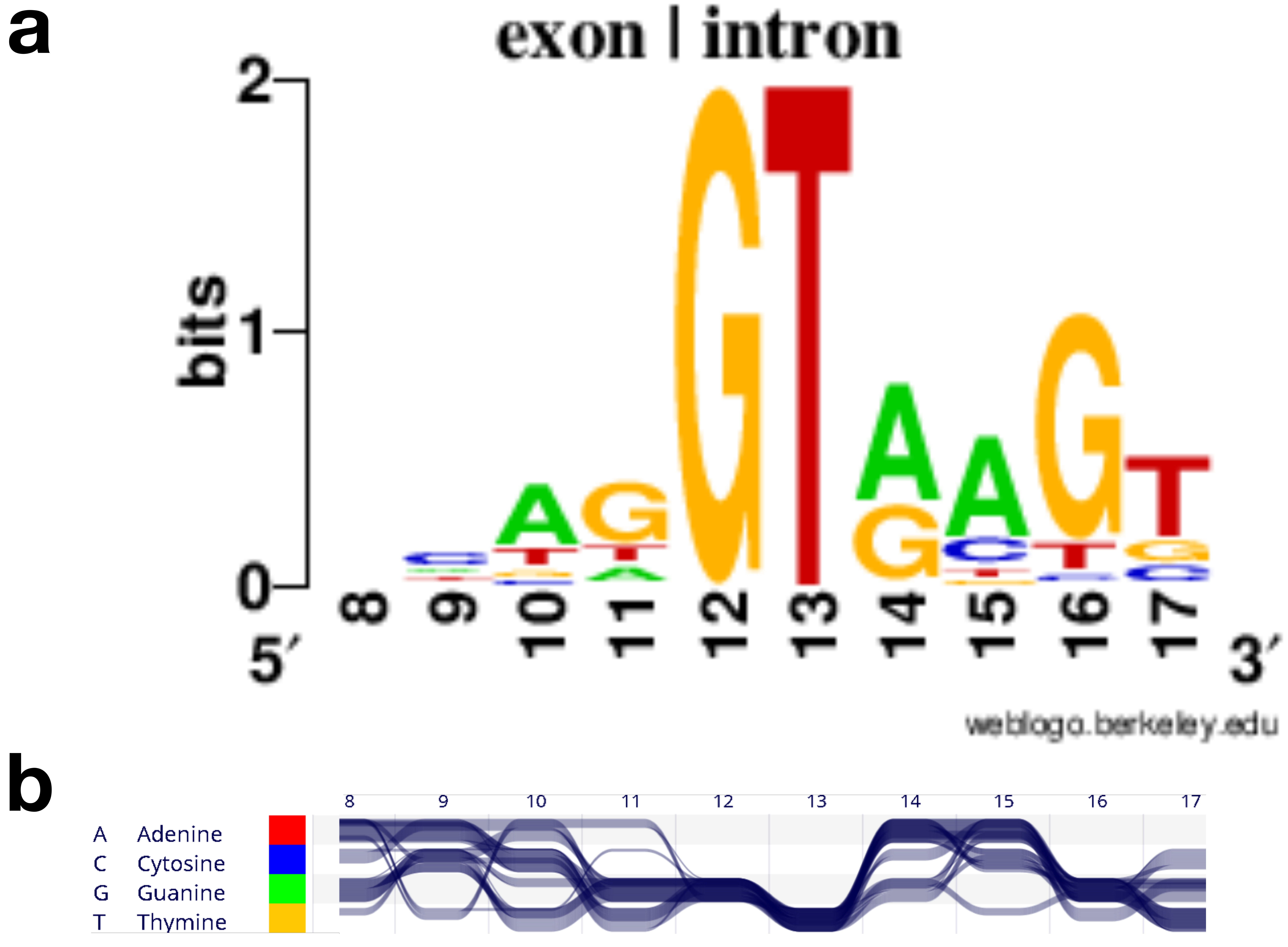}
    \caption{Representations of the same sequence motif for a small sample of human intron-exon splice boundaries by (a) a sequence logo created with WebLogo~\cite{Crooks2004-js} and (b) a sequence bundle created with Alvis~\protect\cite{Schwarz2016-vu}. \protect\\$D=$ \symb{type-contiguous-point}+\symb{interconnection-none};
    $C=$ \symb{layout-linear}+\symb{partition-contiguous}+\symb{abstraction-none};
    $V=$ \symb{view-one}+\symb{scale-one}+\symb{focus-one}}
    \label{fig:LogoBundle}
\end{figure}
In contrast to variant visualizations, motif visualizations do not show base alterations that are evident of a disease or a phenotype. They show the composition of bases at a reoccurring sequence element, such as transcription factor binding sites. Motifs can be found by analyzing sequence patterns within a genome or across species. Motif visualizations are used to explore and summarize the base composition of short reoccurring biological relevant subsequences, or to compare the base composition at different positions within a motif.  

Sequence logos, which are the basis of most motif visualizations, were introduced in 1990~\cite{Schneider1990-up}. In order to construct a motif, the authors of the paper created an ungapped multiple sequence alignment of all possible sequence motifs and a table containing each base and its frequency at each position. For the visualization of the results they stacked the characters at each position on top of each other with the height of each letter corresponding to its entropy. In terms of our visualization taxonomy, this corresponds to a visualization of a single valued point track with features with complex attributes: each point is associated with the distribution of bases. 

One of the first web applications to visualize sequence logos was implemented by Crooks et al. called WebLogo~\cite{Crooks2004-js} (see Figure~\ref{fig:LogoBundle}a). The tool pLogo uses the same basic visual encoding~\cite{OShea2013-xh}, but it displays values of statistical significance for each base at each position. Moreover, in addition to the motif, it shows underrepresented bases in a parallel track. Two Sample Logo has been developed for the comparison of motifs~\cite{Vacic2006-hy}. It visualizes statistically significant differences between motifs in three tracks: one for each track showing overrepresented symbols and one track showing consensus symbols.

Sequence bundles represent an alternative to sequence logos~\cite{Kultys2014-xf}. \new{According to the authors describing sequence bundles, sequence logo visualization suffers from several limitations. Most importantly, these visualizations do not show relationships between residues at different positions.}
Sequence bundles display the features in an unconventional way, by encoding different bases with different positions on the y-axis. For each motif sequence of the multiple sequence alignment, a line is drawn (see Figure~\ref{fig:LogoBundle}b). At conserved positions, the lines will bundle together, while at variable positions, the lines will be more scattered. The tool Alvis is a Java implementation for the creation of sequence bundles~\cite{Schwarz2016-vu}. \new{Although sequence bundles enable visualizing longer sequences and relationships of residues at different positions, sequence logos remain the standard in visualizing motifs. A possible reason for this could be that the visual encoding is easier to interpret and the visualization is less cluttered.}

MEME is a tool for the discovery of motifs and motif visualization \cite{Bailey2006-tr}. It shows the location of found motifs in multiple input sequences as segment tracks  and additional detailed views showing sequence logos for each motif (see n-n-1 view configuration in Table \ref{tab:aggregation}). The logos are associated with strongly connected utility views showing meta data about the motifs. Each motif is encoded with a colored rectangle in the segment tracks. When hovering over a segment the detailed views of the motif is displayed at the position of the cursor.

\section{Multiple Genomic Coordinate Systems}\label{sec:MultiSequence}
\subsection{Genome-Scale Visualizations}
\begin{table*}[t]
\centering
\caption{Layouts (L) (linear \symb{layout-linear}, circular \symb{layout-circular}), Arrangements (A) ( parallel \symb{arrangement-parallel-linear}, serial \symb{arrangement-serial-linear}, orthogonal \symb{arrangement-orthogonal}) and view configurations of tools for the visualization of multi genomic coordinate systems. Tools marked with * apply a form of abstraction.}\label{tab:compVis}
\begin{tabularx}{\textwidth}{lggggh}
\toprule
\rowcolor{White}
L & A & Views & Scales & Foci &  \\
\midrule
\rowcolor{White}
\symb{layout-linear} & - & 1 & 1 & 1 & JalView~\cite{Waterhouse2009-ep}, AliView~\cite{Larsson2014-dh}, MSAViewer~\cite{Yachdav2016-hw},
IRScope~\cite{Amiryousefi2018-ua},
VistaPoint~\cite{Mayor2000-ud},
BactoGenie~\cite{Aurisano2015-qv}*,
\new{SynteBase/SynteView~\cite{Lemoine2008-il}, 3d Genome Browser~\cite{Wang2018-vi}}\\
\cmidrule{2-6}
\rowcolor{White}
& &  n &  1 & 1 &  Sequence Surveyor~\cite{Albers2011-hg}*,   Cinteny~\cite{Sinha2007-uc}, Synteny Explorer~\cite{Bryan2017-zz}, CEpBrowser~\cite{Cao2013-wt}\new{ , Vista Synteny~\cite{Frazer2004-nw}, Edgar Genome Browser~\cite{Blom2016-gk}}\\
 \rowcolor{White}
& &  n &  n & n & GBrowse\_syn~\cite{McKay2010-wi}, Persephone \cite{Persephone_Software_LLC_undated-yy},
Combo~\cite{Engels2006-es}\\
\cmidrule{2-6}
\rowcolor{White}
 & \symb{arrangement-orthogonal} &   1 & 1 &  1 &  Vista Dot~\cite{Frazer2004-nw}, Gepard~\cite{Krumsiek2007-ig}, SynMap2~\cite{Haug-Baltzell2017-lj}, Edgar Synteny Plots~\cite{Blom2016-gk}, Edgar Synteny Matrix~\cite{Blom2016-gk}, Persephone \cite{Persephone_Software_LLC_undated-yy},
 Combo~\cite{Engels2006-es}\\
 \midrule
 \rowcolor{White}
\symb{layout-circular} & - & 1 & 1 & 1 & GenomeRing~\cite{Herbig2012-bn}\\
 \rowcolor{White}
 & \symb{arrangement-serial-linear} & 1 & 1 & 1 & Circos~\cite{Krzywinski2009-yv}, ClicO Free Service~\cite{Cheong2015-lz} \\
\rowcolor{White}
 & &  n &  1 &  1 & Synteny Explorer~\cite{Bryan2017-zz} \\
 \cmidrule{2-6}
  \rowcolor{White}
 & \symb{arrangement-parallel-linear} & 1 & 1 & 1 & Edgar Circular Plot~\cite{Blom2016-gk} \\
   \rowcolor{White}
  &  & n & n & 1 & MizBee~\cite{Meyer2009-bi} \\
\bottomrule
\end{tabularx}
\end{table*}
Sequences are often compared in order to find differences between two species in a field of research called \textit{comparative genomics}.
The evolutionary forces that shape genomes work on the scale of individual nucleotides in the form of mutations, insertions, and deletions, as well as on the scale of chromosomes in the form of duplications, translocations, and inversions. Sequences can be compared on different scales from nucleotide level to whole genomes, most commonly in \new{linear and circular layouts and parallel, orthogonal and serial arrangements}. The goal is to visually encode the differences between the sequences of the different genomes.

Tools for comparisons are mostly categorized in single view (1-1-1) or multi view, single scale, single focus (n-1-1) view configurations (see Table \ref{tab:compVis}), as they show comparisons of one set of species in a single view or multiple comparisons in separate views. The only exceptions we found are GBrowse\_syn~\cite{McKay2010-wi} and Persephone~\cite{Persephone_Software_LLC_undated-yy} as well as MizBee, which is an overview-detail visualization (n-n-1) in a circular layout~\cite{Meyer2009-bi}.

\paragraph*{Linear Layouts}

\begin{figure}
    \centering
    \includegraphics[width=0.9\linewidth]{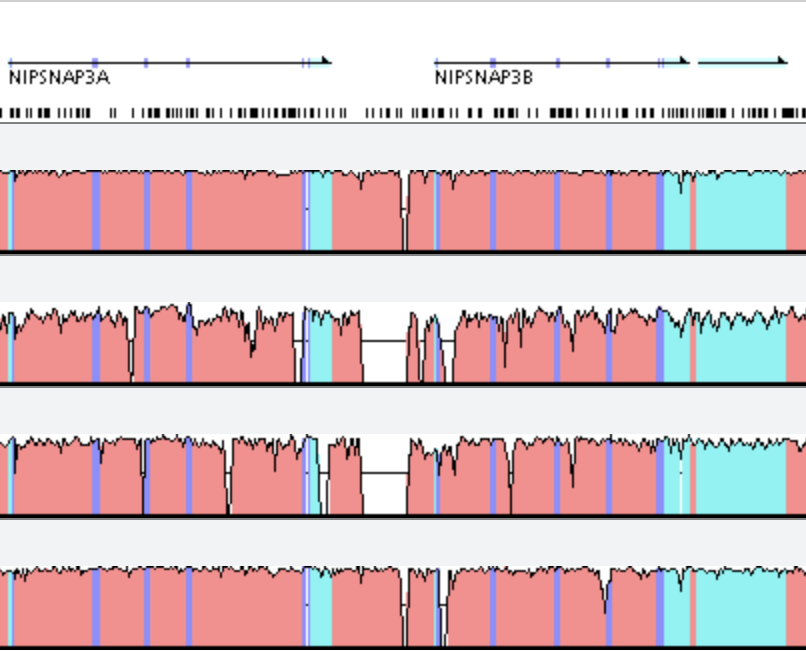}
    \caption{Multiple sequence alignment displayed with area charts using VISTA~\protect\cite{Mayor2000-ud}. Colors indicate if the conserved region belongs to an exon (dark blue), an untranslated region (light blue) or to a non-coding region (red). \protect\\$D=$ \symb{type-contiguous-point}+\symb{interconnection-none};
    $C=$ \symb{layout-linear}+\symb{partition-contiguous}+\symb{abstraction-none};
    $T=$ \symb{alignment-parallel};
    $V=$ \symb{view-one}+\symb{scale-one}+\symb{focus-one}}
    \label{fig:vista}
\end{figure}
In order to display long alignments, the online server VISTA offers tools to display alignments using line charts \cite{Mayor2000-ud}. \new{In general, an alignment is used to identify conserved genomic regions in order to estimate the evolutionary relationship of organisms or finding shared genes. In VISTA each track corresponds to a sequence.} The y-axis represents the percent identity with the reference genome computed with a window-averaged identity score. Regions with a high identity with the reference genome (conserved regions) which are part of an exon are indicated by highlighting the area under the curve dark blue, regions which are part of a non-coding part are colored red, and untranslated regions (UTRs) are colored light-blue. VISTA can display alignments of various lengths by adapting parameters, such as the zoom level, the resolution, and the minimum length of an aligned sequence in order to be displayed. \new{A disadvantage of VISTA and other long alignments is that they cannot display segments that are conserved but have changed position during genome evolution (synteny blocks).}

\begin{figure}[t]
    \centering
    \includegraphics[width=0.9\linewidth]{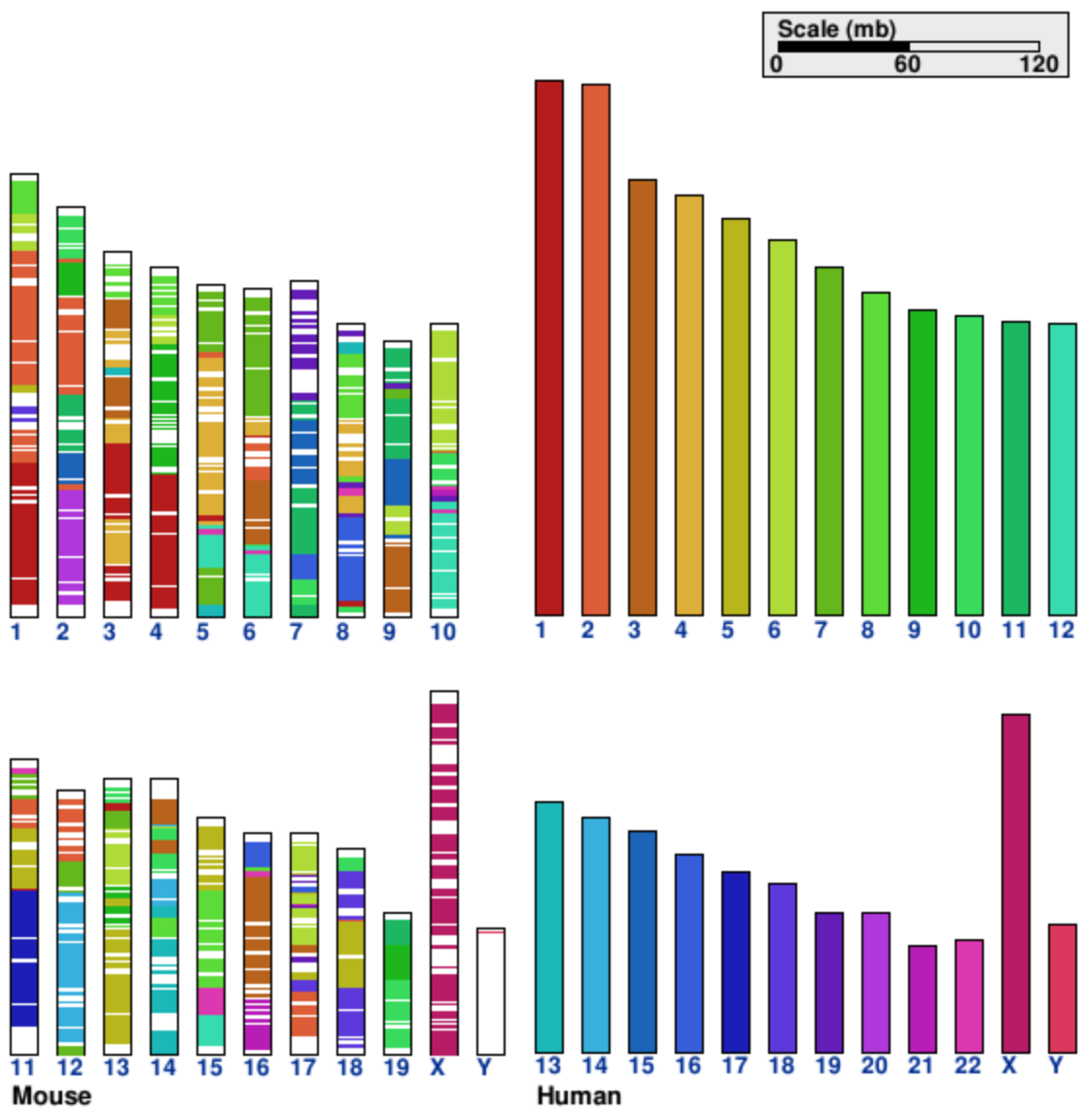}
    \caption{Comparison of the human and mouse genome using Cinteny~\protect\cite{Sinha2007-uc}. Chromosomes are segregated and arranged in a linear separate layout, color encodes for syntenic regions. \protect\\$D=$ \symb{type-contiguous-segment}+\symb{interconnection-between};
    $C=$ \symb{layout-linear}+\symb{partition-segregated}+\symb{abstraction-none}+ \symb{arrangement-parallel-linear};
    $T=$ \symb{alignment-parallel};
    $V=$ \symb{view-one}+\symb{scale-one}+\symb{focus-one}}
    \label{fig:cinteny}
\end{figure}
\new{To highlight syntenic regions}, other types of visualizations display similarities with connections via lines and bands or color. Cinteny~\cite{Sinha2007-uc} applies both approaches but on different scales. On a chromosome scale, Cinteny connects syntenic regions with lines. When the entire genome is compared, chromosomes are displayed in a separate segregated layout (see Figure \ref{fig:cinteny}). One genome acts as a reference, where each chromosome is colored differently. Regions in the chromosomes of the other genome are colored corresponding to their syntenic region in the reference. \new{This visualization demonstrates two of the shortcomings of color encodings for synteny. First, each of the human chromosomes is encoded using a different color. This is problematic since especially in the green spectrum the colors are hard to distinguish. Second, the visualization shows to which chromosome a synteny block of the mouse genome belongs, but not the exact position in that chromosome. The tool Syntenty Explorer implements this technique and two other possibilities of synteny visualization (described in the following paragraphs)~\cite{Bryan2017-zz}. To conquer the problem of identifying the exact positions of synteny blocks in both genomes, Synteny Explorer allows adding lines to connect syntenic regions for one chromosome or an entire genome. However, when lines are added for an entire genome, the visualization becomes cluttered and hard to interpret.}

\new{In addition, Synteny Explorer implements }a unique way of visualizing synteny~\cite{Bryan2017-zz} \new{by using animation to reorder syntenic regions to match the reference.} We could not find this technique in any other sequence comparison tool, presumably because it is impossible to follow the movement of all syntenic regions in a genome \new{or chromosome} at once. However, Synteny Explorer represents an educational tool and animation can illustrate the actual process of rearrangements and inversions.

A straight forward solution for visualizing \new{and comparing} multiple long genomes is to provide more screen space as done by the tool BactoGenie~\cite{Aurisano2015-qv}. BactoGenie visualizes genomic neighborhoods of bacterial strains on large displays to address scalability issues. Coding sequences are displayed as arrows, while the length and position of the sequence is preserved. Similarity is encoded using color, where similar coding sequence receive the same color.

\paragraph*{Circular Layouts}


Synteny can be visualized using Circos plots, where the two sequences are arranged in a circular combined layout. Synteny Explorer provides a visualization in this layout where synteny is shown by drawing bands between syntenic regions \cite{Bryan2017-zz}. 

\begin{figure}[t]
    \centering
    \includegraphics[width=\linewidth]{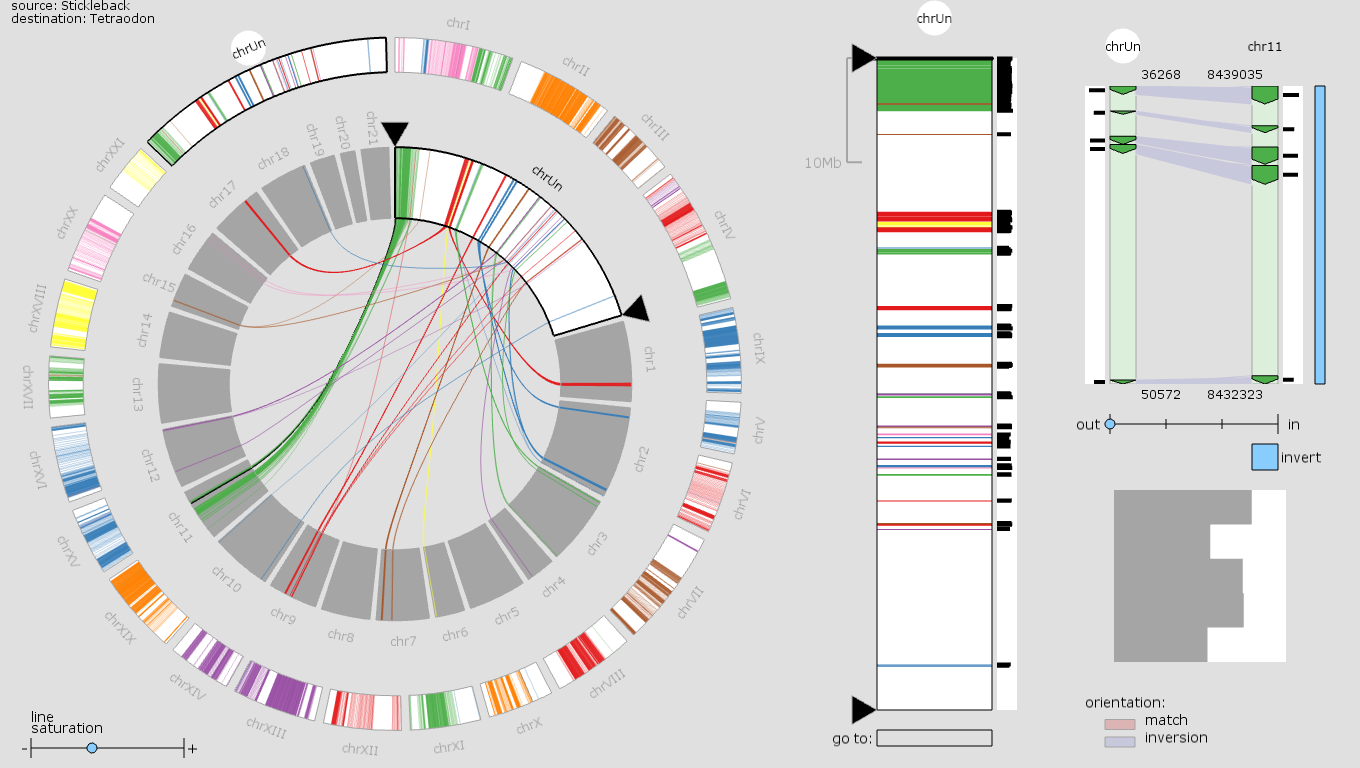}
    \caption{MizBee visualizes synteny on different scales, applying circular and linear layouts and encodings using bands and color~\protect\cite{Meyer2009-bi}.\protect\\$D=$ \symb{type-contiguous-segment}+\symb{interconnection-between};
    $C=$\symb{layout-circular}+\symb{layout-linear}+\symb{partition-contiguous}+\symb{abstraction-none}+ \symb{arrangement-parallel-linear},\symb{arrangement-parallel-circular}; $T=$ \symb{alignment-parallel}; $V=$ \symb{view-many}+\symb{scale-many}+\symb{focus-one}}
    \label{fig:Mizbee}
\end{figure}
In \new{MizBee}~\cite{Meyer2009-bi}, Meyer et al. explores synteny relationships with a visualization using linked views at the genome, chromosome, and block levels (see Figure \ref{fig:Mizbee}). MizBee represents the only comparison tool we found that applies multiple scales and multiple layouts. For the visualization of synteny on the genome level, MizBee uses a circular parallel arrangement, where one set of chromosomes is displayed in the outer ring (source chromosomes), the other set in the inner ring (destination chromosomes). A chromosome in the outer ring can be selected and inserted into the inner ring. Bands connecting the inserted source chromosome with destination chromosomes are drawn to show syntenic regions. Additionally, synteny is encoded using color. Regions and bands in the source chromosomes are colored according to their destination chromosomes. Two additional views in linear layouts show the selected source chromosome in detail, as well as a comparison of the source chromosome and a destination chromosome in a linear separated layout encoding synteny using bands. 

\paragraph*{Orthogonal Layouts}

\begin{figure}[t]
    \centering
    \includegraphics[width=0.9\linewidth]{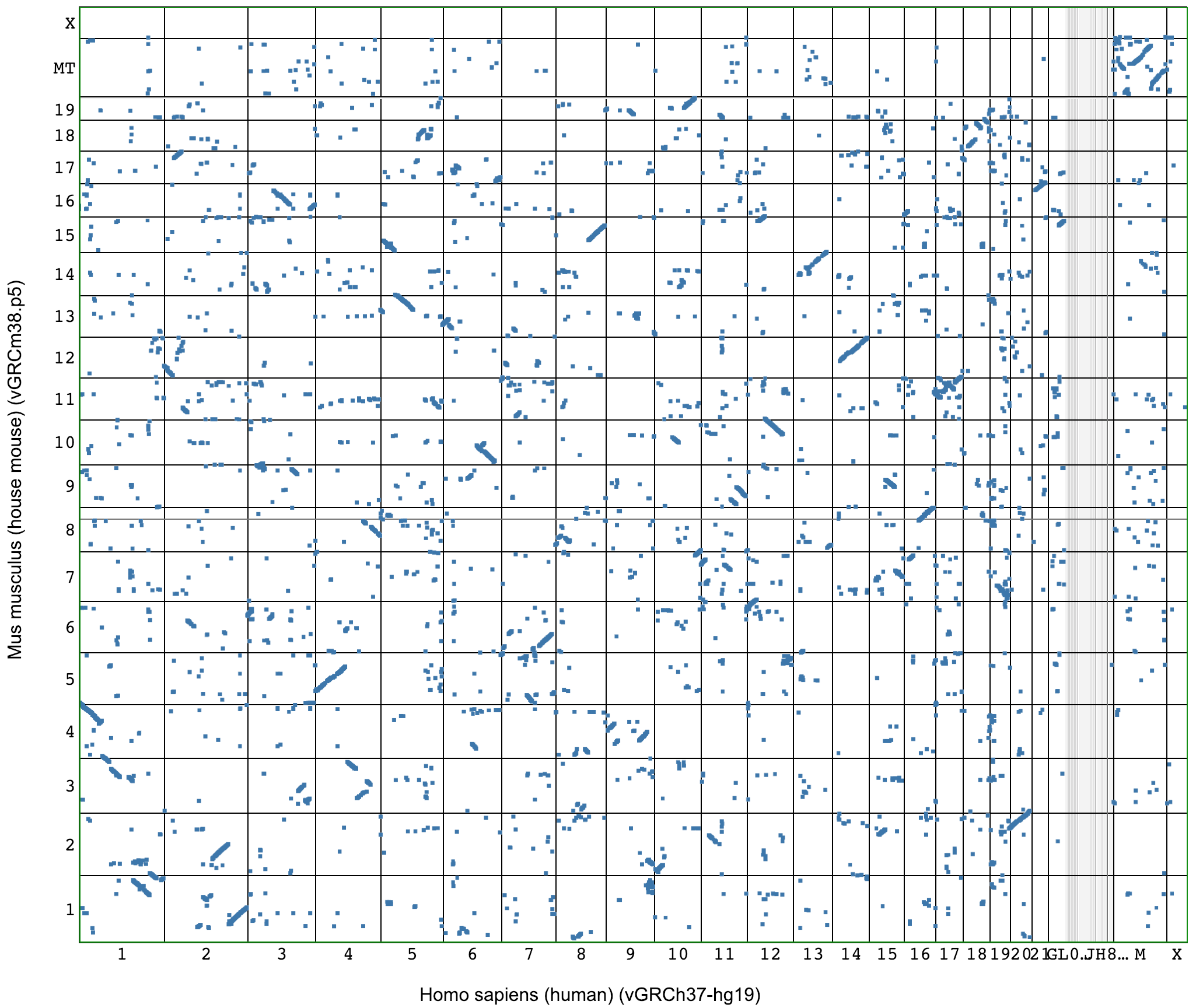}
    \caption{Dotplot created using SynMap2~\protect\cite{Haug-Baltzell2017-lj} showing a comparison of the human and the mouse genome. The sequences are aligned in an orthogonal layout. Diagonal lines of dots imply syntenic regions.\protect\\$D=$ \symb{type-contiguous-segment}+\symb{interconnection-between};
    $C=$ \symb{layout-linear}+\symb{partition-contiguous}+\symb{abstraction-none}+\symb{arrangement-orthogonal};
    $V=$ \symb{view-one}+\symb{scale-one}+\symb{focus-one}}
    \label{fig:DotPlot}
\end{figure}
In dot based approaches, similarities of sequences are indicated by diagonal rows of dots. They are used by tools like Gepard~\cite{Krumsiek2007-ig}, EDGAR~\cite{Blom2016-gk} and SynMap2~\cite{Haug-Baltzell2017-lj}. \new{These tools} distribute the data along two dimensions to show a larger quantity of relationships than possible using a single axis (see Figure \ref{fig:DotPlot}). All tools in an orthogonal layout are in a single view configuration, as most of them only show a single comparison between two genomes. Edgar Synteny Matrix~\cite{Blom2016-gk} shows multiple comparisons. However, all of them are aligned by sequence coordinates, or single view. 

\subsection{Feature-Scale Visualizations}
Comparing shorter sequences diverging by mutations, insertions and deletions is readily accomplished using alignment based visualizations. Tools such as Jalview, AliView and MSAViewer provide convenient implementations of this concept~\cite{Waterhouse2009-ep, Larsson2014-dh, Yachdav2016-hw}. This form of visualization provides a detailed view of every sequence and general patterns of similarity can be identified by applying a categorical color scale to the four nucleotides. Since computationally aligning sequences creates a shared coordinate system, most of these tools are grouped into linear layout, no axis arrangement (see Table \ref{tab:compVis}). 

\new{Not only the sequence itself but also other features can be compared across genomes at conserved regions. With the 3d Genome browser Hi-C data and epigenetic signals for biological relevant regions (such as genes) can be compared across species~\cite{Wang2018-vi}. Since the regions are conserved, tracks for the different species can simply be stacked on the same sequence axis.}

\section{Discussion}

The proposed taxonomies for data types, visualizations, and tasks span a very wide range of applications for genomic data. The most important takeaway is that despite the multi-scale nature of genomic data, not many tools take advantage of multiple linked view configurations that would support efficient navigation and pattern discovery of the space. A limitation of our taxonomy is that we do not distinguish between constant number of views, scales, and foci greater than 1 and a flexible number of views, scales, and foci. Both are currently represented as $n$ in our taxonomy and tool review. Additionally, a more detailed evaluation and assessment of visual encodings that are used in tracks could be helpful in understanding where further visualization research is warranted. The same is true for utility views that are commonly integrated into visualization tools for genomic data. Utility views can apply visualization techniques, such as node-link diagrams and reorderable matrices for genomic data that do not visualize data in sequence context. However, as discussed in the Introduction, we did not include such techniques in our taxonomy given the limited spaces in this survey. 

Our work shows that there are a lot of tools, not because there are many different visualization needs, but because the quantity of tools is driven primarily by the need to access a wide range of incompatible data formats and sources, as well as the need to integrate into common analysis workflows. These are also typically defined by the data formats that they operate on.

Our proposed taxonomies can be used to guide the development of a unified approach for visualizing genomic data, such as a grammar driven approach. However, the concerns about dependency on particular input formats, connections to analysis tools, and other external dependencies, certainly raises the question whether much would be gained by defining a generic visualization approach for genomic data. For such an effort to be successful, it would likely need to support an enormous number of data formats and sources, adding engineering overhead to the visualization challenge.

\section{Opportunities and Challenges}

The rapid progress in the development of new experimental assays to generate genomic data frequently leads to new visualization challenges and opportunities. Many challenges are related to complex genomic data such as 3D genome interactions, temporality, or scale of genomic data, in particular when the number of data sets or feature sets grows into thousands or more.

In particular, 3D genome interaction data present several unsolved challenges. In a recent review, Goodstadt et al. ~\cite{Goodstadt2017-zc} report on challenges for visualizing 3D data in genome browsers. These data are multiscale (from the few microns of the nucleus to the few nanometers of nucleosomes), and multistate (hetero, euchromatin and several other states). They can also be time-dependent and need to capture the order of events describing how the genome structure changes over time due to biological processes. These introduce uncertainty in 3D genome interaction data. Subsequently, the challenges in visualizing these data include abstractions or reductions in data dimensions, variations to show changing events, finding meaningful patterns in the data or having proper interactions. For finding patterns or classification, side-by-side comparisons are common for 2D data, although this increases the cognitive load of the viewer. For 3D data, this is even more challenging as these contain large point-sets that are spatially distributed and may change over time. In certain cases, animation can be useful, although this also provides little insight when there are a large number of changes happening concurrently. 

In addition to temporally resolved 3D genome structure data, it is now also possible to measure not only pairwise but also n-way interactions of regions across the whole genome. This will require novel visualization techniques that enable analysts to navigate this highly complex space and select lower-dimensional slices for visualization. A related challenge is the visualization and integration of imaging data (ranging from standard lightfield microscopy to super-resolution microscopy) with genomics data from chromosome conformation capture assays such as Hi-C. The mapping between image-based and genome-based coordinates and corresponding navigation will be particularly challenging in this context.

The number of genomes and patients in many studies will increase rapidly to thousands and in some cases, millions. This will require novel visualization techniques, as well as infrastructure (APIs, data formats, etc.) that allow visualization tools to access this data in an efficient and secure fashion. In the era of precision medicine, electronic health records and various types of sensor data will also need to be integrated with genomic data and this will be particularly important when viewing data from patient cohorts. These data are noisy, highly heterogeneous, and often have highly variable temporal resolution. Furthermore, driven by new insights into disease mechanisms, discovery of novel drug targets and therapeutics is an area which will benefit from the integration of visualization approaches for molecular data (e.g. MutationAssessor~\cite{Reva2011-zy}) with variant and other genomic data viewers, as well as tools that aid in the prioritization of drugs and compounds. 

There are also several technological challenges, that if addressed, would allow visualization tools to have a more profound impact in the analysis process. For example, tight integration of algorithmic and visual approaches and linking a diverse set of such tools into a coherent interface that reduces the cognitive burden of the investigator and enables seamless data flow across these tools. In particular in this context, more research is needed on evaluating genome visualization tools to guide future efforts and to develop user-centric approaches for building new visualization tools. Crisan et al.~\cite{Crisan2018-os} showed that a human-centered design approach integrating quantitative and qualitative feedback from users is important in creating successful clinical genomics reports, and suggested that this approach can also be useful in building complex bioinformatics data visualization software.
\section{Conclusion}
As our survey demonstrates, the direct impact of the data visualization field on genomic data visualization techniques has been limited to date. There are three possible explanations for the small number of publications on genomic data visualization that originated in the visualization community. First, this might be due to the size and complexity of the data that often require complex client-server architectures, resulting in engineering work that is not rewarded by the visualization community. This issue will become even more pressing in the context of visualizations for patient genomes which require special considerations of privacy and secure data access and storage. Second, genomics is a complex and rapidly evolving field which presents a steep learning curve for researchers who are not actively working in this domain. Third, as we demonstrate, the number of techniques and tasks is much smaller than the large number of published tools implies. Therefore, genomics might not be seen as a fruitful domain for visualization research projects. 

However, the fact that the majority of genomic visualization tools is developed in an \textit{ad hoc} fashion and published without formal task analyses or evaluations, is a clear sign that more visualization research is needed. On one hand, the bioinformatics community needs to become more aware of appropriate design and evaluation approaches for visualization tool development. On the other hand, the visualization community needs to be incentivized and enabled to study genomics visualization problems. This could be achieved by developing an infrastructure to provide convenient access to genomic data and to enable effective integration of new visualization tools into existing analysis frameworks or tool ecosystems. This will also allow researchers in the bioinformatics community to shift focus away from reimplementing basic functionality over and over and instead focus on visualization problems, rather than data access problems. Finally, as the first survey to comprehensively assess the landscape of genomic visualization tools, techniques, and tasks from a visualization point of view, our work should be considered a starting point for future efforts that aim at organizing the biological data visualization literature.


\section{Acknowledgements}

The authors thank Peter Kerpedjiev for his contributions to the initial list of manuscripts and tools that are discussed in this survey. Funding for this project was provided by the National Institutes of Health (R00 HG007583, U01 CA200059).
\bibliographystyle{eg-alpha-doi}
\bibliography{bibliography.bib}

\end{document}